\documentclass[%
 reprint,
 onecolumn,
 amsmath,amssymb,
 aps,
 prd,
]{revtex4-1}

\usepackage{dcolumn}% Align table columns on decimal point
\usepackage{graphics}
\usepackage{amssymb}
\usepackage{amsfonts,amsmath,bm}
\usepackage{mathrsfs}
\usepackage{marginnote}
\usepackage{ifthen}
\usepackage{setspace}

\newcommand{\mymarginnote}[1]{%
\marginnote{%
\begin{spacing}{1}
  \vspace{-\baselineskip}%
  \color{blue}\footnotesize#1
\end{spacing}
}
}
\newboolean{printnotes}
\setboolean{printnotes}{false}
\newcommand{\mytodo}[1]{%
\ifthenelse{\boolean{printnotes}}{
\mymarginnote{#1}
}{}
}

\usepackage{comment}

\usepackage{caption,subcaption}
\usepackage{graphicx}
\usepackage[x11names, rgb]{xcolor}
\usepackage[figuresright]{rotating}
\usepackage[]{pgfplots}

\usepackage{pgfplotstable}
\usepgfplotslibrary{statistics}

\newboolean{usepgf}
\setboolean{usepgf}{false}
\ifthenelse{\boolean{usepgf}}{%

\usepackage[inline]{asymptote}
\usepackage{tikz}
\usetikzlibrary{snakes,arrows,shapes,plotmarks,calc}
\usetikzlibrary{backgrounds}
\usepgfplotslibrary{fillbetween}
\usetikzlibrary{pgfplots.groupplots}
\pgfplotsset{compat=1.8}

\usepgfplotslibrary{external}
\tikzexternalize[prefix=pics/]
}{%
\usepackage{tikzexternal}
\tikzsetexternalprefix{pics/}
\tikzexternalize
}

\usepackage{xspace}

\makeatletter
\newcommand\addplotgraphicsnatural[2][]{%
\begingroup
\pgfqkeys{/pgfplots/plot graphics}{#1}%
\setbox0=\hbox{\includegraphics{#2}}%
\pgfmathparse{\wd0/(\pgfkeysvalueof{/pgfplots/plot graphics/xmax} - \pgfkeysvalueof{/pgfplots/plot graphics/xmin})}%
\let\xunit=\pgfmathresult
\pgfmathparse{\ht0/(\pgfkeysvalueof{/pgfplots/plot graphics/ymax} - \pgfkeysvalueof{/pgfplots/plot graphics/ymin})}%
\let\yunit=\pgfmathresult
\xdef\marshal{%
\noexpand\pgfplotsset{unit vector ratio={\xunit\space \yunit}}%
}%
\endgroup
\marshal
\addplot graphics[#1] {#2};
}
\makeatother
\definecolor{color0}{HTML}{272822}
\definecolor{color1}{HTML}{383830}
\definecolor{color2}{HTML}{49483E}
\definecolor{color3}{HTML}{75715E}
\definecolor{color4}{HTML}{A59F85}
\definecolor{color5}{HTML}{F8F8F2}
\definecolor{color6}{HTML}{F5F4F1}
\definecolor{color7}{HTML}{F9F8F5}
\definecolor{color8}{HTML}{F92672}
\definecolor{color9}{HTML}{FD971F}
\definecolor{color10}{HTML}{F4BF75}
\definecolor{color11}{HTML}{A6E22E}
\definecolor{color12}{HTML}{A1EFE4}
\definecolor{color13}{HTML}{66D9EF}
\definecolor{color14}{HTML}{AE81FF}
\definecolor{color15}{HTML}{CC6633}

\newcommand{\bvec}[1]{\mathbf{#1}}
\newcommand{\ve}[1]{\bvec{#1}}
\newcommand{\gradient}[1]{\cdot\nabla{#1}}
\newcommand{\divergence}[1]{\nabla\cdot{#1}}
\newcommand{\scalarprod}{\cdot}

\DeclareMathOperator*{\argmax}{arg\,max}

\newcommand{\units}[1]{\:\mathrm{#1}} % \, = 3/18\quad | \: = 4/18\quad | \; = 5/18\quad
\newcommand\Mach{\mbox{\textit{Ma}}}  % Mach number according to jfm.cls

\newcommand\TTop{\rule{0pt}{3ex}}
\newcommand\TBot{\rule[-1.5ex]{0pt}{0pt}}

\newcommand{\myincludegraphics}[2][]{%
\begin{tikzpicture}[baseline]
  \node[inner sep=0pt] at (0,0) [pos=0, anchor=south west] {\includegraphics[#1]{#2}};
\end{tikzpicture}
}%

\begin{document}

%%%%%%%%%%%%%%%%%%%%%%%%%%%%%%%%%%%%%%%%%%%%%%%%%%%%%%%%%%%%%%%%%%%%%%%%%%%%%%%
%\begin{frontmatter}

  \title{A Computational Study of the Collapse  of \\ a Cloud with $12'500$ Gas Bubbles in a Liquid}
  %\title{Numerical investigation of collapsing clouds with up to O($10^4$) gas bubbles}


  %%%%%%%%%%%%%%%%%%%%%%%%%%%%%%%%%%%%%%%%%%%%%%%%%%%%%%%%%%%%%%%%%%%%%%%%%%%%%
%  \author[cse]{U. Rasthofer}
%  %\ead{urasthofer@ethz.ch}
%  \author[cse]{F. Wermelinger}
%  %\ead{fabianw@mavt.ethz.ch}
%  \author[cse]{P. Karnakov}
%  %\ead{kpetr@ethz.ch}
%  %\author[eaw]{J. \v{S}ukys}
%  \author[cse]{J. \v{S}ukys}
%  %\ead{jonas.sukys@eawag.ch}
%  \author[cse]{P. Koumoutsakos\corref{cor}}
%  \ead{petros@ethz.ch}
%  \cortext[cor]{Corresponding author.}
%  \address[cse]{Chair of Computational Science and Engineering, ETH Zurich, Clausiusstr. 33, 8092 Zurich, Switzerland}
%  %\address[eaw]{Eawag, Swiss Federal Institute of Aquatic Science and Technology, \"Uberlandstrasse 133, 8600 D\"ubendorf, Switzerland}
  
\author{U. Rasthofer}%
\author{F. Wermelinger}
\author{P. Karnakov}
%\affiliation{%
%Chair of Computational Science and Engineering, ETH Zurich, Clausiusstr. 33, 8092 Zurich, Switzerland
%}
\author{J. \v{S}ukys}
%\affiliation{%
%present address: Eawag, Swiss Federal Institute of Aquatic Science and Technology, \"Uberlandstrasse 133, 8600 D\"ubendorf, Switzerland
%}
\author{P. Koumoutsakos}
 \email{petros@ethz.ch}
\affiliation{%
Chair of Computational Science and Engineering, ETH Zurich, Clausiusstr. 33, 8092 Zurich, Switzerland
}%

  \begin{abstract}
    We investigate the process of cloud cavitation collapse  through  large-scale simulation of a cloud composed of $12'500$ gas bubbles.
A finite volume scheme is used on a structured Cartesian grid to solve the Euler equations, and the bubbles are discretized by a diffuse interface method.  We investigate the propagation of the collapse wave front through the cloud and provide comparisons to simplified models. We analyze the flow field to identify each bubble of the cloud and its associated microjet. We find that the oscillation frequency of the bubbles and the velocity magnitude of the microjets depend on the local strength of the collapse wave and hence on the radial position of the bubbles in the cloud. At the same time, the direction of the microjets is influenced by the distribution of the bubbles in its vicinity.  Finally, an analysis of the pressure pulse spectrum shows that the pressure pulse rate is well captured by an exponential law.
  \end{abstract}
  
  \maketitle

%  \begin{keyword}
%    %% keywords here, in the form: keyword \sep keyword
%    cavitation \sep cloud collapse
%    \sep bubble dynamics \sep microjets \sep pressure pulse spectrum
%    \sep compressible multicomponent flow \sep diffuse interface method
%    %% PACS codes here, in the form: \PACS code \sep code
%
%    %% MSC codes here, in the form: \MSC code \sep code
%    %% or \MSC[2008] code \sep code (2000 is the default)
%
%  \end{keyword}
%
%\end{frontmatter}
%%%%%%%%%%%%%%%%%%%%%%%%%%%%%%%%%%%%%%%%%%%%%%%%%%%%%%%%%%%%%%%%%%%%%%%%%%%%%%%



\section{Introduction}
\label{sec:intro}

Cavitation, i.e., the growth and rapid collapse of bubbles in a liquid subjected to large pressure variations, is often associated with  damage on engineering devices such as  marine propellers, hydroelectric turbines and fuel injectors \cite{Escaler:2006,Kumar:2010,Mitroglou:2017}. At the same time, the destructive power of cavitation is harnessed for non-invasive biomedical procedures such as kidney stone lithotripsy, drug delivery and tissue ablation histotripsy \cite{Ikeda:2006,Coussios:2008,Xu:2008}. The collapse of large numbers of bubbles (``clouds'') is among the most damaging types of cavitation, resulting in  impulsive loads of high amplitude and short duration on surfaces. Such loads may cause local damage to the material known as cavitation erosion, largely attributed to the collapse of individual bubbles near the surface \cite{Kim:2014}. 

Cloud cavitation collapse has been investigated both experimentally and numerically.  Experiments have studied the collapse of a cloud of bubbles via the formation of an inward propagating shock
wave and the geometric focusing of this shock at the center of the cloud \cite{Morch:1980}.  Experimental measurements with hydrofoils
subjected to cloud cavitation, conducted in~\cite{Reisman:1998}, showed that very large pressure pulses occur within the cloud and are radiated outward during the collapse process.  A technique developed in \cite{Bremond:2006}
allowed for controlling the bubble distance within a two-dimensional cloud and thus ensured reproducibility of the cavitation events.  The study revealed the shielding effect of the outer bubbles and showed the microjet formation.  The final stage of the collapse of a hemispherical cloud near a solid surface was investigated using ultra high-speed photography in~\cite{Brujan:2011}.  Cloud
cavitation in a water jet, including erosion tests, was examined in
\cite{Yamamoto:2016}.  Various numerical studies were also reported in
literature; for instance, early ones assuming a potential flow in the liquid in~\cite{Chahine:1992,Wang:1999}.  The recently presented study
in~\cite{Ma:2015} used an Euler-Lagrange approach, combining the Navier-Stokes equations with subgrid-scale spherical bubbles governed by a
Rayleigh-Plesset-like equation, to investigate  spherical clouds collapsing near a rigid wall.  A similar approach was applied in~\cite{Chahine:2014c} to
study the impulsive loads generated by a cloud with $400$ bubbles under an
imposed oscillating pressure field.  Resolved and deforming bubbles were
considered in~\cite{Peng:2015,Adams:2013,Tiwari:2015,Sukys:2017}.  A two-dimensional simulation of the collapse of a small cluster with 7  bubbles
in an incompressible liquid using a front tracking method was presented
in~\cite{Peng:2015}.  The collapse dynamics of a cloud composed of $125$
vapor bubbles with random radii was studied in~\cite{Adams:2013} while  
~\cite{Tiwari:2015} reported the evolution of a hemispherical cloud of $50$ air bubbles.  A
comparison with the results of a homogeneous-mixture model and a coupled system
of Rayleigh-Plesset-like equations revealed that both simplified models
provided a qualitatively different prediction of the pressure field.  A recent study ~\cite{Sukys:2017} addressed uncertainty quantification for the collapse of clouds with $500$ randomly located gas
bubbles. The goal of this paper is to advance the state of the art in studies of cloud cavitation collapse by simulating thousands of bubbles and studying their collective interactions.

Numerical methods for multicomponent flow that resolve both components on the
computational grid may be classified into single-fluid and two-fluid
approaches. In two-fluid methods, each component is governed by an individual
set of conservation equations for mass, momentum and energy, and
discontinuities at the interface are treated explicitly
\cite{Fedkiw:1999,Hu:2009,Lauer:2012,Xu:2017}.  In contrast, single-fluid
methods, such as the diffuse interface method
\cite{Allaire:2002,Saurel:1999a,Saurel:2009,Tiwari:2013} introduce a  zone around each interface where the transition from one component to
the other is smeared over a few grid cells.  In this context, single-fluid models present a compromise between accuracy and computational efficiency;
that is, both components are explicitly distinguished, while the same numerical
scheme can be used throughout the computational domain.  This feature renders
diffuse interface methods particularly appropriate for the large-scale
simulation of flow problems with thousands of bubbles, as demonstrated
by the compressible multicomponent flow solver presented
in~\cite{Rossinelli:2013} which showed a throughput of up to $7\cdot 10^{11}$
computational cells per second on $96$ racks of the IBM Sequoia.

Here, we employ an extended version of this compressible multicomponent flow solver to  simulate the collapse process
of a cloud of $12'500$ resolved gas bubbles. The number of bubbles in
the present simulation is up to two orders of magnitude larger than the ones considered in previous studies.  Clouds of this size recover the
separation of scales, i.e., a cloud of large extent formed by small bubbles.
Therefore, the present cloud complies with the assumptions of simplified models for the propagation of the pressure  wave resulting from the cloud collapse.  At the same time, the large bubble count enables reliable
statistics on the behavior of the individual bubbles and their associated microjets. Furthermore, the present simulation  provides the data necessary to compile the pressure pulse spectrum of cloud cavitation collapse, in terms of the collective effect of individual bubble collapses.

The paper is organized as follows:
Sec.~\ref{sec:comput} summarizes the governing equations together with the computational method and presents
the setup of the cloud collapse problem.  Sec.~\ref{sec:cloud_dyn} reports on
the cloud collapse dynamics from a macroscopic point of view.  In
Sec.~\ref{sec:bubble_dyn}, the dynamical behavior of the bubbles and their
associated microjets are analyzed.  The generated pressure pulse spectrum is
examined in Sec.~\ref{sec:damage}.  Sec.~\ref{sec:conclu} concludes the study.

\section{Governing equations and computational approach}
\label{sec:comput}

In the following, we summarize the governing equations, the employed numerical scheme and the setup of the cloud collapse problem.
The simulation presented in this study is conducted using the open source software Cubism-MPCF \cite{Rossinelli:2013,Hadjidoukas:2015b} and~\cite{cubism_link} for download.
The reader is referred to~\cite{Rasthofer:2017} for the verification and validation of the compressible multicomponent flow solver in two-component shock-tube problems and for single-bubble collapse. A quantitative comparison to the solution of a coupled system of Rayleigh-Plesset-like equations for smaller clouds of 5 to 630 bubbles is presented  in~\cite{Rasthofer:2017b}.
 
 \subsection{Governing equations}
 \label{sec:gover_eq}

We study cloud cavitation collapse through the evolution of gas (air) bubbles in a liquid (water).  The two components (water and air) are assumed immiscible and are captured by the diffuse interface method for compressible multicomponent flows.
%
%The present investigation  involves the collapse of highly non-spherical bubbles that come along with strong microjets.  In this case, the Reynolds number and Weber number of the liquid are in the order of $10^4$
%  (see~\cite{Betney:2015} for similar considerations) such that inertia forces
%  dominate the collapse process while  viscous effects and surface tension are considered negligible.
The present investigation involves the collapse of highly non-spherical bubbles that come along with strong microjets. In the case of strong microjets, inertia forces dominate the collapse process while viscous effects and surface tension may be considered negligible; see~\cite{Betney:2015}.
  Hence, we adopt the Euler equations consisting  of the 
  mass conservation equations for each component, conservation equations
  for momentum and total energy in mixture- (or single-)fluid formulation and a
  transport equation for the volume fraction of one of the two components:
  \begin{align}
    \label{eq:mass_1}
    \frac{\partial \alpha_1\rho_1}{\partial t}
    + \divergence{\left(\alpha_1\rho_1 \ve u \right)} & = 0,\\
    \label{eq:mass_2}
    \frac{\partial \alpha_2\rho_2}{\partial t}
    + \divergence{\left(\alpha_2\rho_2 \ve u \right)} & = 0,\\
    \label{eq:momentum}
    \frac{\partial \left(\rho \ve u \right)}{\partial t} +
    \divergence{\left(\rho \ve u \otimes \ve u + p \ve I \right)} & = \ve 0,\\
    \label{eq:energy}
    \frac{\partial E}{\partial t} + \divergence{\left( \left(E+p \right)\ve
    u \right)} & = 0,\displaybreak[0]\\
    \label{eq:vol_frac}
    \frac{\partial \alpha_2}{\partial t} + \ve u\gradient{\alpha_2}
    & = K\,\divergence{\ve u},
  \end{align}
  where
  \begin{equation}
    \label{eq:var_K}
    K=\frac{\alpha_1\alpha_2(\rho_1 c_1^2 - \rho_2 c_2^2 )}{\alpha_1\rho_2
    c_2^2 + \alpha_2\rho_1 c_1^2};
  \end{equation}
  see \cite{Murrone:2005,Perigaud:2005} for derivation.  In
  Eqs.~\eqref{eq:mass_1}-\eqref{eq:vol_frac}, $\ve u$ denotes the velocity, $p$
  the pressure, $\ve I$ the identity tensor, $\rho$ the (mixture) density, $E$
  the (mixture) total energy $E=\rho e + 1/2 \rho (\ve u \scalarprod \ve u)$,
  where $e$ is the (mixture) specific internal energy.  Moreover, $\rho_k$,
  $\alpha_k$ and $c_k$ with  $k \in \lbrace 1,2 \rbrace$ are density, volume
  fraction and speed of sound of the two components. It holds that $\alpha_1 +
  \alpha_2=1$ as well as $\rho = \alpha_1\rho_1 + \alpha_2\rho_2$ and $\rho e =
  \alpha_1\rho_1e_1 + \alpha_2\rho_2e_2$ for the mixture quantities.  The
  source term on the right-hand side of the transport equation for $\alpha_2$
  was
  originally derived in~\cite{Kapila:2001} % in the context of granular materials
  and is non-zero within the diffuse interface only.  It allows for treating
  the interface zone as a compressible, homogeneous mixture of gas and liquid
  by capturing the reduction of the gas volume fraction when a compression wave
  travels across the mixing region and the increase for an expansion wave.  As
  shown in~\cite{Tiwari:2013,Rasthofer:2017}, the inclusion of this term
  notably increases the accuracy and lowers the resolution requirements.
  Moreover, it allows for a smooth transition to a homogeneous mixture model,
  if the resolution limit is reached by a collapsed bubble.

  The system of Eqs.~\eqref{eq:mass_1}-\eqref{eq:vol_frac} is closed by the
  stiffened equation of state~\cite{Menikoff:1989}:
  \begin{equation}
    p = \left(\gamma_k -1\right)\rho_k e_k - \gamma_k p_{\mathrm{c},k},
  \end{equation}
  where isobaric closure is assumed~\cite{Perigaud:2005}.
  The speed of sound is then given by
  \begin{equation}
    \label{eq:sos}
    \rho_k c_k^2 = \gamma_k \left( p + p_{\mathrm{c},k} \right).
  \end{equation}
  The material parameters $\gamma_k$ and $ p_{\mathrm{c},k}$
  are assumed constant.  Here, the values of~\cite{Tiwari:2015,Saurel:1999a}
  are used, which are given by $\gamma_1=4.4$ and $ p_{\mathrm{c},1}=6.0\cdot
  10^{2}\units{MPa}$ for water and $\gamma_2=1.4$ and $
  p_{\mathrm{c},2}=0.0\units{MPa}$ for air.

\subsection{Numerical method}

The system of governing equations \eqref{eq:mass_1}-\eqref{eq:vol_frac} is expressed in a quasi-conservative form as
 \begin{equation}
 \label{eq:cons_sys}
 \frac{\partial \bvec{Q}}{\partial t} + \divergence{\bvec F} = \bvec{R},
 \end{equation}
 where $\bvec{Q}=(\alpha_1\rho_1,\alpha_2\rho_2,\rho{\bvec{u}},E,\alpha_2)^\mathrm{T}$.
The vector
 $\bvec F=(\bvec F^{(x)}, \bvec F^{(y)}, \bvec F^{(z)})^\mathrm{T}$ combines the fluxes
 $\bvec F^{(x)}=(\alpha_1\rho_1 u_x, \alpha_2\rho_2 u_x, \rho u_x^2 + p, \rho u_y u_x, \rho u_z u_x, (E+p)u_x,\alpha_2 u_x)^\mathrm{T}$, 
 $\bvec F^{(y)}=(\alpha_1\rho_1 u_y, \alpha_2\rho_2 u_y, \rho u_x u_y, \rho u_y^2 + p, \rho u_z u_y, (E+p)u_y, \alpha_2 u_y)^\mathrm{T}$ and \linebreak
 $\bvec F^{(z)}=(\alpha_1\rho_1 u_z, \alpha_2\rho_2 u_z, \rho u_x u_z, \rho u_y u_z, \rho u_z^2 + p, (E+p)u_z, \alpha_2 u_z)^\mathrm{T}$.
 The right-hand-side vector $\bvec{R}=(0,0,0,0,0,0,(K+\alpha_2)\,\divergence{\ve u})^\mathrm{T}$ is zero
 except for the last component which comprises
 the source term of Eq.~\eqref{eq:vol_frac} and a contribution obtained from reformulating
 its convective term.

We solve  Eq.~\eqref{eq:cons_sys} %discretized in space
  using a Godunov-type finite volume method on a uniform Cartesian grid. The choice of a uniform Cartesian grid enables the exploitation of High Performance Computing (HPC) architectures \cite{Rossinelli:2013}.% with cell length $h$.
  The numerical fluxes at the cell faces are computed by an HLLC approximate Riemann solver, originally
  introduced for single-phase flow in~\cite{Toro:1994} and more recently extended to multicomponent flows in \cite{Johnsen:2006,Tiwari:2013,Coralic:2014}.
  The fluxes are based on the primitive variables $\ve u$, $p$,
  $\alpha_1 \rho_1$, $\alpha_2 \rho_2$ and $\alpha_2$  at the cell faces,
  which are reconstructed from the cell average values using a shock-capturing third-order  WENO scheme \cite{Jiang:1996}.
  Primitive variables are used for reconstruction to
  prevent numerical instabilities at the interface ~\cite{Johnsen:2006,Karni:1994}.
  The approach suggested in~\cite{Johnsen:2006} is adopted
  for the application of the HLLC Riemann solver to the evolution of $\alpha_2$.
In summary, the resulting semi-discrete system reads as
  \begin{align}
 \label{eq:ode}
 \frac{d\ve{V}(t)}{dt} &= \mathcal{L}(\ve{V}(t)),
 \end{align}
 where $\ve{V}$ denotes the vector of cell average values and $\mathcal{L}(\cdot)$
 the spatially-discrete forms of divergence and source term in Eq.~\eqref{eq:cons_sys}.
 Eq.~\eqref{eq:ode} is discretized in time by a Total Variation Diminishing (TVD), low-storage, explicit third-order Runge-Kutta scheme ~\cite{Gottlieb:1998} with a the time step dictated by the Courant-Friedrichs-Lewy (CFL) condition.
\subsection{Cloud setup}
  \label{sec:setup}

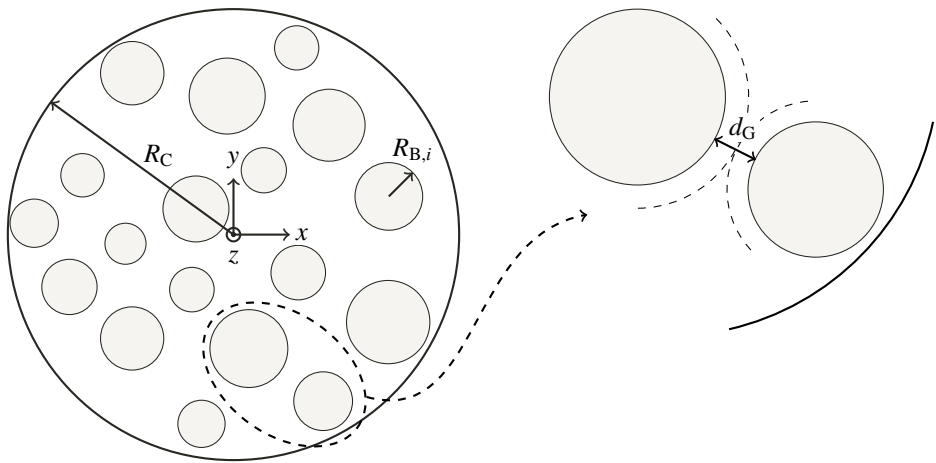
\begin{figure}[bt]
    \centering
    %
% File       : cloud_sketch.tex
% Created    : Wed Mar 22 18:31:27 2017
% Author     : Fabian Wermelinger
% Description: Sketch for cloud setup
% Copyright 2017 ETH Zurich. All Rights Reserved.
\newcommand\bubble[3]{
\fill[fill=#3, draw=color0] #1 circle (#2);
}

\newcommand\bubbletag[3]{
% \draw[->, draw=color0, thick] #1 -- ++(45:#2) node[pos=#3, below, yshift=-4pt, xshift=4pt, fill=white, inner sep=0.3pt] {$R_{\mathrm{B}}$};
\draw[->, draw=color0, thick] #1 -- ++(45:#2) node[pos=#3, below, yshift=-4pt, xshift=4pt] {$R_{\mathrm{B}}$};
}

\newcommand\bubbletagspecial[3]{
% \draw[->, draw=color0, thick] #1 -- ++(45:#2) node[pos=#3, below, yshift=-4pt, xshift=4pt, fill=white, inner sep=0.3pt] {$R_{\mathrm{B}}$};
\draw[->, draw=color0, thick] #1 -- ++(45:#2) node[pos=#3, above] {$R_{{\mathrm{B},i}}$};
}

\newcommand\bubblepos[2]{
\draw[->, draw=color0, thick] (0,0) -- #1 node[pos=#2, above, yshift=3pt, xshift=-3pt, fill=white, inner sep=0.3pt] {$r_{\mathrm{B}}$};
}

\newcommand\cloudtag[2]{
% \draw[->, draw=color0, thick] #1 -- ++(144:#2) node[midway, above right, fill=white, inner sep=0.3pt] {$R_{\mathrm{C}}$};
\draw[->, draw=color0, thick] #1 -- ++(144:#2) node[midway, above right, xshift=-3pt, yshift=-3pt] {$R_{\mathrm{C}}$};
}

\newcommand\frameofreference{
\draw[->, draw=color0, thick] (0,0)  -- (0.25cm,0) node[right, fill=white, inner sep=1.5pt] {$x$};
% \draw[->, draw=color0, thick] (0,0)  -- (0,0.25cm) node[above, fill=white, inner sep=1.5pt] {$y$};
\draw[->, draw=color0, thick] (0,0)  -- (0,0.25cm) node[above] {$y$};
%\draw[->, draw=color0, thick] (0,0)  -- (0.177cm,0.177cm) node[right] {$r$};
\draw[draw=color0, thick] (0,0) circle (0.03cm) node[below, yshift=-4pt, fill=white, inner sep=1.5pt] {$z$};
\draw[draw=color0, fill=color0] (0,0) circle (0.009cm);
}

\begin{tikzpicture}[scale=3.0]
  % \begin{scope}[shift={(30:2)}, shift={(2,0)}, shift={(0,2)}]
  % \node at (1cm,1cm) {\large \mysubfig{a}};
  % \end{scope}

  % cloud hull
  \draw[draw=color0, thick] (0,0) circle (1cm);

  % bubbles
\bubble{(-0.726911cm,-0.232040cm)}{0.122525cm}{color6}
\bubble{(0.068231cm,-0.505412cm)}{0.172586cm}{color6}
%\bubble{(0.685559cm,-0.387975cm)}{0.204655cm}{color6}
\bubble{(0.685559cm,-0.387975cm)}{0.184655cm}{color6}
\bubble{(-0.027936cm,0.614439cm)}{0.168395cm}{color6}
%\bubble{(0.688440cm,0.069362cm)}{0.149939cm}{color6}
\bubble{(0.688440cm,0.169362cm)}{0.149939cm}{color6}
\bubble{(-0.166340cm,0.113766cm)}{0.146180cm}{color6}
\bubble{(-0.448609cm,0.715945cm)}{0.140494cm}{color6}
\bubble{(-0.668975cm,0.263232cm)}{0.096160cm}{color6}
%\bubble{(0.287051cm,-0.123499cm)}{0.150888cm}{color6}
\bubble{(0.287051cm,-0.168499cm)}{0.120888cm}{color6}
\bubble{(0.423599cm,0.485233cm)}{0.159628cm}{color6}
\bubble{(-0.449026cm,-0.460577cm)}{0.140170cm}{color6}
\bubble{(-0.141834cm,-0.839256cm)}{0.104716cm}{color6}
\bubble{(0.397341cm,-0.738709cm)}{0.130199cm}{color6}
%\bubble{(-0.403275cm,0.387695cm)}{0.091771cm}{color6}
\bubble{(-0.184348cm,-0.243969cm)}{0.098718cm}{color6}
\bubble{(0.280072cm,0.828178cm)}{0.098055cm}{color6}
\bubble{(-0.883034cm,0.050521cm)}{0.107057cm}{color6}
\bubble{(0.134458cm,0.285333cm)}{0.100208cm}{color6}
\bubble{(-0.478573cm,-0.040480cm)}{0.091020cm}{color6}

  % bubble tag
  %\bubble{(0.688440cm,0.069362cm)}{0.149939cm}{color6}
  \bubbletagspecial{(0.688440cm,0.169362cm)}{0.149939cm}{1.0}
  %\bubblepos{(0.688440cm,0.169362cm)}{0.5}

  % cloud tag
  \frameofreference
  \cloudtag{(0,0)}{1cm}

  % indication node
  \node[draw, dashed, thick, ellipse, rotate=-36, align=center, minimum height=1.6cm, minimum width=2.4cm] (A) at (-70:0.66cm) {};

  % 2-bubble situation
  \begin{scope}[shift={(2.2cm,0.2cm)}]
  %  \node at (1cm,0.8cm) {\large \mysubfig{b}};

    % arrow connector
    \node (B) at (190:0.60cm) {};
    \draw[->, dashed, thick] (A) [out=0, in=190] to (B);

    % bubble left
    \bubble{(-0.410cm,0.410cm)}{0.39cm}{color6};
    %\bubbletag{(-0.410cm,0.410cm)}{0.39cm}{0.5}
    \draw[dashed, shift={(-0.410cm,0.410cm)}, shift={(45:0.493cm)}] (0,0) arc (45:-90:0.493cm);

    % bubble rigth
    \bubble{(0.380cm,0)}{0.3cm}{color6}
    \draw[dashed, shift={(0.380cm,0)}, shift={(95:0.393cm)}] (0,0) arc (95:225:0.393cm);

    % d_G
    \draw[<->, thick] (-0.07cm,0.225cm) -- (0.115cm,0.132cm) node[midway, yshift=8pt, xshift=3pt, fill=white, inner sep=1.0pt] {$d_{\mathrm{G}}$};
    % \draw[<->, thick, dashed,color8] (-0.410cm,0.410cm) -- (0.380cm,0);

    % partial cloud hull
    \draw[thick, shift={(0.9cm,0.3cm)}, rotate=-22] (0,0) arc (10:-55:1.2cm);
  \end{scope}
\end{tikzpicture}
     \caption{Sketch of spherical cloud with radius
    $R_{\mathrm{C}}$ composed of bubbles with radius $R_{\mathrm{B}}$ (left) as well as close-up of
    two bubbles separated by distance $d_{\mathrm{G}}$ (right).}
    \label{fig:cloud_sketch}
  \end{figure}

\begin{figure}[tb]
    \centering
    \begin{minipage}[c]{0.49\textwidth}
      \centering
\pgfkeys{/pgf/number format/.cd,1000 sep={\,}}
\begin{tikzpicture}[baseline]
  \begin{axis}[
    grid=major,
    width=\textwidth,
    % style={font=\large},
    xlabel=$R_\text{B}\units{[mm]}$,
    ylabel=$n_\text{B}$,
    %    legend columns=2,
    %    legend style={at={(0.6,1.02)},anchor=south,draw=none,font=\normalsize,
    %    /tikz/column 2/.style={column sep=6pt}},
    %    legend cell align=right
    ymin=0,
    xtick={0, 0.5, 0.7, 0.9, 1.1,1.3},
    ]
    \addplot [color15,fill=color15,fill opacity=0.6, thick,
    hist={
    bins=15,
    %bins=20,
    %bins=24,
    data min=0.5,
    data max=1.25
    }
    ] table [y index=4] {./pics/cloud_gen/cloud.dat};
  \end{axis}
\end{tikzpicture}
    \end{minipage}
    \hfill
    \begin{minipage}[c]{0.49\textwidth}
      \centering
      \begin{tikzpicture}[baseline]
        \node[inner sep=0pt] at (0,0) [pos=0, anchor=south west, yshift=0.95cm]
          {\includegraphics[width=5.5cm]{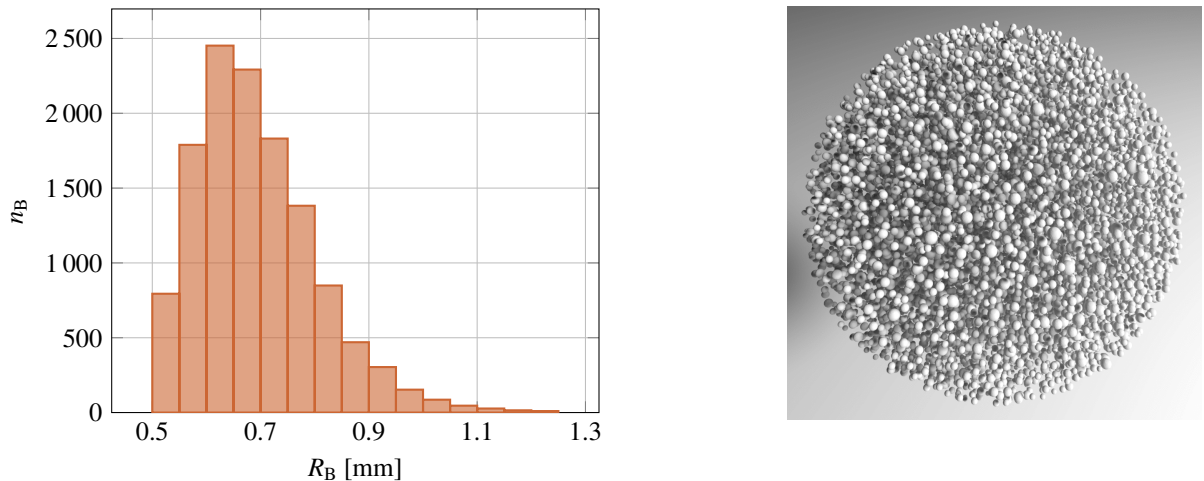}};
      \end{tikzpicture}
    \end{minipage}
    \caption{Distribution of bubble radius (left) and rendering of the initial cloud (right).}
    \label{fig:clouds}
  \end{figure}

 We investigate an initially spherical cloud of radius $R_\mathrm{C}=45\units{mm}$, composed of $n_\mathrm{B}=12'500$ spherical bubbles of radius $R_{\mathrm{B}_i}$
  with $i\in {1,...,n_\mathrm{B}}$.
  The cloud is generated by randomly positioning bubbles within a sphere of
  radius $R_\mathrm{C}$ using a uniform distribution and subject to the
  constraint that the minimum distance between the surfaces of any two bubbles
  is greater than $d_\mathrm{G}=0.4\units{mm}$.
  The radius of the bubbles is  chosen in  the range
  $[R_\mathrm{B,min},R_\mathrm{B,max}]$ using a log-normal probability distribution.
  The minimum and maximum bubble radii values, $R_\mathrm{B,min}=0.5\units{mm}$
  and $R_\mathrm{B,max}=1.25\units{mm}$, 
  are  based on  the respective values suggested in~\cite{Adams:2013,Tiwari:2015}.
  A two-dimensional sketch of the cloud setup is shown in Fig.~\ref{fig:cloud_sketch}.
  The bubble cloud is characterized by the gas volume fraction $\alpha_\mathrm{C}$ and the cloud interaction
  parameter $\beta_\mathrm{C}$, defined as
  \begin{align}
    \alpha_\mathrm{C} = \frac{1}{R_\mathrm{C}^3} \sum\limits_{i=1}^{n_\mathrm{B}} R_{\mathrm{B}_i}^3, \\
    \beta_\mathrm{C} = \alpha_\mathrm{C} \left( \frac{R_\mathrm{C}}{R_\mathrm{B,avg}} \right)^2,
  \end{align}
  where 
  \begin{equation}
    R_\mathrm{B,avg}=\frac{1}{n_\mathrm{B}} \sum_{i=1}^{n_\mathrm{B}} R_{\mathrm{B}_i}
  \end{equation}
  denotes the average bubble radius.
  Higher $\beta_\mathrm{C}$ values indicate stronger interactions among the bubbles ~\cite{Wang:1999,Brennen:1998}.
  For the present cloud, $\alpha_\mathrm{C} = 4.9\% $,  $\beta_\mathrm{C} = 208$, and $R_\mathrm{B,avg}= 0.69\units{mm} $.
  Fig.~\ref{fig:clouds} shows a histogram of the distribution of the bubble radius
  and a visualization of the generated cloud.

  The cloud is centered in a cubic computational domain
  of size $6 R_\mathrm{C} \times 6 R_\mathrm{C} \times 6 R_\mathrm{C}$.
  The domain is uniformly discretized using $6144\times 6144 \times 6144$ cells,
  leading to
  $R_\mathrm{B,min}/h=11.38$ for the minimum bubble resolution
  and $R_\mathrm{B,max}/h=28.44$ for the maximum bubble resolution, where
  the cell length is denoted by $h$.
  Initially, a zero velocity field is assumed.
  The density of water is set to $\rho_1(\ve x, t=0)=\rho_1(0)=1000.0\units{kg/m^3}$
  and of air to $\rho_2(0)=1.0\units{kg/m^3}$.
  Moreover, a smoothed initial pressure field \cite{Tiwari:2015}
  is used which is essential in order to attenuate %/prevents
  the emission of spurious pressure waves caused by the initial conditions.
  The bubble and liquid pressure in the sphere defining the cloud is set to $p_{\mathrm{C}}=0.1\units{MPa}$
  and the ambient pressure to $p_\infty=1.0\units{MPa}$.
Following ~\cite{Tiwari:2015}, the initial pressure field in the liquid
  outside of the cloud is then obtained via
  \begin{equation}
    p(\ve x,t=0) =
    \begin{cases}
      p_\mathrm{C} \quad & \text{if}\; \Vert \ve x - \ve x_\mathrm{C} \Vert \leq R_\mathrm{C}, \\
      p_\mathrm{C} +  \mathrm{tanh}\left( \frac{\Vert \ve x - \ve x_\mathrm{C} \Vert
      - R_\mathrm{C}}{\lambda}\right) \left( p_\infty - p_\mathrm{C} \right)
      \quad & \text{otherwise},
    \end{cases}
  \end{equation}
  where $\ve x_\mathrm{C}$ denotes the center of the cloud.
  Parameter $\lambda$ defines how fast the pressure
  increases from the cloud surface to the ambient and is set
  to $50\units{mm}$.
  Non-reflecting, characteristic-based conditions
  \cite{Thompson:1987,Thompson:1990,Poinsot:1992}
  are applied at the boundaries of the computational domain.
  Additionally, we impose the ambient pressure $p_\infty$ in the
  far-field by adding the term $C_\mathrm{bc} (p-p_\infty)$ to the incoming
  wave \cite{Rudy:1980}. Coefficient
  $C_\mathrm{bc}=\sigma (1-\Mach^2)c_1 / \ell \approx \sigma c_1/\ell$
  depends on a characteristic length $\ell=3R_\mathrm{C}$,
  the speed of sound $c_1$ in the liquid at the boundary, the Mach number
  $\Mach$ at the boundary, which is assumed negligible,
  and a user-defined parameter $\sigma=0.75\units{s}$. Moreover, the CFL number is set to~$0.3$.
 %%%%%%%%%%%%%%%%%%%%%%%%%%%%%%%%%%%%%%%%%%%%%%%%%%%%%%%%%%%%%%%%%%%%%%%%%%%%%%%

%
\section{Cloud collapse dynamics}
\label{sec:cloud_dyn}

In this section, the cloud cavitation collapse is examined from a macroscopic point of view without 
considering the dynamics of the individual  bubbles.
The temporal evolution of characteristic quantities is
provided together with visualizations of the collapsing cloud. Subsequently, the propagation of the collapse wave through the cloud is analyzed and compared to predictions by simplified models.

\subsection{Temporal evolution and visualizations}
\label{sec:temp_stat}

\begin{figure}[tbp]
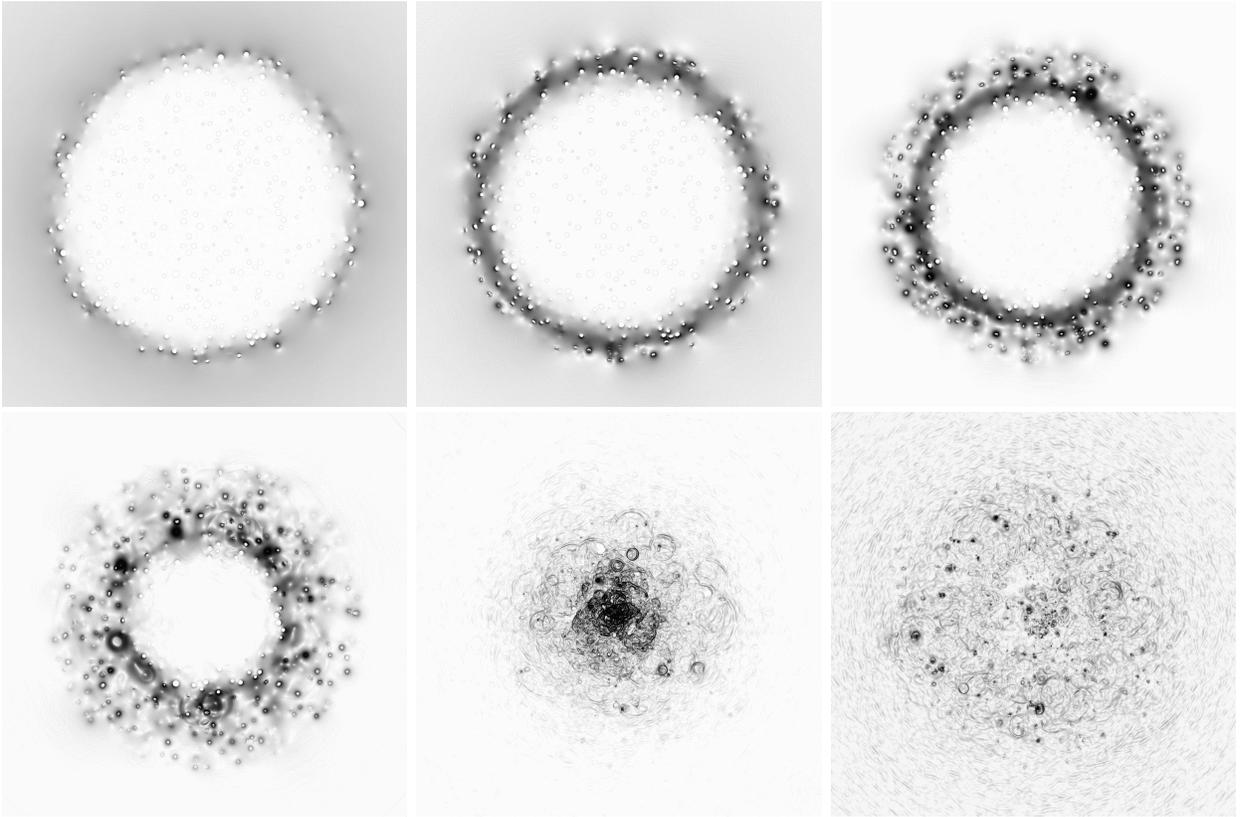

      \centering
      \resizebox{!}{0.3\textwidth}{\myincludegraphics{pics/cloud_vis/schlieren/{data_007644-pressure_slice1.h5-schlieren_8bit_comp90}.png}}
      \resizebox{!}{0.3\textwidth}{\myincludegraphics{pics/cloud_vis/schlieren/{data_015288-pressure_slice1.h5-schlieren_8bit_comp90}.png}}
      \resizebox{!}{0.3\textwidth}{\myincludegraphics{pics/cloud_vis/schlieren/{data_023288-pressure_slice1.h5-schlieren_8bit_comp90}.png}}\\[0.3ex]
      \resizebox{!}{0.3\textwidth}{\myincludegraphics{pics/cloud_vis/schlieren/{data_031109-pressure_slice1.h5-schlieren_8bit_comp90}.png}}
      \resizebox{!}{0.3\textwidth}{\myincludegraphics{pics/cloud_vis/schlieren/{data_039692-pressure_slice1.h5-schlieren_8bit_comp90}.png}}
      \resizebox{!}{0.3\textwidth}{\myincludegraphics{pics/cloud_vis/schlieren/{data_044220-pressure_slice1.h5-schlieren_8bit_comp90}.png}}
    \caption{Temporal evolution of collapsing cloud visualized using numerical schlieren images of the pressure field in the $xy$-plane at $z=0$.}
    \label{fig:schlieren}
  \end{figure}

We quantify the cloud collapse process through  the temporal evolution
  of a number of local and global quantities.
  Fig.~\ref{fig:cloud_temp_stat_12500} shows
  the development of the gas volume $V_2/V_2(0)$,
  the point-wise maximum pressure $p_\mathrm{max}/p_\mathrm{peak}$ within the computational domain,
  the average pressure $p_\mathrm{C}/p_\mathrm{C,peak}$ within the cloud,
  the average pressure $p_\mathrm{S}/p_\mathrm{S,peak}$ within a sensor at the center of the cloud,
  further described below, and
  the total kinetic energy $E_\mathrm{kin,C}/E_\mathrm{kin,C,peak}$ within the cloud.
  All quantities are normalized by their peak (i.e., maximum) values.
  The symbols on top of the curve for the gas volume coincide with the time instants for
  which three-dimensional visualizations of the cloud together with the pressure iso-surface at 
  $p_\text{iso}=0.15\units{MPa}$ are shown in Fig.~\ref{fig:3d_vis}
  and numerical schlieren of the pressure field in the
  $xy$-plane at $z=0$ in Fig.~\ref{fig:schlieren}.
  The last two symbols correspond to the time of peak pressure $p_\mathrm{S,peak}$
  within the sensor and the time of minimum gas volume, respectively. The remaining symbols
  are spaced evenly between $t=0$ and the time of occurrence of $p_\mathrm{S,peak}$.
\begin{figure}[bt]
  \centering
\pgfkeys{/pgf/number format/.cd,1000 sep={\,}}
\begin{tikzpicture}[baseline]
\begin{axis}[
  grid=major,
  width=0.49\textwidth,
  %style={font=\large},
  xmin=0, %xmax=20,
  ymin=0,
  xlabel=$t\units{[\mu s]}$,
  legend columns=3,
  legend style={at={(0.5,1.02)},anchor=south,draw=none,font=\small,
  /tikz/column 2/.style={column sep=10pt},
  /tikz/column 4/.style={column sep=10pt}},
  legend cell align=left,
  try min ticks=5,
  ]
  
   \addplot[mark=none, color0, thick, densely dashed] table [x index=0, y index=1]
  {./pics/cloud_vis/cloud_bubbles_12500_beta_28_pratio_10_normalized_stat_time.dat};
  \addlegendentry{$\frac{V_2}{V_2(0)}$}

  \addplot[mark=none, color8, thick] table [x index=0, y index=2]
  {./pics/cloud_vis/cloud_bubbles_12500_beta_28_pratio_10_normalized_stat_time.dat};
  \addlegendentry{$\frac{p_\text{max}}{p_\text{peak}}$}
  
   \addplot[mark=none, color15, thick, dash dot dot] table [x index=0, y index=5]
  {./pics/cloud_vis/cloud_bubbles_12500_beta_28_pratio_10_normalized_stat_time.dat};
  \addlegendentry{$\frac{E_\text{kin,C}}{E_\text{kin,C,peak}}$}
  
  \addplot[only marks, mark=*, mark size=2.0pt, color0,mark options={solid,thick,fill=color0}]
  coordinates {(5.692100e+01, 0.945069424583)};
  \addplot[only marks, mark=star, mark size=3.0pt, color0,mark options={solid,very thick,fill=color0}]
  coordinates {(1.138420e+02, 0.792600169516)};
  \addplot[only marks, mark=pentagon*, mark size=2.5pt, color0,mark options={solid,thick,fill=color0}]
  coordinates {(1.723441e+02, 0.590988833546)};
  \addplot[only marks, mark=diamond*, mark size=2.5pt, color0,mark options={solid,thick,fill=color0}]
  coordinates {(2.292651e+02, 0.374298990574)};
  \addplot[only marks, mark=square*, mark size=2.0pt, color0,mark options={solid,thick,fill=color0}]
  coordinates {(3.099032e+02, 0.1443572)};
  \addplot[only marks, mark=triangle*, mark size=2.0pt, color0,mark options={solid,thick,fill=color0}]
  coordinates {(3.431071e+02, 0.1194119)};

\end{axis}
\end{tikzpicture}
   \hfill
\pgfkeys{/pgf/number format/.cd,1000 sep={\,}}
\begin{tikzpicture}[baseline]
\begin{axis}[
grid=major,
width=0.49\textwidth,
%style={font=\large},
xmin=0,
ymin=0,
xlabel=$t\units{[\mu s]}$,
legend columns=3,
legend style={at={(0.5,1.02)},anchor=south,draw=none,font=\small,
/tikz/column 2/.style={column sep=10pt},
/tikz/column 4/.style={column sep=10pt}},
legend cell align=left,
try min ticks=5,
]

\addplot[mark=none, color0, thick,densely dashed] table [x index=0, y index=1]
{./pics/cloud_vis/cloud_bubbles_12500_beta_28_pratio_10_normalized_stat_time.dat};
\addlegendentry{$\frac{V_2}{V_2(0)}$}

\addplot[mark=none, color8, thick] table [x index=0, y index=4]
{./pics/cloud_vis/cloud_bubbles_12500_beta_28_pratio_10_normalized_stat_time.dat};
\addlegendentry{$\frac{p_\text{S}}{p_\text{S,peak}}$}

\addplot[mark=none, color15, thick, dash dot dot] table [x index=0, y index=3]
{./pics/cloud_vis/cloud_bubbles_12500_beta_28_pratio_10_normalized_stat_time.dat};
\addlegendentry{$\frac{p_\text{C}}{p_\text{C,peak}}$}

\addplot[only marks, mark=*, mark size=2.0pt, color0,mark options={solid,thick,fill=color0}]
coordinates {(5.692100e+01, 0.945069424583)};
\addplot[only marks, mark=star, mark size=3.0pt, color0,mark options={solid,very thick,fill=color0}]
coordinates {(1.138420e+02, 0.792600169516)};
\addplot[only marks, mark=pentagon*, mark size=2.5pt, color0,mark options={solid,thick,fill=color0}]
coordinates {(1.723441e+02, 0.590988833546)};
\addplot[only marks, mark=diamond*, mark size=2.5pt, color0,mark options={solid,thick,fill=color0}]
coordinates {(2.292651e+02, 0.374298990574)};
\addplot[only marks, mark=square*, mark size=2.0pt, color0,mark options={solid,thick,fill=color0}]
coordinates {(3.099032e+02, 0.1443572)};
\addplot[only marks, mark=triangle*, mark size=2.0pt, color0,mark options={solid,thick,fill=color0}]
coordinates {(3.431071e+02, 0.1194119)};

\end{axis}
\end{tikzpicture}
   \caption{
  Temporal evolution of gas volume $V_2/V_2(0)$ together with
  point-wise maximum pressure $p_\mathrm{max}/p_\mathrm{peak}$ within domain and
  average kinetic energy $E_\mathrm{kin,C}/E_\mathrm{kin,C,peak}$ within cloud (left)
  as well as $V_2/V_2(0)$ together with
  average pressure $p_\mathrm{C}/p_\mathrm{C,peak}$ within cloud and
  average pressure $p_\mathrm{S}/p_\mathrm{S,peak}$ within sensor at cloud center (right).
  All quantities are normalized by their peak values.
  Symbols mark time instants for three-dimensional visualizations and numerical schlieren.
  }
  \label{fig:cloud_temp_stat_12500}
\end{figure}
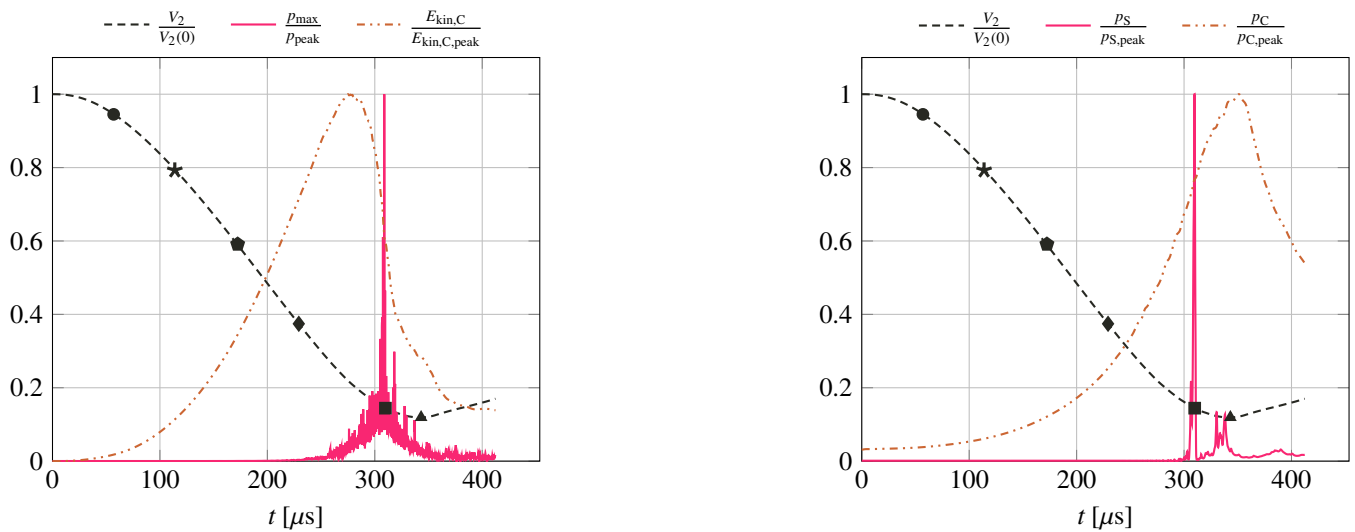
  \begin{figure}[tbp]
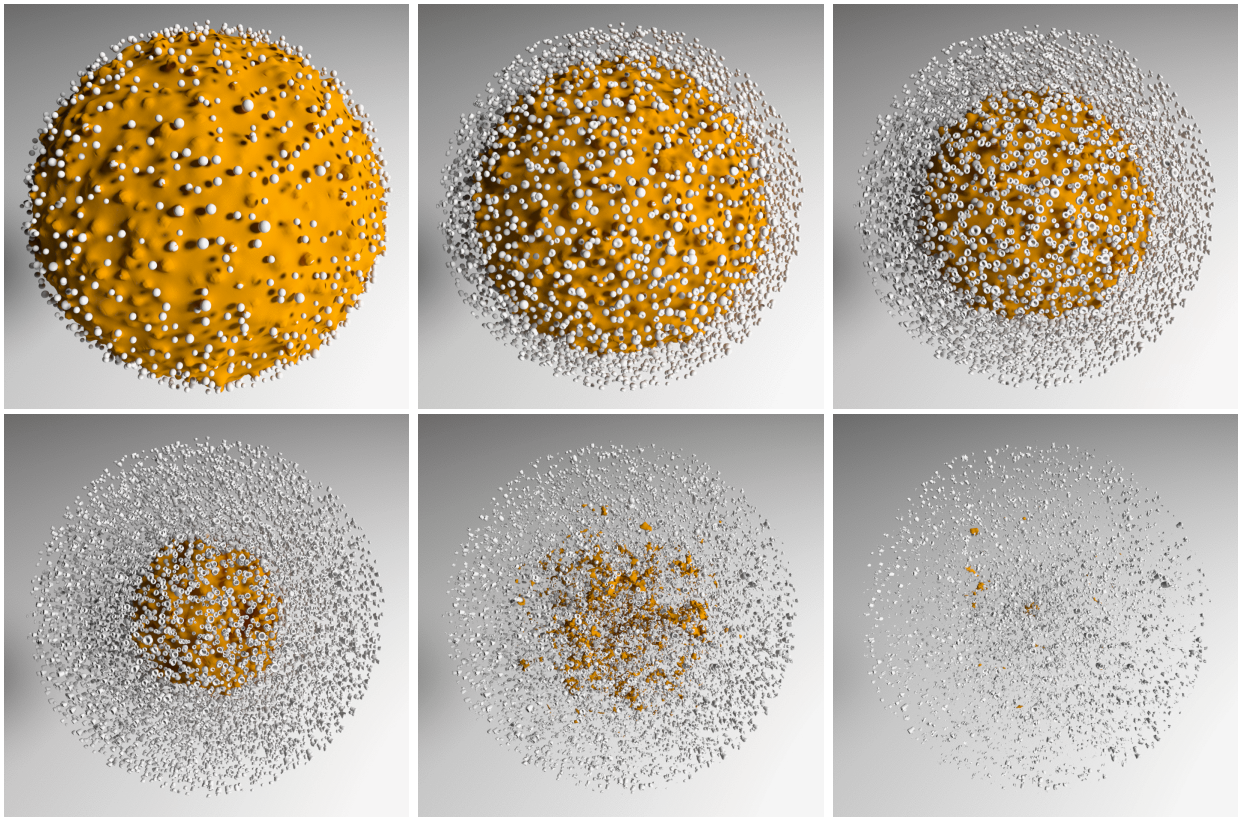

      \centering
      \resizebox{!}{0.3\textwidth}{\myincludegraphics{pics/cloud_vis/3D/{datawavelet007644_StreamerGridPointIterative_channel5_cubeScale3_isoval0_5_8bit_comp90}.png}}
      \resizebox{!}{0.3\textwidth}{\myincludegraphics{pics/cloud_vis/3D/{datawavelet015288_StreamerGridPointIterative_channel5_cubeScale3_isoval0_5_8bit_comp90}.png}}
      \resizebox{!}{0.3\textwidth}{\myincludegraphics{pics/cloud_vis/3D/{datawavelet023288_StreamerGridPointIterative_channel5_cubeScale3_isoval0_5_8bit_comp90}.png}}\\[0.3ex]
      \resizebox{!}{0.3\textwidth}{\myincludegraphics{pics/cloud_vis/3D/{datawavelet031109_StreamerGridPointIterative_channel5_cubeScale3_isoval0_5_8bit_comp90}.png}}
      \resizebox{!}{0.3\textwidth}{\myincludegraphics{pics/cloud_vis/3D/{datawavelet039692_StreamerGridPointIterative_channel5_cubeScale3_isoval0_5_8bit_comp90}.png}}
      \resizebox{!}{0.3\textwidth}{\myincludegraphics{pics/cloud_vis/3D/{datawavelet044220_StreamerGridPointIterative_channel5_cubeScale3_isoval0_5_8bit_comp90}.png}}
      \caption{Temporal evolution of collapsing cloud with pressure iso-surface at $p_\text{iso}=0.15\units{MPa}$.}
    \label{fig:3d_vis}
  \end{figure}

  The minimum gas volume is reached at time $t_\mathrm{C}=343.9\units{\mu s}$, which is
  referred to as the cloud collapse time in the following.  At this time, the gas volume
  is reduced by $88$\% compared to its initial value.
  The point-wise maximum pressure $p_\mathrm{max}$ is a highly fluctuating quantity.
  Its peak $p_\mathrm{peak}=3.41\units{GPa}$ is detected at time $t/t_\mathrm{C}=0.898$
  and occurs before the minimum gas volume is encountered.
  A similar observation was made  in~\cite{Yamamoto:2016}.
  To capture the behavior in the core of the cloud, we center a spherical pressure sensor
  of radius $R_\mathrm{S}=1\units{mm}$ at the center of the cloud.
  The sensor measures the average pressure $p_\mathrm{S}$ over its domain.
  The maximum value of $p_\mathrm{S}$ amounts to $p_\mathrm{S,peak} =89.5\units{MPa}$
  and is observed at time $t/t_\mathrm{C}=0.901$.
  The pressure curve of the sensor reveals the shielding effect
 \cite{dAgostino:1989, Brennen:2005}
  of the outer bubbles in the cloud.
  Although a broad time interval of high pressures is observed for $p_\mathrm{max}$,
  merely the major peak and one smaller peak are detected by the sensor.
  Strong pressure waves emitted away from the immediate surrounding of the sensor
  are absorbed by bubbles between the source of the pressure wave and the sensor
  by contributing to the compression of these bubbles.
 The maximum value of the average pressure within the cloud is $p_\mathrm{C,peak}=3.69\units{MPa}$ and
 significantly smaller than $p_\mathrm{S,peak}$. Furthermore, it is encountered at a later time
 $t/t_\mathrm{C}= 1.021$, which is almost exactly the time of minimum gas volume.
  The kinetic energy of the mixture in the cloud region increases until
  it reaches its peak value of $E_\mathrm{kin,C,peak} = 3.69\units{J}$ at $t/t_\mathrm{C}=0.800$,
  which is before the occurrence of $p_\mathrm{peak}$. At time $t_\mathrm{C}$,
  the kinetic energy is already reduced by $72$\%.
  
  Fig.~\ref{fig:3d_vis} illustrates the deformation of the bubbles, which is caused by the formation of microjets. As the collapse of the cloud progresses, the extracted pressure iso-surface is moving inward.
  Accordingly, an evolving circular front is detected by the numerical schlieren of the pressure field
  shown in Fig.~\ref{fig:schlieren}.
  Figs.~\ref{fig:3d_vis} and~\ref{fig:schlieren} thus reveal an inward-propagating
  spherical collapse wave and the aforementioned shielding effect. While the bubbles behind
  the front are subject to a collapse process, bubbles ahead of the front remain at their initial state.
  From the fourth to the fifth frame, a break-down of the shielding effect is observed.
  Furthermore, strong spherical pressure waves emitted from individual bubble collapses are
  clearly visible in the fifth numerical schlieren frame.
 %\clearpage
% 3.2 collapse wave propagation
%
\subsection{Collapse wave propagation}
\label{sec:wave}

\begin{figure}[bt]
  \centering
\pgfkeys{/pgf/number format/.cd,1000 sep={\,}}
\begin{tikzpicture}[baseline]
\tikzset{mark size=1.5}
\begin{axis}[
  grid=major,
  height=0.37\textwidth,
  %style={font=\normalsize},
  xmin=0, xmax=335,
  ymin=0, ymax=47.5,
  xlabel=$t\units{[\mu s]}$,
  ylabel=$r\units{[mm]}$,
  colorbar,
  colorbar style={
  ylabel= {$\frac{\bar{\alpha_2}}{\alpha_\mathrm{C}}$},
  },
  %colormap={mine}{rgb255=(166,226,46);rgb255=(249,38,114)},
  colormap={mine}{rgb255=(165, 159, 133);rgb255=(249, 248, 245)},
  %colormap={mine}{rgb255=(117, 113, 94);rgb255=(249, 248, 245)},
  %colormap={mine}{rgb255=(73, 72, 62);rgb255=(249, 248, 245)},
  domain = 1:2,
  view = {0}{90},
  xtick={0, 50, 100, 150, 200, 250, 300},
  ytick={0, 10, 20, 30, 40},
  ]
  
  %\addplot3[scatter,surf,z buffer=sort]
  \addplot3[surf,shader=interp]
  %table[x index=0,y index=1,z index=3] {./pics/cloud12500_ascii_profile_trajectory/test.dat};
  table[x index=0,y index=1,z index=2] {./pics/cloud_wave/data/cloud12500__contour_combinded.dat};

\end{axis}

\begin{axis}[
  grid=major,
  height=0.37\textwidth,
  %style={font=\normalsize},
  xmin=0, xmax=335,
  ymin=0, ymax=47.5,
  hide x axis,
  hide y axis,
  legend columns=3,
  legend style={at={(0.5,1.02)},anchor=south,draw=none,font=\small,
  /tikz/column 2/.style={column sep=6pt},
  /tikz/column 4/.style={column sep=6pt}},
  legend cell align=left,
  domain = 1:2,
  view = {0}{90},
  ]
  
   \addplot[mark=*, color8, thick] table [x index=0, y index=1]
  {./pics/cloud_wave/data/cloud12500__trajectory_cubism.dat};
  \addlegendentry{bubbles}

  \addplot[mark=none, color0, thick,dashed] table [x index=0, y index=1]
  {./pics/cloud_wave/data/cloud12500__trajectory_morch05.dat};
  \addlegendentry{M\o rch}
  
   \addplot[mark=square*, color0, thick,solid] table [x index=0, y index=1]
  {./pics/cloud_wave/data/cloud12500__trajectory_homo.dat};
  \addlegendentry{mixture}
 
\end{axis}
\end{tikzpicture}
  \hfill%
\pgfkeys{/pgf/number format/.cd,1000 sep={\,}}
\begin{tikzpicture}[baseline]
\tikzset{mark size=1.5}
\begin{axis}[
  grid=major,
  height=0.37\textwidth,
  %style={font=\normalsize},
  xmin=0, %xmax=0.08,
  ymin=0, ymax=875,
  xlabel=$t\units{[\mu s]}$,
  ylabel=$\dot R_\mathrm{F}\units{[m/s]}$,
  legend columns=3,
  legend style={at={(0.5,1.02)},anchor=south,draw=none,font=\small,
  /tikz/column 2/.style={column sep=6pt},
  /tikz/column 4/.style={column sep=6pt}},
  legend cell align=left,
  try min ticks=6,
  ]

  \addplot[mark=*, color8, thick] table [x index=0, y index=1]
  {./pics/cloud_wave/data/cloud12500__speed_cubism.dat};
  \addlegendentry{bubbles}

  \addplot[mark=none, color0, thick,dashed] table [x index=0, y index=1]
  {./pics/cloud_wave/data/cloud12500__speed_morch05.dat};
  \addlegendentry{M\o rch}
  
  \addplot[mark=square*, color0, thick] table [x index=0, y index=1]
  {./pics/cloud_wave/data/cloud12500__speed_homo.dat};
  \addlegendentry{mixture}

\end{axis}
\end{tikzpicture}
   \caption{Front trajectory of collapse wave on $\bar\alpha_2$ contour plot (left) and front speed (right).
  Results obtained with the M\o rch model and a homogeneous mixture approach are included for
  comparison.}
  \label{fig:trajectory12500}
\end{figure}
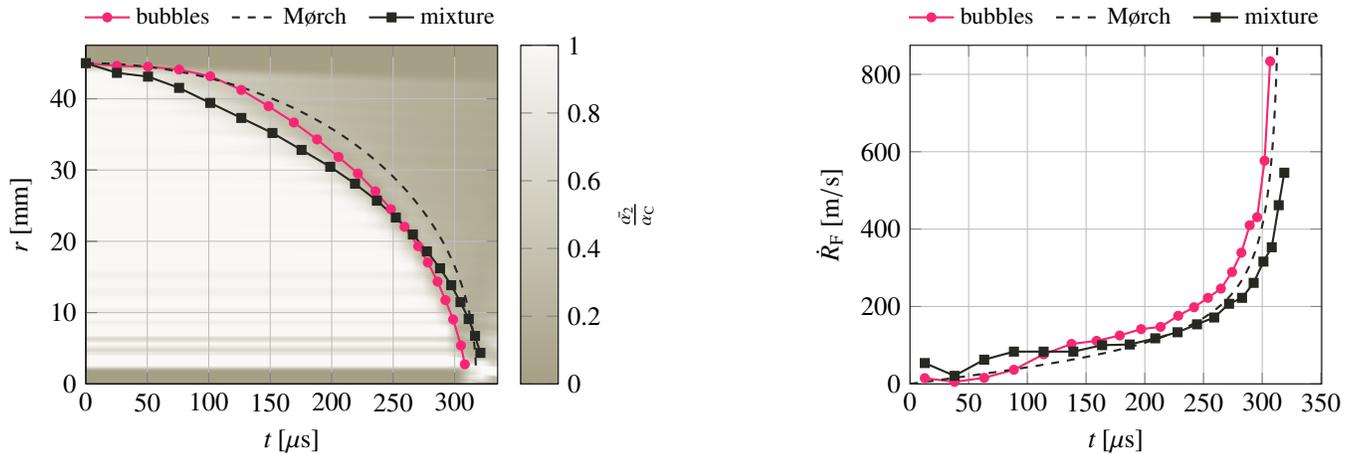

The large number of bubbles in the cloud
renders the macroscopic flow spherically symmetric
and allows for analyzing the collapse wave observed in the
previous section.
Therefore, spherical averages $\bar\alpha_2(r,t)$, $\bar p(r,t)$ and $\bar u(r,t)$ 
of the gas volume fraction, the pressure and the velocity magnitude
are computed over spheres with radius $r$ centered at the cloud center.
The radial position of the collapse wave front is defined by
the location of the maximum average velocity magnitude as
\begin{equation}
  \label{eqn:front_criterion}
  R_\mathrm{F}(t) = \argmax_r \bar u(r,t).
\end{equation}
Fig.~\ref{fig:trajectory12500} shows the front trajectory in the $r$-$t$-space
on top of a contour plot of $\bar\alpha_2(r,t)$ as well as the evolution of the
front speed $\dot R_\mathrm{F}$.
Apart from these curves, labeled ``bubbles'', predictions by simplified models
which are further addressed below are also included.
The propagation of the front starts immediately. The front gradually accelerates
so that the front speed reaches
$100\units{m/s}$ at $t=150\units{\mu s}$ and $200\units{m/s}$ at $t=240\units{\mu s}$.  
These velocities are lower than the speed of sound in both pure fluids
which amounts to $1625\units{m/s}$ for water and 
to $374\units{m/s}$ for air under pressure $p_\mathrm{C}=0.1\units{MPa}$. 
Eventually, the front  reaches the speed of sound of air
at approximately $t=270\units{\mu s}$.
At about the same time, the kinetic energy of the mixture in the cloud
starts to decrease and pressure disturbances 
penetrate the front despite the shielding effect; see Fig.~\ref{fig:cloud_temp_stat_12500}.

\begin{figure}[btp]
  \centering
\pgfkeys{/pgf/number format/.cd,1000 sep={\,}}
\begin{tikzpicture}[baseline]
\begin{axis}[
  grid=major,
  width=0.45\textwidth,
  style={font=\normalsize},
  xmin=-35, xmax=35,
  ymin=0, ymax=1.23,
  xlabel=$r-R_\text{F}\units{[mm]}$,
  ylabel=$\frac{\bar\alpha_2}{\alpha_\text{C}}$,
  try min ticks=7,
  ]

  \addplot[mark=none, color0, thick] table [x index=0, y index=1]
  {./pics/cloud_wave/data/cloud12500_wavelet__profile_a2_0.dat};

  \addplot[mark=none, color8, thick] table [x index=0, y index=1]
  {./pics/cloud_wave/data/cloud12500_wavelet__profile_a2_1.dat};
  
    \addplot[mark=none, color9, thick] table [x index=0, y index=1]
  {./pics/cloud_wave/data/cloud12500_wavelet__profile_a2_2.dat};
  
    \addplot[mark=none, color11, thick] table [x index=0, y index=1]
  {./pics/cloud_wave/data/cloud12500_wavelet__profile_a2_3.dat};
  
    \addplot[mark=none, color13, thick] table [x index=0, y index=1]
  {./pics/cloud_wave/data/cloud12500_wavelet__profile_a2_4.dat};
  
    \addplot[mark=none, color14, thick] table [x index=0, y index=1]
  {./pics/cloud_wave/data/cloud12500_wavelet__profile_a2_5.dat};
  
    \addplot[mark=none, color15, thick] table [x index=0, y index=1]
  {./pics/cloud_wave/data/cloud12500_wavelet__profile_a2_6.dat};
  
   \node[anchor=east] (A) at (axis cs:0, 0.08){increasing $t$};
   \node[anchor=west] (B) at (axis cs:34, 0.08){};
   \draw[->,thick] (A) -- (B);

\end{axis}
\end{tikzpicture}
  \hfill%
\pgfkeys{/pgf/number format/.cd,1000 sep={\,}}
\begin{tikzpicture}[baseline]
\begin{axis}[
  grid=major,
  width=0.45\textwidth,
  style={font=\normalsize},
  xmin=-35, xmax=35,
  ymin=0, ymax=1.23,
  xlabel=$r-R_\text{F}\units{[mm]}$,
  ylabel=$\frac{\bar\alpha_2}{\alpha_\text{C}}$,
  try min ticks=7,
%      yticklabel style={
%            /pgf/number format/fixed,
%            /pgf/number format/precision=2,
%            /pgf/number format/fixed zerofill
%        },
%        scaled y ticks=false
%  legend columns=2,
%  legend style={at={(0.5,1.02)},anchor=south,draw=none,font=\normalsize,
%  /tikz/column 2/.style={column sep=6pt}},
%  legend cell align=left
  ]

  \addplot[mark=none, color0, thick] table [x index=0, y index=1]
  {./pics/cloud_wave/data/cloud12500_homo_slice__profile_a2_0.dat};

  \addplot[mark=none, color8, thick] table [x index=0, y index=1]
  {./pics/cloud_wave/data/cloud12500_homo_slice__profile_a2_1.dat};
  
    \addplot[mark=none, color9, thick] table [x index=0, y index=1]
  {./pics/cloud_wave/data/cloud12500_homo_slice__profile_a2_2.dat};
  
    \addplot[mark=none, color11, thick] table [x index=0, y index=1]
  {./pics/cloud_wave/data/cloud12500_homo_slice__profile_a2_3.dat};
  
    \addplot[mark=none, color13, thick] table [x index=0, y index=1]
  {./pics/cloud_wave/data/cloud12500_homo_slice__profile_a2_4.dat};
  
    \addplot[mark=none, color14, thick] table [x index=0, y index=1]
  {./pics/cloud_wave/data/cloud12500_homo_slice__profile_a2_5.dat};
  
    \addplot[mark=none, color15, thick] table [x index=0, y index=1]
  {./pics/cloud_wave/data/cloud12500_homo_slice__profile_a2_6.dat};
  
   \node[anchor=east] (A) at (axis cs:0, 0.08){};
   \node[anchor=west] (B) at (axis cs:34, 0.08){};
   \draw[->,thick] (A) -- (B);

\end{axis}
\end{tikzpicture}
\\
\pgfkeys{/pgf/number format/.cd,1000 sep={\,}}
\begin{tikzpicture}[baseline]
\begin{axis}[
  grid=major,
  width=0.45\textwidth,
  style={font=\normalsize},
  xmin=-35, xmax=35,
  ymin=0, ymax=1.35,
  xlabel=$r-R_\text{F}\units{[mm]}$,
  ylabel=$\frac{\bar p-p_\text{C}}{p_\text{F}-p_\text{C}}$,
  try min ticks=7,
  ]

  \addplot[mark=none, color0, thick] table [x index=0, y index=1]
  {./pics/cloud_wave/data/cloud12500_wavelet__profile_p_0.dat};

  \addplot[mark=none, color8, thick] table [x index=0, y index=1]
  {./pics/cloud_wave/data/cloud12500_wavelet__profile_p_1.dat};
  
    \addplot[mark=none, color9, thick] table [x index=0, y index=1]
  {./pics/cloud_wave/data/cloud12500_wavelet__profile_p_2.dat};
  
    \addplot[mark=none, color11, thick] table [x index=0, y index=1]
  {./pics/cloud_wave/data/cloud12500_wavelet__profile_p_3.dat};
  
    \addplot[mark=none, color13, thick] table [x index=0, y index=1]
  {./pics/cloud_wave/data/cloud12500_wavelet__profile_p_4.dat};
  
    \addplot[mark=none, color14, thick] table [x index=0, y index=1]
  {./pics/cloud_wave/data/cloud12500_wavelet__profile_p_5.dat};
  
    \addplot[mark=none, color15, thick] table [x index=0, y index=1]
  {./pics/cloud_wave/data/cloud12500_wavelet__profile_p_6.dat};
  
   \node[anchor=east] (A) at (axis cs:0.75, 1.24){};
   \node[anchor=west] (B) at (axis cs:15, 0.3){};
   \draw[->,thick] (A) -- (B);

\end{axis}
\end{tikzpicture}
  \hfill%
\pgfkeys{/pgf/number format/.cd,1000 sep={\,}}
\begin{tikzpicture}[baseline]
\begin{axis}[
  grid=major,
  width=0.45\textwidth,
  style={font=\normalsize},
  xmin=-35, xmax=35,
  ymin=0, ymax=1.35,
  xlabel=$r-R_\text{F}\units{[mm]}$,
  ylabel=$\frac{\bar p-p_\text{C}}{p_\text{F}-p_\text{C}}$,
  try min ticks=7,
  ]

  \addplot[mark=none, color0, thick] table [x index=0, y index=1]
  {./pics/cloud_wave/data/cloud12500_homo_slice__profile_p_0.dat};

  \addplot[mark=none, color8, thick] table [x index=0, y index=1]
  {./pics/cloud_wave/data/cloud12500_homo_slice__profile_p_1.dat};
  
    \addplot[mark=none, color9, thick] table [x index=0, y index=1]
  {./pics/cloud_wave/data/cloud12500_homo_slice__profile_p_2.dat};
  
    \addplot[mark=none, color11, thick] table [x index=0, y index=1]
  {./pics/cloud_wave/data/cloud12500_homo_slice__profile_p_3.dat};
  
    \addplot[mark=none, color13, thick] table [x index=0, y index=1]
  {./pics/cloud_wave/data/cloud12500_homo_slice__profile_p_4.dat};
  
    \addplot[mark=none, color14, thick] table [x index=0, y index=1]
  {./pics/cloud_wave/data/cloud12500_homo_slice__profile_p_5.dat};
  
    \addplot[mark=none, color15, thick] table [x index=0, y index=1]
  {./pics/cloud_wave/data/cloud12500_homo_slice__profile_p_6.dat};

   \node[anchor=east] (A) at (axis cs:0.75, 1.24){};
   \node[anchor=west] (B) at (axis cs:15, 0.3){};
   \draw[->,thick] (A) -- (B);

\end{axis}
\end{tikzpicture}
\\
\pgfkeys{/pgf/number format/.cd,1000 sep={\,}}
\begin{tikzpicture}[baseline]
\begin{axis}[
  grid=major,
  width=0.45\textwidth,
  style={font=\normalsize},
  xmin=-35, xmax=35,
  ymin=0, ymax=1.23,
  xlabel=$r-R_\text{F}\units{[mm]}$,
  ylabel=$\frac{\bar{u}}{u_\text{F}}$,
  try min ticks=7,
  ]

  \addplot[mark=none, color0, thick] table [x index=0, y index=1]
  {./pics/cloud_wave/data/cloud12500_wavelet__profile_m_0.dat};

  \addplot[mark=none, color8, thick] table [x index=0, y index=1]
  {./pics/cloud_wave/data/cloud12500_wavelet__profile_m_1.dat};
  
    \addplot[mark=none, color9, thick] table [x index=0, y index=1]
  {./pics/cloud_wave/data/cloud12500_wavelet__profile_m_2.dat};
  
    \addplot[mark=none, color11, thick] table [x index=0, y index=1]
  {./pics/cloud_wave/data/cloud12500_wavelet__profile_m_3.dat};
  
    \addplot[mark=none, color13, thick] table [x index=0, y index=1]
  {./pics/cloud_wave/data/cloud12500_wavelet__profile_m_4.dat};
  
    \addplot[mark=none, color14, thick] table [x index=0, y index=1]
  {./pics/cloud_wave/data/cloud12500_wavelet__profile_m_5.dat};
  
    \addplot[mark=none, color15, thick] table [x index=0, y index=1]
  {./pics/cloud_wave/data/cloud12500_wavelet__profile_m_6.dat};
  
   \node[anchor=east] (A) at (axis cs:1, 1.18){};
   \node[anchor=west] (B) at (axis cs:12, 0.15){};
   \draw[->,thick] (A) -- (B);

\end{axis}
\end{tikzpicture}
  \hfill%
\pgfkeys{/pgf/number format/.cd,1000 sep={\,}}
\begin{tikzpicture}[baseline]
\begin{axis}[
  grid=major,
  width=0.45\textwidth,
  style={font=\normalsize},
  xmin=-35, xmax=35,
  ymin=0, ymax=1.23,
  xlabel=$r-R_\text{F}\units{[mm]}$,
  ylabel=$\frac{\bar{u}}{u_\text{F}}$,
  try min ticks=7,
  ]

  \addplot[mark=none, color0, thick] table [x index=0, y index=1]
  {./pics/cloud_wave/data/cloud12500_homo_slice__profile_m_0.dat};

  \addplot[mark=none, color8, thick] table [x index=0, y index=1]
  {./pics/cloud_wave/data/cloud12500_homo_slice__profile_m_1.dat};
  
    \addplot[mark=none, color9, thick] table [x index=0, y index=1]
  {./pics/cloud_wave/data/cloud12500_homo_slice__profile_m_2.dat};
  
    \addplot[mark=none, color11, thick] table [x index=0, y index=1]
  {./pics/cloud_wave/data/cloud12500_homo_slice__profile_m_3.dat};
  
    \addplot[mark=none, color13, thick] table [x index=0, y index=1]
  {./pics/cloud_wave/data/cloud12500_homo_slice__profile_m_4.dat};
  
    \addplot[mark=none, color14, thick] table [x index=0, y index=1]
  {./pics/cloud_wave/data/cloud12500_homo_slice__profile_m_5.dat};
  
    \addplot[mark=none, color15, thick] table [x index=0, y index=1]
  {./pics/cloud_wave/data/cloud12500_homo_slice__profile_m_6.dat};
  
   \node[anchor=east] (A) at (axis cs:1, 1.18){};
   \node[anchor=west] (B) at (axis cs:12, 0.15){};
   \draw[->,thick] (A) -- (B);

\end{axis}
\end{tikzpicture}
   \caption{Normalized profiles of spherical averages of the gas volume fraction, pressure and
    velocity magnitude corresponding to $R_\mathrm{F}=40,35,30,25,20,15$ and $10\units{mm}$.
    Simulation with resolved bubbles (left) and homogeneous mixture approach (right).  Arrows
    indicate increasing time.
    }
  \label{fig:profiles12500}
\end{figure}
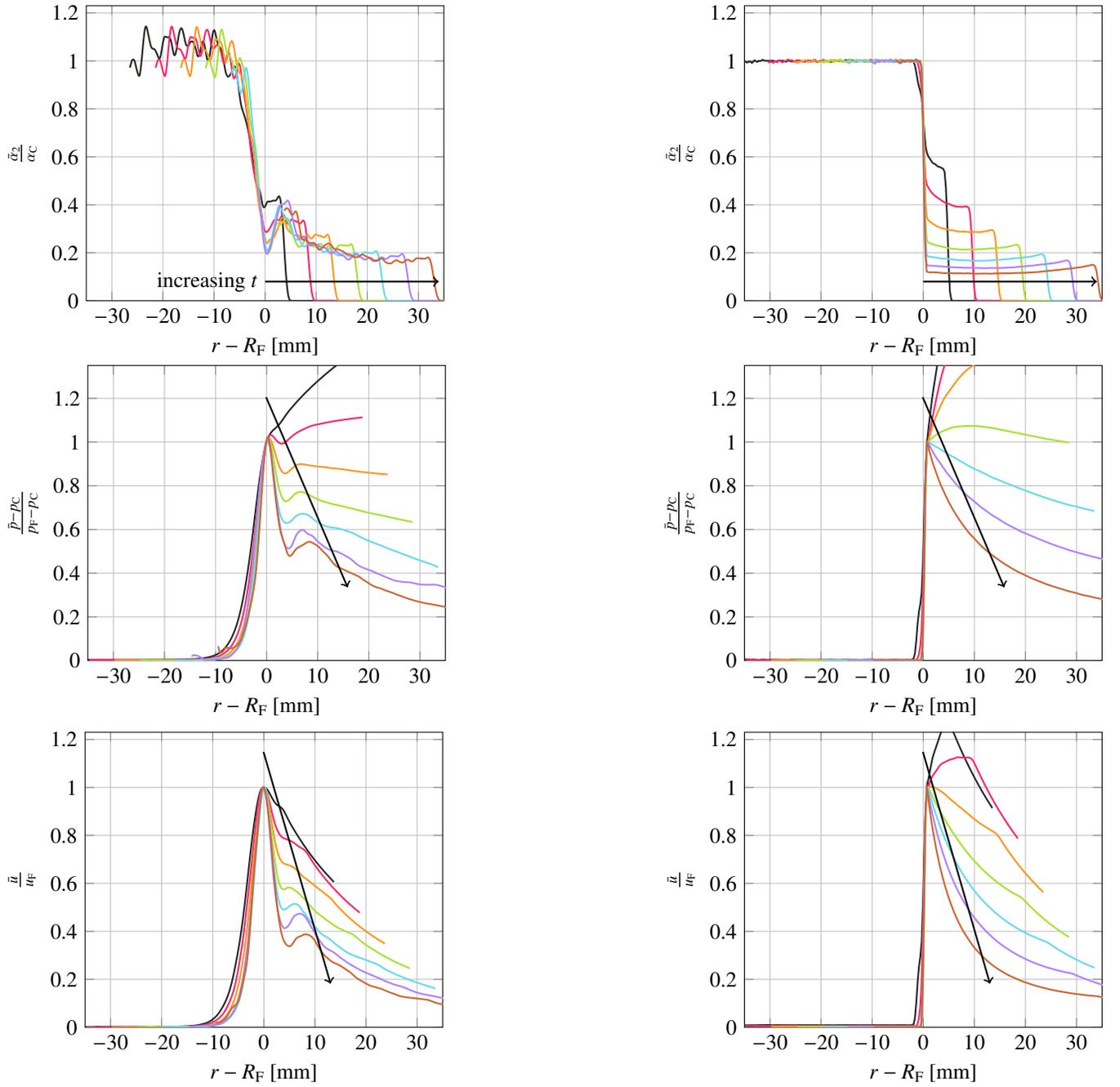

Profiles of the spherical averages
at various time instants
$t=139,183,218,245,267,285$ and $297\units{\mu s}$
corresponding to $R_\mathrm{F}=40,35,30,25,20,15$ and $10\units{mm}$
are shown in Fig.~\ref{fig:profiles12500}.
The profiles are normalized and plotted in the frame of reference of the front,
i.e., depending on the relative radial location $r-R_\mathrm{F}(t)$.
The normalized gas volume fraction, pressure and velocity are defined as 
${\bar\alpha_2}/{\alpha_\mathrm{C}}$, $(\bar p-p_\mathrm{C})/(\bar p_\mathrm{F}-p_\mathrm{C})$ and ${\bar u}/{\bar u_\mathrm{F}}$,
where $\bar p_\mathrm{F}(t)= \bar p(R_\mathrm{F}(t), t)$ and $\bar u_\mathrm{F}(t)= \bar u(R_\mathrm{F}(t), t)$
are pressure and velocity at the front.
The gas volume fraction shows some oscillations 
which decay towards the cloud surface
as more bubbles contribute to the averages with increasing~$r$.
The normalization of the radial profiles reveals their self-similarity in the vicinity of the front.
The collapse wave, or bubbly shock, does not exhibit a sharp front,
but has a finite thickness which is related to the dynamics of the individual
collapsing bubbles (see \cite{Wijngaarden:1970,Ganesh:2016} and references
therein). Consistent with the observations of the aforementioned studies, the
thickness of the front 
is of the size of a few bubble length scales.
From  the velocity profiles in Fig.~\ref{fig:profiles12500}, we obtain a front
thickness of approximately $10\units{mm}$, which is about seven bubble diameters.
Owing to the shielding effect by the outer bubbles, 
all fields remain at their initial values ahead of the front,
i.e., for $r-R_\mathrm{F}<-10\units{mm}$.
Closer to the front, the gas volume fraction gradually 
decreases to $\alpha_2/\alpha_\mathrm{C} \approx 0.2$ at the front,
while the pressure and the velocity grow towards their peak values.
Behind the front, the gas volume fraction rebounds and reaches a value of
$\alpha_2/\alpha_\mathrm{C} \approx 0.4$ at a
distance of $r-R_\mathrm{F} \approx 3\units{mm}$.
The gas volume fraction rebound behind the front 
~\cite{Brennen:2005} is accompanied by a drop in the pressure and velocity.
Farther outward from the cloud center, all profiles keep declining.
At the cloud surface, the gas volume fraction drops to zero in a sharp fashion
whereas pressure and velocity decrease smoothly to their prescribed far field values.

\begin{figure}[bt]
  \centering
\pgfkeys{/pgf/number format/.cd,1000 sep={\,}}
\begin{tikzpicture}[baseline]
\tikzset{mark size=1.5}
\begin{axis}[
  grid=major,
  width=0.49\textwidth,
  style={font=\normalsize},
  xlabel=$t\units{[\mu s]}$,
  ylabel=$p_\text{F}\units{[MPa]}$,
  legend columns=3,
  legend style={at={(0.5,1.02)},anchor=south,draw=none,font=\small,
  /tikz/column 2/.style={column sep=10pt},
  /tikz/column 4/.style={column sep=10pt}},
  legend cell align=left
  ]

  \addplot[mark=*, color8, thick] table [x index=0, y index=1]
  {./pics/cloud_wave/data/cloud12500__front__p_resolved.dat};
  \addlegendentry{bubbles}

  \addplot[mark=none, color0, thick,dashed] table [x index=0, y index=1]
  {./pics/cloud_wave/data/cloud12500__front__p_resolved_fit.dat};
  \addlegendentry{fit}
  
  \addplot[mark=square*, color0, thick] table [x index=0, y index=1]
  {./pics/cloud_wave/data/cloud12500__front__p_homogeneous.dat};
  \addlegendentry{mixture}

\end{axis}
\end{tikzpicture}
  \hfill%
\pgfkeys{/pgf/number format/.cd,1000 sep={\,}}
\begin{tikzpicture}[baseline]
\tikzset{mark size=1.5}
\begin{axis}[
  grid=major,
  width=0.49\textwidth,
  style={font=\normalsize},
  xlabel=$t\units{[\mu s]}$,
  ylabel=$u_\text{F}\units{[m/s]}$,
  legend columns=3,
%  legend style={at={(0.5,1.02)},anchor=south,draw=none,font=\normalsize,
  legend style={at={(0.5,1.02)},anchor=south,draw=none,font=\small,
  /tikz/column 2/.style={column sep=10pt},
  /tikz/column 4/.style={column sep=10pt}},
  legend cell align=left
  ]

  \addplot[mark=*, color8, thick] table [x index=0, y index=1]
  {./pics/cloud_wave/data/cloud12500__front__m_resolved.dat};
  \addlegendentry{bubbles}

  \addplot[mark=none, color0, thick,dashed] table [x index=0, y index=1]
  {./pics/cloud_wave/data/cloud12500__front__m_resolved_fit.dat};
  \addlegendentry{fit}
  
  \addplot[mark=square*, color0, thick] table [x index=0, y index=1]
  {./pics/cloud_wave/data/cloud12500__front__m_homogeneous.dat};
  \addlegendentry{mixture}

\end{axis}
\end{tikzpicture}
   \caption{Temporal evolution of average pressure (left) and average velocity magnitude (right) at the front.
  }
  \label{fig:front12500}
\end{figure}
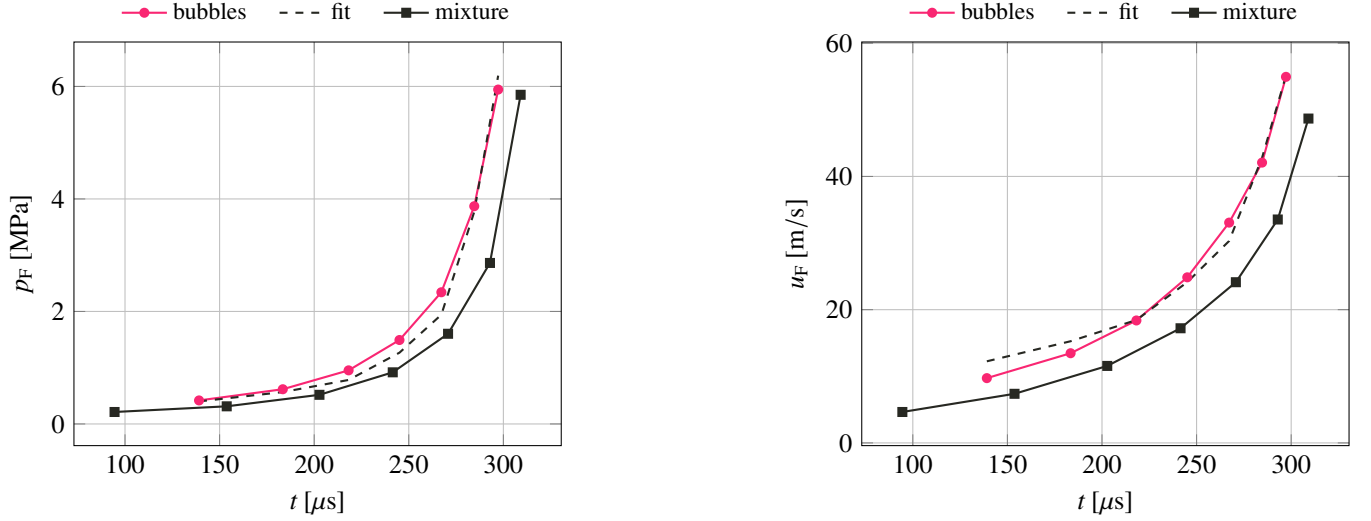

The values of the pressure and velocity at the front increase
as seen from their temporal evolution shown in Fig.~\ref{fig:front12500}.
As derived from mass and momentum balance \cite{Wijngaarden:1970,Morch:1989}, 
$p_\mathrm{F}$ and $u_\mathrm{F}$ are related to the front speed.
Approximate relations for these quantities near the front are given by
\begin{align}
  \label{eqn:front_speed_relation_pres}
    p_\mathrm{F} - p_\mathrm{C} & \sim  \rho_1 (1-\alpha_\mathrm{C}) \alpha_\mathrm{C}  \dot{R}_\mathrm{F}^2,
    \\
   \label{eqn:front_speed_relation_vel}
    u_\mathrm{F} & \sim \alpha_\mathrm{C} \dot{R}_\mathrm{F}.
\end{align}
up to a scaling factor which depends on the definition of the front location.
Fitting these relations to the simulation data results in
\begin{align}
  \label{eqn:front_speed_relation_fit_pres}
    p_\mathrm{F} - p_\mathrm{C} &= 5.95 \rho_1 (1-\alpha_\mathrm{C}) \alpha_\mathrm{C} \dot{R}_\mathrm{F}^2 ,
    \\
  \label{eqn:front_speed_relation_fit_vel}
   u_\mathrm{F} &=  2.39 \alpha_\mathrm{C} \dot{R}_\mathrm{F}
\end{align}
and provides a good approximation to the present results; see Fig.~\ref{fig:front12500}.

A model proposed by M\o rch in \cite{Morch:1989} describes the collapse of a spherical cloud
of vapor bubbles in the form of a Rayleigh-Plesset-like equation:
\begin{equation}
  \label{eqn:morch}
    R_\mathrm{F}\ddot{R}_\mathrm{F} + \left(\frac{3}{2} -\frac{1}{2}(1-\psi)(1-\alpha_\mathrm{C}) \right)\dot{R}_\mathrm{F}^2
    =
    -\frac{p_\infty-p_v}{\alpha_\mathrm{C}\rho_1},
\end{equation}
where
$p_v$ denotes the vapor pressure of the liquid and
$\psi$ an energy conservation factor. 
The energy conservation factor accounts for energy losses due to the radiation of acoustic waves
and dissipation. A larger value leads to a higher front speed. According to \cite{Morch:1989},
the energy conservation factor should be in the range $0\leq\psi\leq0.5$.
The model assumes that the bubbles are small compared to the cloud radius and that
the vapor volume fraction is sufficiently high.
In contrast to the present simulation of a cloud of gas bubbles,
the M\o rch model is derived for vapor bubbles which means that the pressure inside the bubbles 
remains constant during the collapse and that the bubbles collapse completely without any rebound stage.
When setting $p_v=p_\mathrm{C}$, the M\o rch model also provides a reasonable prediction
for the front trajectory and speed
of the present case, as can be seen from Fig.~\ref{fig:trajectory12500}
where the respective curves are labeled ``M\o rch''.
For the curves shown in Fig.~\ref{fig:trajectory12500}, the energy conservation factor, which is only of minor
influence, is set to $\psi =0.5$.

Furthermore, results obtained by a homogeneous mixture approach are included for comparison.
The mathematical description
introduced in Sec.~\ref{sec:gover_eq} may also be used to describe a homogeneous
mixture of gas and liquid owing to the right-hand-side term of Eq.~\eqref{eq:vol_frac}.
Instead of initially prescribing a cloud composed of individual bubbles,
a uniform gas volume fraction $\alpha_2=\alpha_\mathrm{C}$ is set within the sphere of radius $R_\mathrm{C}$.
The initial conditions for the velocity and the pressure as well as the applied boundary conditions
remain unchanged compared to the case with resolved bubbles.
A similar approach was used in \cite{Tiwari:2015}.
For the homogeneous mixture approach, the computational domain  is discretized by
$1024$ cells per spatial direction.
Spherically averaged profiles for $R_\mathrm{F}=40,35,30,25,20,15$ and $10\units{mm}$
corresponding to $t=94,154,203,242,271,293$ and $309\units{\mu s}$,
are shown in Fig.~\ref{fig:profiles12500}. 
In contrast to the case with resolved bubbles,
the radial profiles are discontinuous at the front 
and do not demonstrate features such as
the gas volume fraction rebound behind the front or the gradual transition of the profiles 
ahead of the front.
Therefore, the location of the collapse wave front for the homogeneous mixture case
is determined from the gas volume fraction via
\begin{equation}
  R_\mathrm{F}(t) = \argmax_r \left| \frac{\partial \bar\alpha_2}{\partial t}(r,t)\right|,
\end{equation}
which detects the discontinuity in $\bar\alpha_2$.
The front trajectory and speed, shown in Fig.~\ref{fig:trajectory12500} by
the curves labeled ``mixture'',
are qualitatively similar to the ones of the resolved simulation. 
However, the front speed is underestimated starting from $t=150\units{\mu s}$,
and the deviation grows in time 
reaching about $50\units{m/s}$ at $t=250\units{\mu s}$.
The temporal evolution of the pressure and the velocity at the front are included
in Fig.~\ref{fig:front12500}.
The values observed with the homogeneous mixture approach are about $30\%$
lower compared to the resolved simulation.
In summary, our results indicate that the front trajectory and speed observed in the simulation with
large numbers of bubbles  are well captured by simplified models. 
The evolution of the pressure and the velocity near the front 
matches the theoretical relations and in turn validates  the present numerical results.

 %\clearpage
%%%%%%%%%%%%%%%%%%%%%%%%%%%%%%%%%%%%%%%%%%%%%%%%%%%%%%%%%%%%%%%%%%%%%%%%%%%%%%%

%%%%%%%%%%%%%%%%%%%%%%%%%%%%%%%%%%%%%%%%%%%%%%%%%%%%%%%%%%%%%%%%%%%%%%%%%%%%%%%
%% 4. bubble dynamics
% 4.1 bubble oscillations
%
\section{Bubble dynamics}
\label{sec:bubble_dyn}

Next, the evolution of the bubbles in the cloud is examined. Their oscillation frequencies as well
as the microjets leading to their deformation are investigated.

\subsection{Bubble oscillations}

The shape of the bubbles is implicitly described by the gas-volume-fraction field $\alpha_2$,
which is sampled at a frequency of $0.63\units{MHz}$.
The center $\mathbf x_{\mathrm{B}_i}(t)$ and the equivalent radius
$R_{\mathrm{B}_i}(t)$ of bubble $i$ are calculated as
\begin{align}
  \mathbf x_{\mathrm{B}_i}(t) &= \frac{1}{V_{\mathrm{B}_i}(t)} 
      \int\limits_{\Omega_{\mathrm{B}_i}} \alpha_2\mathbf x\:\mathrm{d}V,\\
  R_{\mathrm{B}_i}(t) &= \left(\frac{3}{4\pi}V_{\mathrm{B}_i}(t)\right)^{1/3},
\end{align}
where 
\begin{equation}
V_{\mathrm{B}_i}(t) = \int\limits_{\Omega_{\mathrm{B}_i}} \alpha_2\:\mathrm{d}V
\end{equation}
is the bubble volume.
The integration is performed over a spherical domain $\Omega_{\mathrm{B}_i}$
concentric with the bubble center of the previous time sample
and
with a radius equal to the initial bubble radius $R_{\mathrm{B}_i}(0)$.
In order to improve the accuracy of peak detection, 
the function $R_{\mathrm{B}_i}(t)$ is interpolated in time 
with a cubic spline.

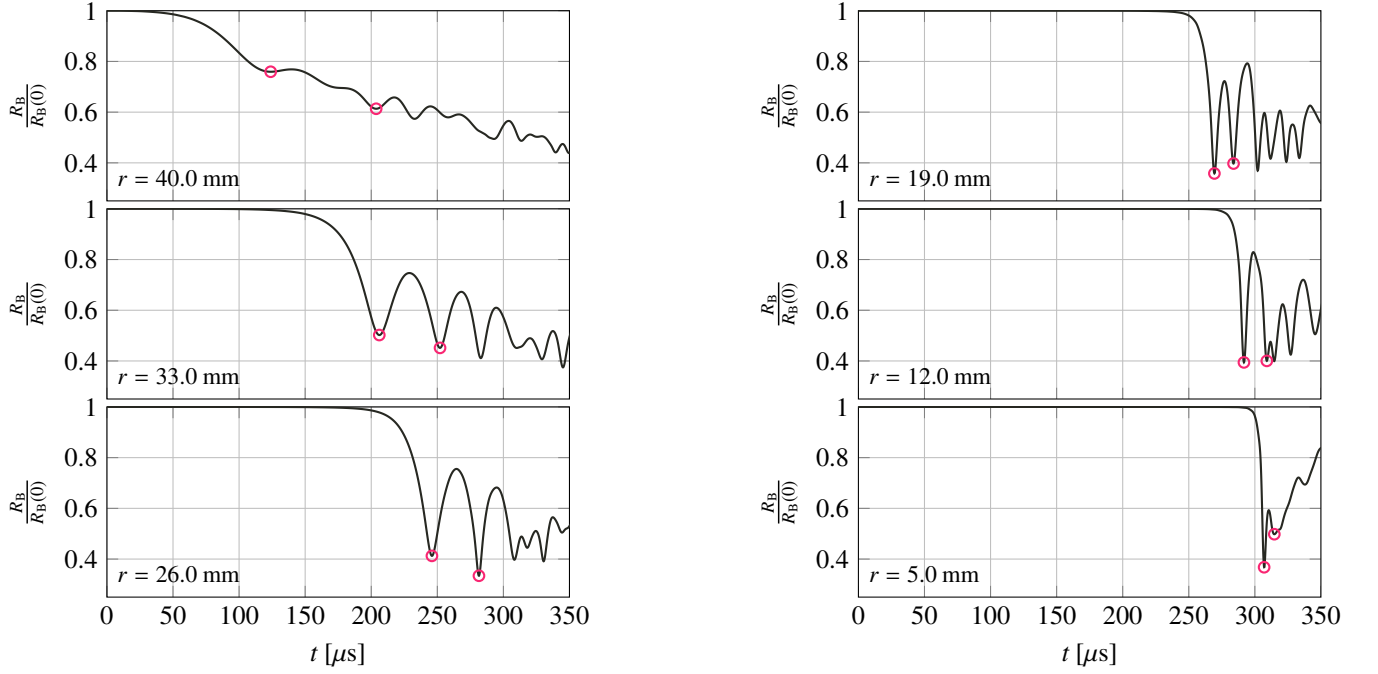
\begin{figure}[tb]
  \centering
  %
% requires \usetikzlibrary{pgfplots.groupplots}
\begin{tikzpicture}[scale=1.0]
  \tikzset{mark size=2.0}
  \pgfplotsset{mylinestyle/.style={mark=none, thick, line join=round}}
  %\pgfplotsset{every tick label/.append style={font=\large}}
  \begin{groupplot}[
    group style={
    group name=my plots,
    group size=1 by 3,
    xlabels at=edge bottom,
    xticklabels at=edge bottom,
    vertical sep=3pt
    },
    grid=major,
    width=0.47\textwidth,
    height=0.25\textwidth,
    xlabel=$t\units{[\mu s]}$,
    ylabel=$\frac{R_\text{B}}{R_\text{B}(0)}$,
    xmin=0.0, xmax=350,
    xtick={0, 50, 100, 150, 200, 250, 300, 350},
    ]

    \nextgroupplot[ymin=0.25, ymax=1.0,
    ytick={0.4,0.6,0.8,1.0}, yticklabels={$0.4$,$0.6$,$0.8$, $1$}]

    \addplot[mark=none, color0, thick,solid] table [x index=0, y index=1]
    {./pics/oscillation/data/cloud12500_track_evolution_radius_0_normalized.dat};
    
    \addplot[mark=o, only marks, color8, thick,solid] table [x index=0, y index=1]
    {./pics/oscillation/data/min_cloud12500_track_evolution_radius_0_normalized.dat};
    
    \node[anchor=west, font=\small] at (axis cs:1,0.34) {$r=40.0\units{mm}$};

    \nextgroupplot[ymin=0.25, ymax=1.0,
    ytick={0.4,0.6,0.8,1.0}, yticklabels={$0.4$,$0.6$,$0.8$, $1$}]

    \addplot[mark=none, color0, thick,solid] table [x index=0, y index=1]
    {./pics/oscillation/data/cloud12500_track_evolution_radius_1_normalized.dat};
    
    \addplot[mark=o, only marks, color8, thick,solid] table [x index=0, y index=1]
    {./pics/oscillation/data/min_cloud12500_track_evolution_radius_1_normalized.dat};
    
    \node[anchor=west, font=\small] at (axis cs:1,0.34) {$r=33.0\units{mm}$};
    
    \nextgroupplot[ymin=0.25, ymax=1.0,
    ytick={0.4,0.6,0.8,1.0}, yticklabels={$0.4$,$0.6$,$0.8$, $1$}]

    \addplot[mark=none, color0, thick,solid] table [x index=0, y index=1]
    {./pics/oscillation/data/cloud12500_track_evolution_radius_2_normalized.dat};
    
    \addplot[mark=o, only marks, color8, thick,solid] table [x index=0, y index=1]
    {./pics/oscillation/data/min_cloud12500_track_evolution_radius_2_normalized.dat};

    \node[anchor=west, font=\small] at (axis cs:1,0.34) {$r=26.0\units{mm}$};

  \end{groupplot}
\end{tikzpicture}
  \hfill%
  %
% requires \usetikzlibrary{pgfplots.groupplots}
\begin{tikzpicture}[scale=1.0]
  \tikzset{mark size=2.0}
  \pgfplotsset{mylinestyle/.style={mark=none, thick, line join=round}}
  %\pgfplotsset{every tick label/.append style={font=\large}}
  \begin{groupplot}[
    group style={
    group name=my plots,
    group size=1 by 3,
    xlabels at=edge bottom,
    xticklabels at=edge bottom,
    vertical sep=3pt
    },
    grid=major,
    width=0.47\textwidth,
    height=0.25\textwidth,
    xlabel=$t\units{[\mu s]}$,
    ylabel=$\frac{R_\text{B}}{R_\text{B}(0)}$,
    xmin=0.0, xmax=350,
    xtick={0, 50, 100, 150, 200, 250, 300, 350},
    ]

    \nextgroupplot[ymin=0.25, ymax=1.0,
    ytick={0.4,0.6,0.8,1.0}, yticklabels={$0.4$,$0.6$,$0.8$, $1$}]

    \addplot[mark=none, color0, thick,solid] table [x index=0, y index=1]
    {./pics/oscillation/data/cloud12500_track_evolution_radius_3_normalized.dat};
    
     \addplot[mark=o, only marks, color8, thick,solid] table [x index=0, y index=1]
    {./pics/oscillation/data/min_cloud12500_track_evolution_radius_3_normalized.dat};
    
    \node[anchor=west, font=\small] at (axis cs:1,0.34) {$r=19.0\units{mm}$};
    
    \nextgroupplot[ymin=0.25, ymax=1.0,
    ytick={0.4,0.6,0.8,1.0}, yticklabels={$0.4$,$0.6$,$0.8$, $1$}]

    \addplot[mark=none, color0, thick,solid] table [x index=0, y index=1]
    {./pics/oscillation/data/cloud12500_track_evolution_radius_4_normalized.dat};
    
    \addplot[mark=o, only marks, color8, thick,solid] table [x index=0, y index=1]
    {./pics/oscillation/data/min_cloud12500_track_evolution_radius_4_normalized.dat};
    
    \node[anchor=west, font=\small] at (axis cs:1,0.34) {$r=12.0\units{mm}$};

    \nextgroupplot[ymin=0.25, ymax=1.0,
    ytick={0.4,0.6,0.8,1.0}, yticklabels={$0.4$,$0.6$,$0.8$, $1$}]

    \addplot[mark=none, color0, thick,solid] table [x index=0, y index=1]
    {./pics/oscillation/data/cloud12500_track_evolution_radius_5_normalized.dat};
    
    \addplot[mark=o, only marks, color8, thick,solid] table [x index=0, y index=1]
    {./pics/oscillation/data/min_cloud12500_track_evolution_radius_5_normalized.dat};
    
    \node[anchor=west, font=\small] at (axis cs:1,0.34) {$r=5.0\units{mm}$};

  \end{groupplot}
\end{tikzpicture}
   \caption{
    Temporal evolution of equivalent radius of selected  bubbles at various radial locations
    $r=40.0,33.0,26.0,19.0,12.0$ and $5.0\units{mm}$ (from top to bottom and left to right).
    Circles mark the minima for estimation of the oscillation frequency.  All curves are
    normalized by the corresponding initial radius.
  }
  \label{fig:oscillation_selected}
\end{figure}

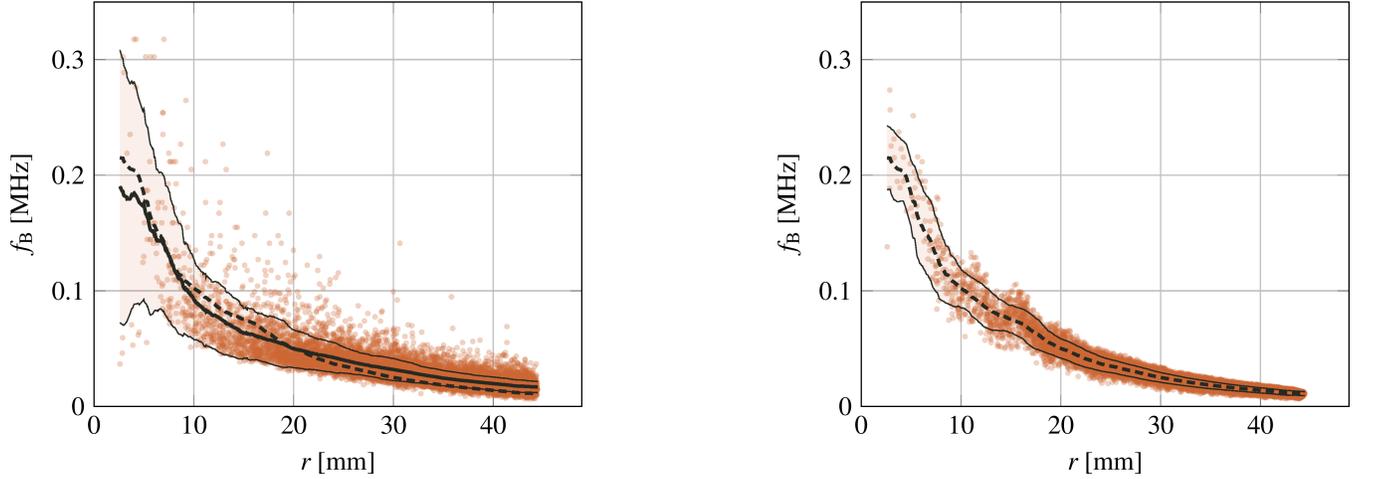
\begin{figure}[tb]
  \centering
\pgfkeys{/pgf/number format/.cd,1000 sep={\,}}
\begin{tikzpicture}[baseline]
\pgfplotsset{set layers}% using layers
\begin{axis}[
  set layers,
  grid=major,
  width=0.49\textwidth,
  style={font=\normalsize},
  xmin=0, %xmax=20,
  ymin=0, ymax=0.35,
xlabel=$r\units{[mm]}$,
ylabel=$f_\text{B}\units{[MHz]}$,
  legend columns=2,
  legend style={at={(0.5,1.02)},anchor=south,draw=none,font=\small,
  /tikz/column 2/.style={column sep=6pt}},
  legend cell align=left
  ]

  %\begin{scope}[on background layer]
      \addplot[only marks, mark=*, mark size=1.0pt, color15, %mark options={solid,thick,fill=color4, opacity=0.6},
      line width= 0pt,
      opacity=0.3,
%      line width= 0pt,
%          scatter,
%    scatter/use mapped color=
%        {draw=none,fill=color3,opacity=0.3}
      ]
      table [x index=0,y index=1]
      {./pics/oscillation/data/cloud12500_oscfreq_sim_scatter.dat};
  %\end{scope}

\begin{pgfonlayer}{axis foreground}
    \addplot[color0, thin, solid, name path=A] table[x index=0,y index=2]
    {./pics/oscillation/data/cloud12500_oscfreq_red_sim_avg.dat};

    \addplot[color0, thin, solid, name path=B] table[x index=0,y index=3]
    {./pics/oscillation/data/cloud12500_oscfreq_red_sim_avg.dat};

    \addplot [fill=color15, fill opacity=0.1] fill between[of=A and B];

    \addplot[mark=none, color0, very thick,solid] table [x index=0, y index=1]
    {./pics/oscillation/data/cloud12500_oscfreq_red_sim_avg.dat};

    \addplot[mark=none, color0, very thick,densely dashed] table [x index=0, y index=1]
    {./pics/oscillation/data/cloud12500_oscfreq_red_theory_avg.dat};    
    
\end{pgfonlayer}

\end{axis}

\end{tikzpicture}
  \hfill%
\pgfkeys{/pgf/number format/.cd,1000 sep={\,}}
\begin{tikzpicture}[baseline]
\pgfplotsset{set layers}% using layers
\begin{axis}[
  set layers,
  grid=major,
  width=0.49\textwidth,
  style={font=\normalsize},
  xmin=0, %xmax=20,
  ymin=0, ymax=0.35,
xlabel=$r\units{[mm]}$,
ylabel=$f_\text{B}\units{[MHz]}$,
  legend columns=2,
  legend style={at={(0.5,1.02)},anchor=south,draw=none,font=\small,
  /tikz/column 2/.style={column sep=6pt}},
  legend cell align=left
  ]

  %\begin{scope}[on background layer]
      \addplot[only marks, mark=*, mark size=1.0pt, color15,% mark options={solid,thick,fill=color15, opacity=0.6},
      line width= 0pt,
      opacity=0.3,
      ]
      table [x index=0,y index=1]
      {./pics/oscillation/data/cloud12500_oscfreq_theory_scatter.dat};
 % \end{scope}

\begin{pgfonlayer}{axis foreground}
    \addplot[color0, thin, solid, name path=A] table[x index=0,y index=2]
    {./pics/oscillation/data/cloud12500_oscfreq_red_theory_avg.dat};

    \addplot[color0, thin, solid, name path=B] table[x index=0,y index=3]
    {./pics/oscillation/data/cloud12500_oscfreq_red_theory_avg.dat};

    \addplot [fill=color15, fill opacity=0.1] fill between[of=A and B];

    \addplot[mark=none, color0, very thick,densely dashed] table [x index=0, y index=1]
    {./pics/oscillation/data/cloud12500_oscfreq_red_theory_avg.dat};
\end{pgfonlayer}

\end{axis}

\end{tikzpicture}
   \caption{
  Scatter plot of bubble oscillation frequencies with respect to  their radial location.
  Detailed simulation (left) and theoretical relation for a single bubble in an
  infinite liquid (right).  Moving averages of the simulation data (solid line)
  and theoretical data (dashed line) are shown.  Color shades indicate the
  standard deviation.
  }
  \label{fig:oscillation_freq}
\end{figure}

Fig.~\ref{fig:oscillation_selected} shows the evolution of the
equivalent bubble radius for a few bubbles selected at various radial locations.
All curves are normalized by the initial bubble radius.
A bubble starts to oscillate once it is overtaken by the collapse induced wave.
The oscillation frequency of each bubble in the cloud
is calculated from the time between the first and second minimum of its equivalent radius, as marked in Fig.~\ref{fig:oscillation_selected}.
A scatter plot of the bubble oscillation frequencies
depending on the radial location is shown in Fig.~\ref{fig:oscillation_freq}
along with the moving average and the corresponding standard deviation
computed with a window length of $4\units{mm}$.
The oscillation frequency is higher for bubbles closer to the cloud center.
The natural frequency of a single bubble oscillating in an
unbounded liquid  ~\cite{Franc:2004} depends on the ambient pressure and 
the  equilibrium radius of the bubble.
Using $p_{{\mathrm{B}_i},\mathrm{F}}$, which is the average pressure $\bar p(r,t)$
at the front when it reaches bubble $i$, and the
associated equilibrium radius obtained via
\begin{equation}
  R_{{\mathrm{B}_i},\mathrm{F}} = 
  R_{\mathrm{B}_i}(0)\left( \frac{p_C}{p_{{\mathrm{B}_i},\mathrm{F}}} \right)^{\frac{1}{3\gamma_2}},
\end{equation}
the frequency of each bubble in the cloud
is estimated as
\begin{equation}
\label{eq:freq_bubb}
f_{\mathrm{B}_i} = \frac{1}{2\pi R_{{\mathrm{B}_i},\mathrm{F}}} 
\sqrt{\frac{3\gamma_2p_{{\mathrm{B}_i},\mathrm{F}}}{\rho_1}}.
\end{equation}
The resulting frequencies are likewise shown in
Fig.~\ref{fig:oscillation_freq}.
Although Eq.~\eqref{eq:freq_bubb} does not
account for the influence of the other bubbles in the cloud,
this estimation is in a  reasonable agreement with  the simulation results.
In particular, this result shows  that the oscillation
frequency of a bubble is governed by the front pressure
and therefore, given Eq.~\eqref{eqn:front_speed_relation_pres}, by the front speed.

 % 4.2 microjet formation
%
\subsection{Microjet formation}

The evolving pressure gradient along the bubble surface leads to the formation of 
a localized liquid jet of high velocity which notably deforms the bubble and eventually
pierces though it.
Following \cite{Jayaprakash:2012}, the tip $\mathbf x_{\mathrm{tip}_i}$ of
the microjet associated with bubble $i$ is identified as the 
location of minimum curvature on the bubble surface.
Here, the interface is represented by the iso-surface $\alpha_2=0.5$
of the gas-volume-fraction field. The curvature of any iso-contour
of $\alpha_2$ can be calculated from the gas-volume-fraction field via
$\kappa = -\nabla \cdot \frac{\nabla \alpha_2}{|\nabla \alpha_2|}$.

\begin{figure}[tb]
  \centering
\begin{tikzpicture}[baseline]
  \begin{axis}[
    grid=major,
    width=14.0cm,
    height=6.0cm,
    view = {0}{90},
    %style={font=\large},
    xmin=79.1402170474, xmax=199.140217047,
    ymin=-1.0, ymax=1.0,
    xlabel=$t\units{[\mu s]}$,
    %ylabel=$y$,
    legend columns=5,
    legend style={at={(0.5,1.02)},anchor=south,draw=none,font=\small,
    /tikz/column 2/.style={column sep=7pt},
    /tikz/column 4/.style={column sep=7pt},
    /tikz/column 6/.style={column sep=7pt},
    /tikz/column 8/.style={column sep=7pt},
    name=leg,
    },
    legend cell align=left,
    ]
    \addplot[mark=*, mark size=1.3, solid, color13, thick] % x-xb
    table[x index=0,y index=1]
    {./pics/microjets/data/microjet_trajectory_10135.dat};
    \addlegendentry{$x_\mathrm{tip}-x_\mathrm{B}$}
    \addplot[mark=triangle*, mark size=1.5, solid, color9, thick] % y-yb
    table[x index=0,y index=2]
    {./pics/microjets/data/microjet_trajectory_10135.dat};
    \addlegendentry{$y_\mathrm{tip}-y_\mathrm{B}$}
    \addplot[mark=square*, mark size=1.3, solid, color11, thick] % z-zb
    table[x index=0,y index=3]
    {./pics/microjets/data/microjet_trajectory_10135.dat};
    \addlegendentry{$z_\mathrm{tip}-z_\mathrm{B}$}
    \addplot[solid, color8, thick] % |X-Xb|
    table[x index=0,y index=4]
    {./pics/microjets/data/microjet_trajectory_10135.dat};
    \addlegendentry{$|\mathbf{x}_\mathrm{tip}-\mathbf{x}_\mathrm{B}|$}
    \addplot[solid, very thick, color0] % rb
    table[x index=0,y index=5]
    {./pics/microjets/data/microjet_trajectory_10135.dat};
    \addlegendentry{$R_\mathrm{B}$}
    \addplot[dashed, thick, color13, thick] % xfit
    table[x index=0,y index=1]
    {./pics/microjets/data/microjet_trajectory_10135_fit.dat};
    \addplot[dashed, thick, color9, thick] % yfit
    table[x index=0,y index=2]
    {./pics/microjets/data/microjet_trajectory_10135_fit.dat};
    \addplot[dashed, thick, color11, thick] % zfit
    table[x index=0,y index=3]
    {./pics/microjets/data/microjet_trajectory_10135_fit.dat};
    \addplot[dashed, thick, color8, thick] % |Xfit|
    table[x index=0,y index=4]
    {./pics/microjets/data/microjet_trajectory_10135_fit.dat};
    \addplot [mark=none, color0, very thick, dashed] coordinates {(119.1314511916286,-1) (119.1314511916286,1)};
    \addplot [mark=none, color0, very thick] coordinates {(117.00427342623004,-1) (117.00427342623004,1)};
    \addplot [mark=none, color0, very thick] coordinates {(161.27616066858735,-1) (161.27616066858735,1)};
    \addplot [mark=none, color0, dotted] coordinates {(88.54377448471463,-1) (88.54377448471463,1)};
    \addplot [mark=none, color0, dotted] coordinates {(99.61174629530396,-1) (99.61174629530396,1)};
    \addplot [mark=none, color0, dotted] coordinates {(109.09857927580909,-1) (109.09857927580909,1)};
    \addplot [mark=none, color0, dotted] coordinates {(118.58541225631423,-1) (118.58541225631423,1)};
    \addplot [mark=none, color0, dotted] coordinates {(129.65338406690356,-1) (129.65338406690356,1)};
    \addplot [mark=none, color0, dotted] coordinates {(139.1402170474087,-1) (139.1402170474087,1)};
    \addplot [mark=none, color0, dotted] coordinates {(148.62705002791384,-1) (148.62705002791384,1)};
    \addplot [mark=none, color0, dotted] coordinates {(159.69502183850315,-1) (159.69502183850315,1)};
    \addplot [mark=none, color0, dotted] coordinates {(169.1818548190083,-1) (169.1818548190083,1)};
    \addplot [mark=none, color0, dotted] coordinates {(178.66868779951344,-1) (178.66868779951344,1)};
    \addplot [mark=none, color0, dotted] coordinates {(189.73665961010278,-1) (189.73665961010278,1)};

    \coordinate (iso01) at (axis cs:88.54377448471463,-1.0);
    \coordinate (iso02) at (axis cs:99.61174629530396,-1.0);
    \coordinate (iso03) at (axis cs:109.09857927580909,-1.0);
    \coordinate (iso04) at (axis cs:118.58541225631423,-1.0);
    \coordinate (iso05) at (axis cs:129.65338406690356,-1.0);
    \coordinate (iso06) at (axis cs:139.1402170474087,-1.0);
    \coordinate (iso07) at (axis cs:148.62705002791384,-1.0);
    \coordinate (iso08) at (axis cs:159.69502183850315,-1.0);
    \coordinate (iso09) at (axis cs:169.1818548190083,-1.0);
    \coordinate (iso10) at (axis cs:178.66868779951344,-1.0);
    \coordinate (iso11) at (axis cs:189.73665961010278,-1.0);

  \end{axis}

  % marker with bubble index
  \node[align=center, font=\small, color0, anchor=west]
  at (leg.east)
  {\quad\raisebox{2.5pt}{\pgfuseplotmark{triangle*}}\hspace{3pt}1};

  \begin{axis}[
    view = {0}{90},
    hide axis, at=(iso01), anchor=south, scale only axis,
    height=1.10833333333cm, width=1.10833333333cm,
    xmin=-0.05,xmax=1.05, ymin=-0.05,ymax=1.05]
    \addplot3[contour gnuplot={levels={0.5}, labels=false, draw color=color0},
    mesh/rows=48,mesh/cols=48,thick]
    table {./pics/microjets/data/microjet_trajectory_10135_contour_010976.dat};
  \end{axis}
  \begin{axis}[
    view = {0}{90},
    hide axis, at=(iso02), anchor=south, scale only axis,
    height=1.10833333333cm, width=1.10833333333cm,
    xmin=-0.05,xmax=1.05, ymin=-0.05,ymax=1.05]
    \addplot3[contour gnuplot={levels={0.5}, labels=false, draw color=color0},
    mesh/rows=48,mesh/cols=48,thick]
    table {./pics/microjets/data/microjet_trajectory_10135_contour_012348.dat};
  \end{axis}
  \begin{axis}[
    view = {0}{90},
    hide axis, at=(iso03), anchor=south, scale only axis,
    height=1.10833333333cm, width=1.10833333333cm,
    xmin=-0.05,xmax=1.05, ymin=-0.05,ymax=1.05]
    \addplot3[contour gnuplot={levels={0.5}, labels=false, draw color=color0},
    mesh/rows=48,mesh/cols=48,thick]
    table {./pics/microjets/data/microjet_trajectory_10135_contour_013524.dat};
  \end{axis}
  \begin{axis}[
    view = {0}{90},
    hide axis, at=(iso04), anchor=south, scale only axis,
    height=1.10833333333cm, width=1.10833333333cm,
    xmin=-0.05,xmax=1.05, ymin=-0.05,ymax=1.05]
    \addplot3[contour gnuplot={levels={0.5}, labels=false, draw color=color0},
    mesh/rows=48,mesh/cols=48,thick]
    table {./pics/microjets/data/microjet_trajectory_10135_contour_014700.dat};
  \end{axis}
  \begin{axis}[
    view = {0}{90},
    hide axis, at=(iso05), anchor=south, scale only axis,
    height=1.10833333333cm, width=1.10833333333cm,
    xmin=-0.05,xmax=1.05, ymin=-0.05,ymax=1.05]
    \addplot3[contour gnuplot={levels={0.5}, labels=false, draw color=color0},
    mesh/rows=48,mesh/cols=48,thick]
    table {./pics/microjets/data/microjet_trajectory_10135_contour_016088.dat};
  \end{axis}
  \begin{axis}[
    view = {0}{90},
    hide axis, at=(iso06), anchor=south, scale only axis,
    height=1.10833333333cm, width=1.10833333333cm,
    xmin=-0.05,xmax=1.05, ymin=-0.05,ymax=1.05]
    \addplot3[contour gnuplot={levels={0.5}, labels=false, draw color=color0},
    mesh/rows=48,mesh/cols=48,thick]
    table {./pics/microjets/data/microjet_trajectory_10135_contour_017288.dat};
  \end{axis}
  \begin{axis}[
    view = {0}{90},
    hide axis, at=(iso07), anchor=south, scale only axis,
    height=1.10833333333cm, width=1.10833333333cm,
    xmin=-0.05,xmax=1.05, ymin=-0.05,ymax=1.05]
    \addplot3[contour gnuplot={levels={0.5}, labels=false, draw color=color0},
    mesh/rows=48,mesh/cols=48,thick]
    table {./pics/microjets/data/microjet_trajectory_10135_contour_018488.dat};
  \end{axis}
  \begin{axis}[
    view = {0}{90},
    hide axis, at=(iso08), anchor=south, scale only axis,
    height=1.10833333333cm, width=1.10833333333cm,
    xmin=-0.05,xmax=1.05, ymin=-0.05,ymax=1.05]
    \addplot3[contour gnuplot={levels={0.5}, labels=false, draw color=color0},
    mesh/rows=48,mesh/cols=48,thick]
    table {./pics/microjets/data/microjet_trajectory_10135_contour_019888.dat};
  \end{axis}
  \begin{axis}[
    view = {0}{90},
    hide axis, at=(iso09), anchor=south, scale only axis,
    height=1.10833333333cm, width=1.10833333333cm,
    xmin=-0.05,xmax=1.05, ymin=-0.05,ymax=1.05]
    \addplot3[contour gnuplot={levels={0.5}, labels=false, draw color=color0},
    mesh/rows=48,mesh/cols=48,thick]
    table {./pics/microjets/data/microjet_trajectory_10135_contour_021088.dat};
  \end{axis}
  \begin{axis}[
    view = {0}{90},
    hide axis, at=(iso10), anchor=south, scale only axis,
    height=1.10833333333cm, width=1.10833333333cm,
    xmin=-0.05,xmax=1.05, ymin=-0.05,ymax=1.05]
    \addplot3[contour gnuplot={levels={0.5}, labels=false, draw color=color0},
    mesh/rows=48,mesh/cols=48,thick]
    table {./pics/microjets/data/microjet_trajectory_10135_contour_022288.dat};
  \end{axis}
  \begin{axis}[
    view = {0}{90},
    hide axis, at=(iso11), anchor=south, scale only axis,
    height=1.10833333333cm, width=1.10833333333cm,
    xmin=-0.05,xmax=1.05, ymin=-0.05,ymax=1.05]
    \addplot3[contour gnuplot={levels={0.5}, labels=false, draw color=color0},
    mesh/rows=48,mesh/cols=48,thick]
    table {./pics/microjets/data/microjet_trajectory_10135_contour_023688.dat};
  \end{axis}

\end{tikzpicture}

\begin{tikzpicture}[baseline]
  \begin{axis}[
    grid=major,
    width=14.0cm,
    height=6.0cm,
    view = {0}{90},
    %style={font=\large},
    xmin=87.8364806129, xmax=207.836480613,
    ymin=-1.0, ymax=1.0,
    xlabel=$t\units{[\mu s]}$,
    %ylabel=$y$,
    legend columns=5,
    legend style={at={(0.5,1.02)},anchor=south,draw=none,font=\small,
    /tikz/column 2/.style={column sep=7pt},
    /tikz/column 4/.style={column sep=7pt},
    /tikz/column 6/.style={column sep=7pt},
    /tikz/column 8/.style={column sep=7pt},
    name=leg,
    },
    legend cell align=left,
    ]
    \addplot[mark=*, mark size=1.3, solid, color13, thick] % x-xb
    table[x index=0,y index=1]
    {./pics/microjets/data/microjet_trajectory_6236.dat};
    \addlegendentry{$x_\mathrm{tip}-x_\mathrm{B}$}
    \addplot[mark=triangle*, mark size=1.5, solid, color9, thick] % y-yb
    table[x index=0,y index=2]
    {./pics/microjets/data/microjet_trajectory_6236.dat};
    \addlegendentry{$y_\mathrm{tip}-y_\mathrm{B}$}
    \addplot[mark=square*, mark size=1.3, solid, color11, thick] % z-zb
    table[x index=0,y index=3]
    {./pics/microjets/data/microjet_trajectory_6236.dat};
    \addlegendentry{$z_\mathrm{tip}-z_\mathrm{B}$}
    \addplot[solid, color8, thick] % |X-Xb|
    table[x index=0,y index=4]
    {./pics/microjets/data/microjet_trajectory_6236.dat};
    \addlegendentry{$|\mathbf{x}_\mathrm{tip}-\mathbf{x}_\mathrm{B}|$}
    \addplot[solid, very thick, color0] % rb
    table[x index=0,y index=5]
    {./pics/microjets/data/microjet_trajectory_6236.dat};
    \addlegendentry{$R_\mathrm{B}$}
    \addplot[dashed, thick, color13, thick] % xfit
    table[x index=0,y index=1]
    {./pics/microjets/data/microjet_trajectory_6236_fit.dat};
    \addplot[dashed, thick, color9, thick] % yfit
    table[x index=0,y index=2]
    {./pics/microjets/data/microjet_trajectory_6236_fit.dat};
    \addplot[dashed, thick, color11, thick] % zfit
    table[x index=0,y index=3]
    {./pics/microjets/data/microjet_trajectory_6236_fit.dat};
    \addplot[dashed, thick, color8, thick] % |Xfit|
    table[x index=0,y index=4]
    {./pics/microjets/data/microjet_trajectory_6236_fit.dat};
    \addplot [mark=none, color0, very thick, dashed] coordinates {(125.30406524300832,-1) (125.30406524300832,1)};
    \addplot [mark=none, color0, very thick] coordinates {(124.90996757665098,-1) (124.90996757665098,1)};
    \addplot [mark=none, color0, very thick] coordinates {(170.7629936490925,-1) (170.7629936490925,1)};
    \addplot [mark=none, color0, dotted] coordinates {(98.03060746521976,-1) (98.03060746521976,1)};
    \addplot [mark=none, color0, dotted] coordinates {(107.5174404457249,-1) (107.5174404457249,1)};
    \addplot [mark=none, color0, dotted] coordinates {(118.58541225631423,-1) (118.58541225631423,1)};
    \addplot [mark=none, color0, dotted] coordinates {(128.07224523681938,-1) (128.07224523681938,1)};
    \addplot [mark=none, color0, dotted] coordinates {(137.5590782173245,-1) (137.5590782173245,1)};
    \addplot [mark=none, color0, dotted] coordinates {(148.62705002791384,-1) (148.62705002791384,1)};
    \addplot [mark=none, color0, dotted] coordinates {(158.11388300841898,-1) (158.11388300841898,1)};
    \addplot [mark=none, color0, dotted] coordinates {(167.60071598892412,-1) (167.60071598892412,1)};
    \addplot [mark=none, color0, dotted] coordinates {(177.08754896942926,-1) (177.08754896942926,1)};
    \addplot [mark=none, color0, dotted] coordinates {(188.15552078001858,-1) (188.15552078001858,1)};
    \addplot [mark=none, color0, dotted] coordinates {(197.64235376052372,-1) (197.64235376052372,1)};

    \coordinate (iso01) at (axis cs:98.03060746521976,-1.0);
    \coordinate (iso02) at (axis cs:107.5174404457249,-1.0);
    \coordinate (iso03) at (axis cs:118.58541225631423,-1.0);
    \coordinate (iso04) at (axis cs:128.07224523681938,-1.0);
    \coordinate (iso05) at (axis cs:137.5590782173245,-1.0);
    \coordinate (iso06) at (axis cs:148.62705002791384,-1.0);
    \coordinate (iso07) at (axis cs:158.11388300841898,-1.0);
    \coordinate (iso08) at (axis cs:167.60071598892412,-1.0);
    \coordinate (iso09) at (axis cs:177.08754896942926,-1.0);
    \coordinate (iso10) at (axis cs:188.15552078001858,-1.0);
    \coordinate (iso11) at (axis cs:197.64235376052372,-1.0);

  \end{axis}

  % marker with bubble index
  \node[align=center, font=\small, color0, anchor=west]
  at (leg.east)
  {\quad\raisebox{2.5pt}{\pgfuseplotmark{triangle*}}\hspace{3pt}2};

  \begin{axis}[
    view = {0}{90},
    hide axis, at=(iso01), anchor=south, scale only axis,
    height=1.10833333333cm, width=1.10833333333cm,
    xmin=-0.05,xmax=1.05, ymin=-0.05,ymax=1.05]
    \addplot3[contour gnuplot={levels={0.5}, labels=false, draw color=color0},
    mesh/rows=48,mesh/cols=48,thick]
    table {./pics/microjets/data/microjet_trajectory_6236_contour_012152.dat};
  \end{axis}
  \begin{axis}[
    view = {0}{90},
    hide axis, at=(iso02), anchor=south, scale only axis,
    height=1.10833333333cm, width=1.10833333333cm,
    xmin=-0.05,xmax=1.05, ymin=-0.05,ymax=1.05]
    \addplot3[contour gnuplot={levels={0.5}, labels=false, draw color=color0},
    mesh/rows=48,mesh/cols=48,thick]
    table {./pics/microjets/data/microjet_trajectory_6236_contour_013328.dat};
  \end{axis}
  \begin{axis}[
    view = {0}{90},
    hide axis, at=(iso03), anchor=south, scale only axis,
    height=1.10833333333cm, width=1.10833333333cm,
    xmin=-0.05,xmax=1.05, ymin=-0.05,ymax=1.05]
    \addplot3[contour gnuplot={levels={0.5}, labels=false, draw color=color0},
    mesh/rows=48,mesh/cols=48,thick]
    table {./pics/microjets/data/microjet_trajectory_6236_contour_014700.dat};
  \end{axis}
  \begin{axis}[
    view = {0}{90},
    hide axis, at=(iso04), anchor=south, scale only axis,
    height=1.10833333333cm, width=1.10833333333cm,
    xmin=-0.05,xmax=1.05, ymin=-0.05,ymax=1.05]
    \addplot3[contour gnuplot={levels={0.5}, labels=false, draw color=color0},
    mesh/rows=48,mesh/cols=48,thick]
    table {./pics/microjets/data/microjet_trajectory_6236_contour_015888.dat};
  \end{axis}
  \begin{axis}[
    view = {0}{90},
    hide axis, at=(iso05), anchor=south, scale only axis,
    height=1.10833333333cm, width=1.10833333333cm,
    xmin=-0.05,xmax=1.05, ymin=-0.05,ymax=1.05]
    \addplot3[contour gnuplot={levels={0.5}, labels=false, draw color=color0},
    mesh/rows=48,mesh/cols=48,thick]
    table {./pics/microjets/data/microjet_trajectory_6236_contour_017088.dat};
  \end{axis}
  \begin{axis}[
    view = {0}{90},
    hide axis, at=(iso06), anchor=south, scale only axis,
    height=1.10833333333cm, width=1.10833333333cm,
    xmin=-0.05,xmax=1.05, ymin=-0.05,ymax=1.05]
    \addplot3[contour gnuplot={levels={0.5}, labels=false, draw color=color0},
    mesh/rows=48,mesh/cols=48,thick]
    table {./pics/microjets/data/microjet_trajectory_6236_contour_018488.dat};
  \end{axis}
  \begin{axis}[
    view = {0}{90},
    hide axis, at=(iso07), anchor=south, scale only axis,
    height=1.10833333333cm, width=1.10833333333cm,
    xmin=-0.05,xmax=1.05, ymin=-0.05,ymax=1.05]
    \addplot3[contour gnuplot={levels={0.5}, labels=false, draw color=color0},
    mesh/rows=48,mesh/cols=48,thick]
    table {./pics/microjets/data/microjet_trajectory_6236_contour_019688.dat};
  \end{axis}
  \begin{axis}[
    view = {0}{90},
    hide axis, at=(iso08), anchor=south, scale only axis,
    height=1.10833333333cm, width=1.10833333333cm,
    xmin=-0.05,xmax=1.05, ymin=-0.05,ymax=1.05]
    \addplot3[contour gnuplot={levels={0.5}, labels=false, draw color=color0},
    mesh/rows=48,mesh/cols=48,thick]
    table {./pics/microjets/data/microjet_trajectory_6236_contour_020888.dat};
  \end{axis}
  \begin{axis}[
    view = {0}{90},
    hide axis, at=(iso09), anchor=south, scale only axis,
    height=1.10833333333cm, width=1.10833333333cm,
    xmin=-0.05,xmax=1.05, ymin=-0.05,ymax=1.05]
    \addplot3[contour gnuplot={levels={0.5}, labels=false, draw color=color0},
    mesh/rows=48,mesh/cols=48,thick]
    table {./pics/microjets/data/microjet_trajectory_6236_contour_022088.dat};
  \end{axis}
  \begin{axis}[
    view = {0}{90},
    hide axis, at=(iso10), anchor=south, scale only axis,
    height=1.10833333333cm, width=1.10833333333cm,
    xmin=-0.05,xmax=1.05, ymin=-0.05,ymax=1.05]
    \addplot3[contour gnuplot={levels={0.5}, labels=false, draw color=color0},
    mesh/rows=48,mesh/cols=48,thick]
    table {./pics/microjets/data/microjet_trajectory_6236_contour_023488.dat};
  \end{axis}
  \begin{axis}[
    view = {0}{90},
    hide axis, at=(iso11), anchor=south, scale only axis,
    height=1.10833333333cm, width=1.10833333333cm,
    xmin=-0.05,xmax=1.05, ymin=-0.05,ymax=1.05]
    \addplot3[contour gnuplot={levels={0.5}, labels=false, draw color=color0},
    mesh/rows=48,mesh/cols=48,thick]
    table {./pics/microjets/data/microjet_trajectory_6236_contour_024688.dat};
  \end{axis}

\end{tikzpicture}

\begin{tikzpicture}[baseline]
  \begin{axis}[
    grid=major,
    width=14.0cm,
    height=6.0cm,
    view = {0}{90},
    %style={font=\large},
    xmin=143.176339666, xmax=263.176339666,
    ymin=-1.0, ymax=1.0,
    xlabel=$t\units{[\mu s]}$,
    %ylabel=$y$,
    legend columns=5,
    legend style={at={(0.5,1.02)},anchor=south,draw=none,font=\small,
    /tikz/column 2/.style={column sep=7pt},
    /tikz/column 4/.style={column sep=7pt},
    /tikz/column 6/.style={column sep=7pt},
    /tikz/column 8/.style={column sep=7pt},
    name=leg,
    },
    legend cell align=left,
    ]
    \addplot[mark=*, mark size=1.3, solid, color13, thick] % x-xb
    table[x index=0,y index=1]
    {./pics/microjets/data/microjet_trajectory_255.dat};
    \addlegendentry{$x_\mathrm{tip}-x_\mathrm{B}$}
    \addplot[mark=triangle*, mark size=1.5, solid, color9, thick] % y-yb
    table[x index=0,y index=2]
    {./pics/microjets/data/microjet_trajectory_255.dat};
    \addlegendentry{$y_\mathrm{tip}-y_\mathrm{B}$}
    \addplot[mark=square*, mark size=1.3, solid, color11, thick] % z-zb
    table[x index=0,y index=3]
    {./pics/microjets/data/microjet_trajectory_255.dat};
    \addlegendentry{$z_\mathrm{tip}-z_\mathrm{B}$}
    \addplot[solid, color8, thick] % |X-Xb|
    table[x index=0,y index=4]
    {./pics/microjets/data/microjet_trajectory_255.dat};
    \addlegendentry{$|\mathbf{x}_\mathrm{tip}-\mathbf{x}_\mathrm{B}|$}
    \addplot[solid, very thick, color0] % rb
    table[x index=0,y index=5]
    {./pics/microjets/data/microjet_trajectory_255.dat};
    \addlegendentry{$R_\mathrm{B}$}
    \addplot[dashed, thick, color13, thick] % xfit
    table[x index=0,y index=1]
    {./pics/microjets/data/microjet_trajectory_255_fit.dat};
    \addplot[dashed, thick, color9, thick] % yfit
    table[x index=0,y index=2]
    {./pics/microjets/data/microjet_trajectory_255_fit.dat};
    \addplot[dashed, thick, color11, thick] % zfit
    table[x index=0,y index=3]
    {./pics/microjets/data/microjet_trajectory_255_fit.dat};
    \addplot[dashed, thick, color8, thick] % |Xfit|
    table[x index=0,y index=4]
    {./pics/microjets/data/microjet_trajectory_255_fit.dat};
    \addplot [mark=none, color0, very thick, dashed] coordinates {(189.8078820799264,-1) (189.8078820799264,1)};
    \addplot [mark=none, color0, very thick] coordinates {(196.06121493043952,-1) (196.06121493043952,1)};
    \addplot [mark=none, color0, very thick] coordinates {(210.29146440119723,-1) (210.29146440119723,1)};
    \addplot [mark=none, color0, dotted] coordinates {(153.3704665181664,-1) (153.3704665181664,1)};
    \addplot [mark=none, color0, dotted] coordinates {(162.85729949867155,-1) (162.85729949867155,1)};
    \addplot [mark=none, color0, dotted] coordinates {(173.92527130926086,-1) (173.92527130926086,1)};
    \addplot [mark=none, color0, dotted] coordinates {(183.412104289766,-1) (183.412104289766,1)};
    \addplot [mark=none, color0, dotted] coordinates {(192.89893727027115,-1) (192.89893727027115,1)};
    \addplot [mark=none, color0, dotted] coordinates {(202.3857702507763,-1) (202.3857702507763,1)};
    \addplot [mark=none, color0, dotted] coordinates {(213.45374206136563,-1) (213.45374206136563,1)};
    \addplot [mark=none, color0, dotted] coordinates {(222.94057504187074,-1) (222.94057504187074,1)};
    \addplot [mark=none, color0, dotted] coordinates {(232.4274080223759,-1) (232.4274080223759,1)};
    \addplot [mark=none, color0, dotted] coordinates {(243.49537983296523,-1) (243.49537983296523,1)};
    \addplot [mark=none, color0, dotted] coordinates {(252.98221281347037,-1) (252.98221281347037,1)};

    \coordinate (iso01) at (axis cs:153.3704665181664,-1.0);
    \coordinate (iso02) at (axis cs:162.85729949867155,-1.0);
    \coordinate (iso03) at (axis cs:173.92527130926086,-1.0);
    \coordinate (iso04) at (axis cs:183.412104289766,-1.0);
    \coordinate (iso05) at (axis cs:192.89893727027115,-1.0);
    \coordinate (iso06) at (axis cs:202.3857702507763,-1.0);
    \coordinate (iso07) at (axis cs:213.45374206136563,-1.0);
    \coordinate (iso08) at (axis cs:222.94057504187074,-1.0);
    \coordinate (iso09) at (axis cs:232.4274080223759,-1.0);
    \coordinate (iso10) at (axis cs:243.49537983296523,-1.0);
    \coordinate (iso11) at (axis cs:252.98221281347037,-1.0);

  \end{axis}

  % marker with bubble index
  \node[align=center, font=\small, color0, anchor=west]
  at (leg.east)
  {\quad\raisebox{2.5pt}{\pgfuseplotmark{triangle*}}\hspace{3pt}3};

  \begin{axis}[
    view = {0}{90},
    hide axis, at=(iso01), anchor=south, scale only axis,
    height=1.10833333333cm, width=1.10833333333cm,
    xmin=-0.05,xmax=1.05, ymin=-0.05,ymax=1.05]
    \addplot3[contour gnuplot={levels={0.5}, labels=false, draw color=color0},
    mesh/rows=48,mesh/cols=48,thick]
    table {./pics/microjets/data/microjet_trajectory_255_contour_019088.dat};
  \end{axis}
  \begin{axis}[
    view = {0}{90},
    hide axis, at=(iso02), anchor=south, scale only axis,
    height=1.10833333333cm, width=1.10833333333cm,
    xmin=-0.05,xmax=1.05, ymin=-0.05,ymax=1.05]
    \addplot3[contour gnuplot={levels={0.5}, labels=false, draw color=color0},
    mesh/rows=48,mesh/cols=48,thick]
    table {./pics/microjets/data/microjet_trajectory_255_contour_020288.dat};
  \end{axis}
  \begin{axis}[
    view = {0}{90},
    hide axis, at=(iso03), anchor=south, scale only axis,
    height=1.10833333333cm, width=1.10833333333cm,
    xmin=-0.05,xmax=1.05, ymin=-0.05,ymax=1.05]
    \addplot3[contour gnuplot={levels={0.5}, labels=false, draw color=color0},
    mesh/rows=48,mesh/cols=48,thick]
    table {./pics/microjets/data/microjet_trajectory_255_contour_021688.dat};
  \end{axis}
  \begin{axis}[
    view = {0}{90},
    hide axis, at=(iso04), anchor=south, scale only axis,
    height=1.10833333333cm, width=1.10833333333cm,
    xmin=-0.05,xmax=1.05, ymin=-0.05,ymax=1.05]
    \addplot3[contour gnuplot={levels={0.5}, labels=false, draw color=color0},
    mesh/rows=48,mesh/cols=48,thick]
    table {./pics/microjets/data/microjet_trajectory_255_contour_022888.dat};
  \end{axis}
  \begin{axis}[
    view = {0}{90},
    hide axis, at=(iso05), anchor=south, scale only axis,
    height=1.10833333333cm, width=1.10833333333cm,
    xmin=-0.05,xmax=1.05, ymin=-0.05,ymax=1.05]
    \addplot3[contour gnuplot={levels={0.5}, labels=false, draw color=color0},
    mesh/rows=48,mesh/cols=48,thick]
    table {./pics/microjets/data/microjet_trajectory_255_contour_024088.dat};
  \end{axis}
  \begin{axis}[
    view = {0}{90},
    hide axis, at=(iso06), anchor=south, scale only axis,
    height=1.10833333333cm, width=1.10833333333cm,
    xmin=-0.05,xmax=1.05, ymin=-0.05,ymax=1.05]
    \addplot3[contour gnuplot={levels={0.5}, labels=false, draw color=color0},
    mesh/rows=48,mesh/cols=48,thick]
    table {./pics/microjets/data/microjet_trajectory_255_contour_025288.dat};
  \end{axis}
  \begin{axis}[
    view = {0}{90},
    hide axis, at=(iso07), anchor=south, scale only axis,
    height=1.10833333333cm, width=1.10833333333cm,
    xmin=-0.05,xmax=1.05, ymin=-0.05,ymax=1.05]
    \addplot3[contour gnuplot={levels={0.5}, labels=false, draw color=color0},
    mesh/rows=48,mesh/cols=48,thick]
    table {./pics/microjets/data/microjet_trajectory_255_contour_026688.dat};
  \end{axis}
  \begin{axis}[
    view = {0}{90},
    hide axis, at=(iso08), anchor=south, scale only axis,
    height=1.10833333333cm, width=1.10833333333cm,
    xmin=-0.05,xmax=1.05, ymin=-0.05,ymax=1.05]
    \addplot3[contour gnuplot={levels={0.5}, labels=false, draw color=color0},
    mesh/rows=48,mesh/cols=48,thick]
    table {./pics/microjets/data/microjet_trajectory_255_contour_027888.dat};
  \end{axis}
  \begin{axis}[
    view = {0}{90},
    hide axis, at=(iso09), anchor=south, scale only axis,
    height=1.10833333333cm, width=1.10833333333cm,
    xmin=-0.05,xmax=1.05, ymin=-0.05,ymax=1.05]
    \addplot3[contour gnuplot={levels={0.5}, labels=false, draw color=color0},
    mesh/rows=48,mesh/cols=48,thick]
    table {./pics/microjets/data/microjet_trajectory_255_contour_029088.dat};
  \end{axis}
  \begin{axis}[
    view = {0}{90},
    hide axis, at=(iso10), anchor=south, scale only axis,
    height=1.10833333333cm, width=1.10833333333cm,
    xmin=-0.05,xmax=1.05, ymin=-0.05,ymax=1.05]
    \addplot3[contour gnuplot={levels={0.5}, labels=false, draw color=color0},
    mesh/rows=48,mesh/cols=48,thick]
    table {./pics/microjets/data/microjet_trajectory_255_contour_030501.dat};
  \end{axis}
  \begin{axis}[
    view = {0}{90},
    hide axis, at=(iso11), anchor=south, scale only axis,
    height=1.10833333333cm, width=1.10833333333cm,
    xmin=-0.05,xmax=1.05, ymin=-0.05,ymax=1.05]
    \addplot3[contour gnuplot={levels={0.5}, labels=false, draw color=color0},
    mesh/rows=48,mesh/cols=48,thick]
    table {./pics/microjets/data/microjet_trajectory_255_contour_031721.dat};
  \end{axis}

\end{tikzpicture}

   \caption{
  Temporal evolution of microjets for three selected bubbles.
  Trajectory of microjet tip relative to the bubble center (solid lines),
  linear fit (dashed lines) and equivalent radius (black solid line).
  All quantities are normalized by the corresponding initial radius.
  Fitting range $[t_{\mathrm{tip},i}, t_{\mathrm{imp},i}]$ (vertical solid lines), 
  collapse wave arrival $t_\mathrm{F}$ (vertical dashed line) and 
  intervals of $10\units{\mu s}$ with corresponding iso-lines of $\alpha_2=0.5$ at the bottom
  (vertical dotted lines).
  }
  \label{fig:microjet_traj}
\end{figure}
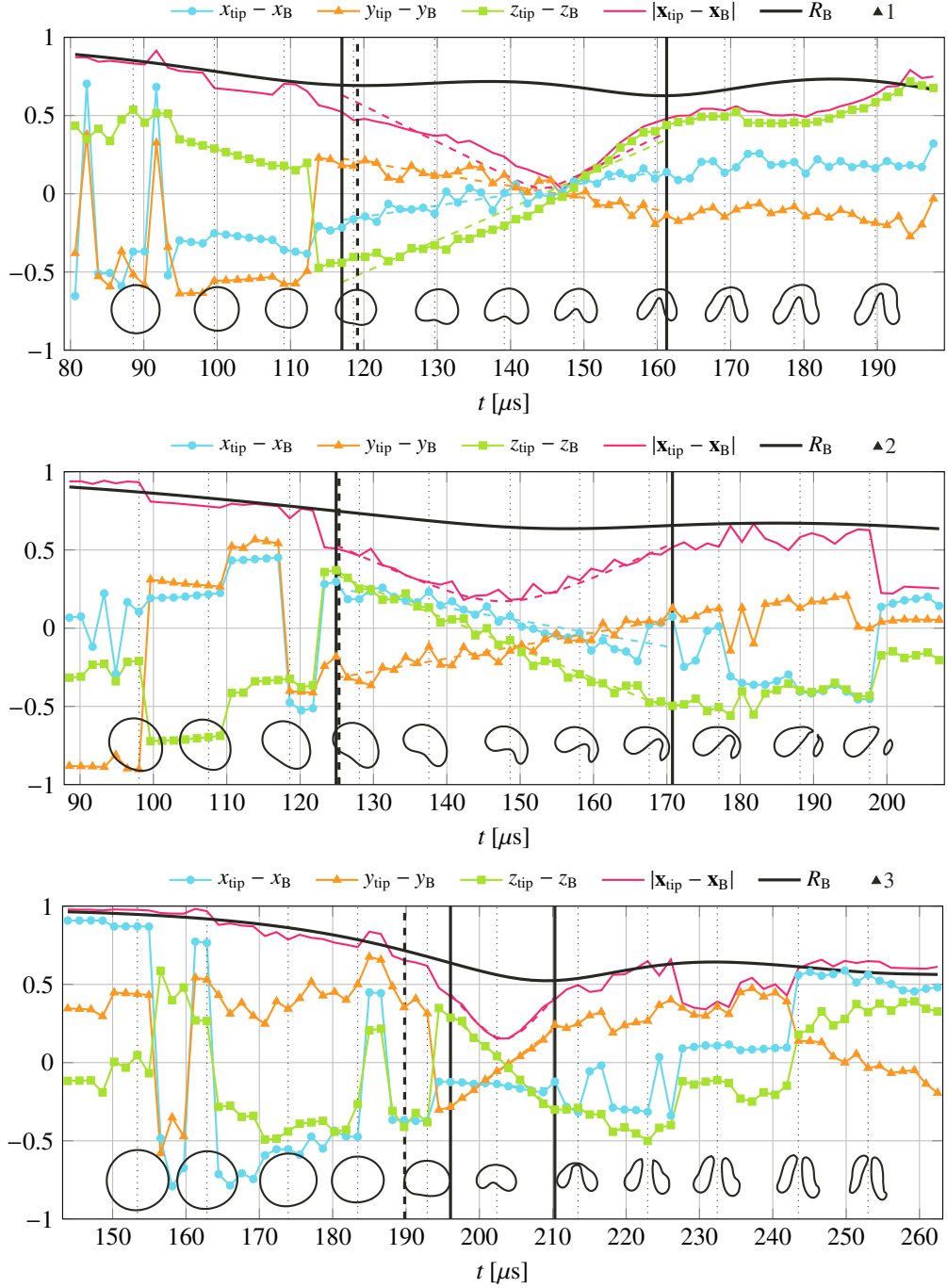

Fig.~\ref{fig:microjet_traj} illustrates the evolution of
the microjet for three bubbles. 
The relative location of the tip, $\ve x_{\mathrm{tip},i} - \ve x_{\mathrm{B}_i}$,
as well as the bubble radius $R_{\mathrm{B}_i}$
are displayed as a function of time. Additionally,
bubble shapes are shown for selected time instants.
At the beginning of the collapse process,
the bubble surface is largely spherical and possesses a positive curvature.
Therefore, the distance between the location of minimum curvature
and the bubble center is approximately equal to the equivalent radius,
but the location itself is not well-defined and thus bounces from one point to another.
Once the microjet starts to form, the curvature changes its sign.
The location of minimum curvature then identifies the tip of the microjet.
The microjet deforms the bubble into a cap-like shape until
it pierces through the bubble on the opposite surface; see Fig.~\ref{fig:microjet_traj}.
At this time, the distance between the location of minimum curvature and the bubble center 
again approximately equals the equivalent radius.
Hence, the characteristic quantities of the microjets are evaluated during the time
interval $[t_{\mathrm{tip},i}, t_{\mathrm{imp},i}]$ for which
\begin{equation}
  |\mathbf x_{\mathrm{B}_i} - \mathbf x_{\mathrm{tip},i}| < 0.75 R_{\mathrm{B}_i}
  \label{eqn:microjet_criterion}
\end{equation}
holds.
As observed in Fig.~\ref{fig:microjet_traj}, the relative trajectory
$\mathbf x_{\mathrm{tip},i}-\mathbf x_{\mathrm{B}_i}$ of the tip of the
microjet travels with approximately a constant velocity  within this interval. 
The microjet velocity $\mathbf u_{\mathrm{tip},i}$ is  defined
by the time derivative of a linear fit of $\mathbf x_{\mathrm{tip},i}-\mathbf x_{\mathrm{B}_i}$ in
the time interval $[t_{\mathrm{tip},i}, t_{\mathrm{imp},i}]$.
In order to obtain reliable statistics,
the fitting range is required to comprise at least six samples in time
(i.e., has duration of at least $10\units{\mu s}$)
and the root-mean-square error of the fitting has to be below $0.1\,R_{\mathrm{B}_i}(0)$.
Due to the limited data sampling frequency and the complexity
of the microjet tip trajectories,
not all bubbles satisfy these requirements.
Such bubbles are excluded from the subsequent analysis of the microjets,
leaving about $7500$ bubbles (i.e., $60\%$ of the bubbles) for further evaluation.

\begin{figure}[tb]
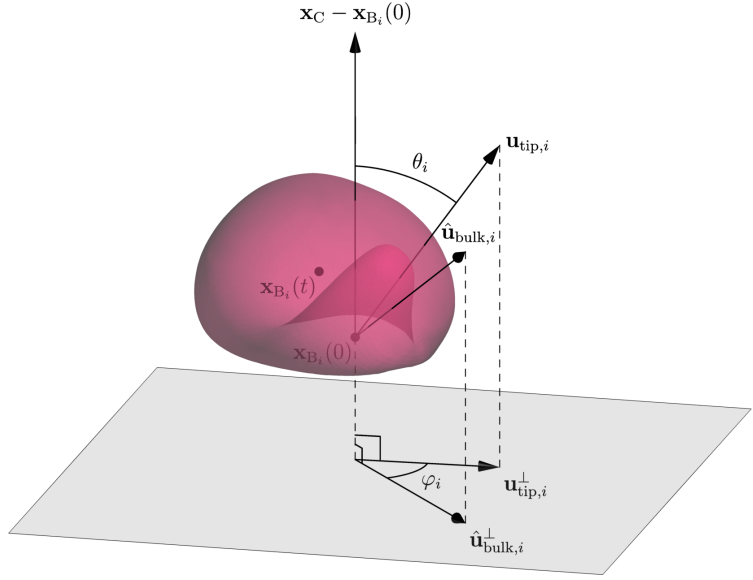

  \centering
  \myincludegraphics[width=0.6\textwidth,keepaspectratio]{./pics/microjets/asymmetric_bubble.pdf}
  \caption{Bubble surface with microjet velocity $\mathbf u_{\mathrm{tip},i}$,
  bulk velocity indicator $\hat{\mathbf u}_{\mathrm{bulk},i}$ as well as their
  projections $\mathbf{u}^\perp_{\mathrm{tip},i}$ and
  $\hat{\mathbf{u}}^\perp_{\mathrm{bulk},i}$ onto a plane perpendicular to
  the radial direction.}
  \label{fig:microjet_sketch}
\end{figure}

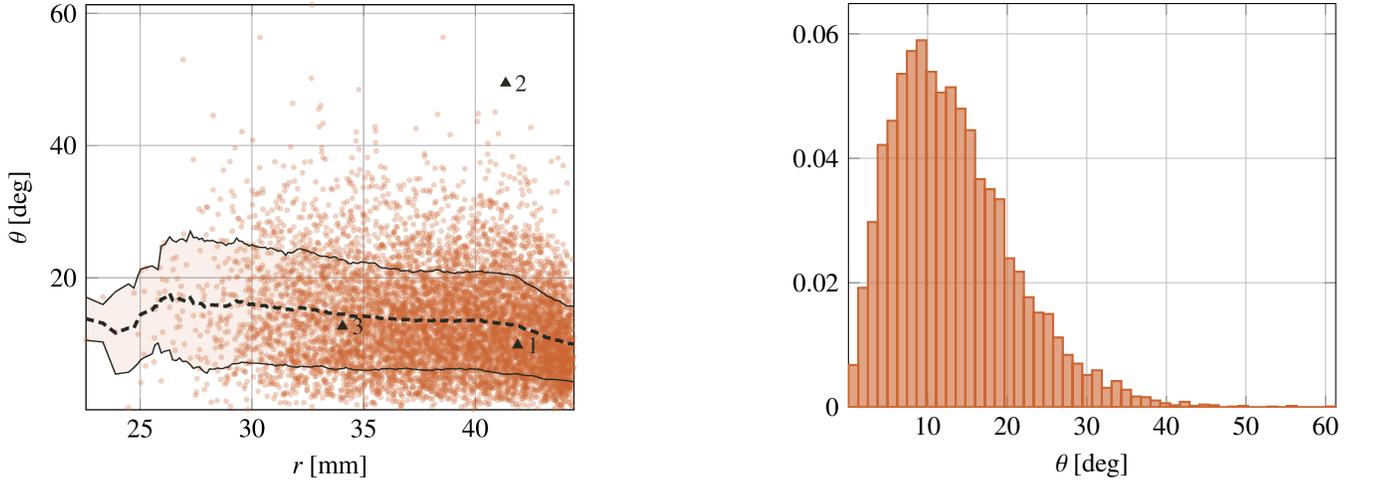
\begin{figure}[tb]
  \centering
\begin{tikzpicture}[baseline]
  \pgfplotsset{set layers}% using layers
  \begin{axis}[
    set layers,
    grid=major,
    width=0.49\textwidth,
    %style={font=\large},
    xmin=22.5720926146, xmax=44.4122255804,
    ymin=0.049150334753, ymax=61.2821487841,
    xlabel=$r\units{[mm]}$,
    ylabel=$\theta\units{[deg]}$,
    clip mode=individual,
    ]
    \addplot[only marks, mark=*, mark size=1.0pt, color15, %mark options={solid,thick,fill=color4, opacity=0.6},
    line width= 0pt,
    opacity=0.3,
    ]
    table [x index=0,y index=1]
    {./pics/microjets/data/cloud12500_theta_vs_radial_scatter.dat};
    \addplot+[nodes near coords, nodes near coords style={anchor=west,font=\small},
    only marks, point meta=explicit symbolic, mark=triangle*, mark options={color0}]
    table [meta=label] {
    x y label
    41.89763704554231 9.794704416973353 1
    41.35693072811375 49.42719293291682 2
    34.053310264936066 12.643623377410279 3
    };

    \addplot[color0, thin, solid, name path=A] table[x index=0,y index=2]
    {./pics/microjets/data/cloud12500_theta_vs_radial_avg_tikz.dat};
    \addplot[color0, thin, solid, name path=B] table[x index=0,y index=3]
    {./pics/microjets/data/cloud12500_theta_vs_radial_avg_tikz.dat};
    \addplot [fill=color15, fill opacity=0.1] fill between[of=A and B];
    \addplot[mark=none, color0, very thick, densely dashed] table [x index=0, y index=1]
    {./pics/microjets/data/cloud12500_theta_vs_radial_avg_tikz.dat};
  \end{axis}
\end{tikzpicture}
  \hfill%
\begin{tikzpicture}[baseline]
  \begin{axis}[
    grid=major,
    width=0.49\textwidth,
    xlabel=$\theta\units{[deg]}$,
    ymin=0,
    xmin=0.049150334753, xmax=61.2821487841,
    scaled ticks=false, tick label style={/pgf/number format/fixed},
    ]
    \addplot[color15,fill=color15,fill opacity=0.6, thick,
    hist={bins=50,density=true},
    ]
    table [y index=0] {./pics/microjets/data/cloud12500_theta_hist.dat};
  \end{axis}
\end{tikzpicture}
  \caption{Microjet inclination angle $\theta_i$ depending on
  radial location (left) and PDF of the inclination angle (right).}
  \label{fig:microjet_theta}
\end{figure}

As reported in preceding studies on cloud cavitation collapse%
~\cite{Bremond:2006,Tiwari:2015},
the microjets point towards the core of the cloud. As shown in the present work, the axes of these microjets
are not perfectly aligned with the radial direction
$\mathbf x_\mathrm{C}-\mathbf x_{\mathrm{B}_i}(0)$
from the initial bubble center to the cloud center.
The inclination angle $\theta_i$ denotes the angle between 
the radial direction and the direction of the microjet velocity
corresponding to bubble $i$
as illustrated in Fig.~\ref{fig:microjet_sketch}.
A microjet with $\theta_i=0^\circ$ is directed towards the cloud center.
Values of the inclination angle for bubbles shown 
in Fig.~\ref{fig:microjet_traj} 
are given 
in Table~\ref{tab:microjet_table} where
the microjet of bubble ``2'' is distinguished by stronger inclination.
Fig.~\ref{fig:microjet_theta} depicts a scatter plot of the inclination
angle $\theta_i$ versus the radial distance $r$.
All scatter plots
shown in this subsection also contain the moving average and the standard deviation
computed with a window length equal to $10\%$ of the corresponding horizontal axis range.
The bubbles selected in Fig.~\ref{fig:microjet_traj} are also marked.
Furthermore, Fig.~\ref{fig:microjet_theta} depicts the Probability Density
Function (PDF) of the inclination angle.
The average inclination angle for the present cloud collapse process is $13.2^\circ$.
Furthermore, $90\%$ of the bubbles exhibit an inclination angle smaller than $24^\circ$.
Local mean values of the inclination angle range from $10^\circ$ at $r=45\units{mm}$ to
$18^\circ$ at $r=26\units{mm}$.
As a result, the microjet inclination angle  increases slightly 
towards the cloud center indicating a weak dependence on the collapse wave speed,
which strongly depends on $r$.
Very large inclination angles in the range of $35^\circ$ to $61^\circ$ are
observed for $1\%$ of the bubbles.
Closer examination of these microjets 
reveals that the microjet inclination is affected by the surrounding bubbles.
Fig.~\ref{fig:microjet_surrounding} shows the neighborhood of
a bubble with an inclination angle of $50^\circ$.
\begin{figure}[!tbp]
  \centering
  \begin{minipage}[c]{0.49\textwidth}
    \centering
    \begin{tikzpicture}[baseline]
      \node[inner sep=0pt] at (0,0) [pos=0, anchor=south west, yshift=-7cm]
      {\includegraphics[height=6.4cm]{pics/microjets/bubble_06236_surround_pv_greyGrad_scaled.png}};
    \end{tikzpicture}
    \caption{
    Neighborhood of a small bubble (red) with a large inclination angle of
    $50^\circ$ that is attracted towards a significantly larger bubble nearby (brown).
    }
    \label{fig:microjet_surrounding}
  \end{minipage}
  \hfill
  \begin{minipage}[c]{0.49\textwidth}
    \centering
\begin{tikzpicture}[baseline]
  \begin{axis}[
    grid=major,
    width=\textwidth, % since this guy lives inside a minipage
    xlabel=$\varphi\units{[deg]}$,
    ymin=0,
    xmin=0.00041740014744, xmax=179.834406011,
    scaled ticks=false, tick label style={/pgf/number format/fixed},
    y tick label style={
    /pgf/number format/.cd, precision=3, /tikz/.cd
    },
    ]
    \addplot[color15,fill=color15,fill opacity=0.6, thick,
    hist={bins=50,density=true},
    ]
    table [y index=0] {./pics/microjets/data/cloud12500_dtau_hist.dat};
  \end{axis}
\end{tikzpicture}
     \caption{
    PDF of angle $\varphi_i$ between $\mathbf{u}^\perp_{\mathrm{tip},i}$ and
    $\hat{\mathbf{u}}^\perp_{\mathrm{bulk},i}$.
    }
    \label{fig:pdf_phi}
  \end{minipage}
\end{figure}
The microjet is inclined towards one specific neighboring bubble that
has a significantly larger size than the considered bubble as well
as all the other bubbles in its vicinity.
This observation suggests that the microjet inclination mainly depends on the
geometrical arrangement of the bubbles.
Larger bubbles have a stronger influence on the liquid flow.
Assuming potential flow
away from the bubbles, the velocity in the surrounding liquid is given by ~\cite{Mettin:1997}:
\begin{equation}
\label{eq:potential_flow}
  \mathbf u(\mathbf x, t) = \sum\limits_{j=1}^{n_B}
  \frac{R_{\mathrm{B}_j}^2\dot R_{\mathrm{B}_j}}
  {|\mathbf x - \mathbf x_{\mathrm{B}_j}|^3}(\mathbf x - \mathbf x_{\mathrm{B}_j}).
\end{equation}
Furthermore, the bubble compression rate $\dot
R_{\mathrm{B}_j}$ in Eq.~\eqref{eq:potential_flow} is taken to be constant and negative
leading to a non-dimensional bulk velocity
\begin{equation}
  \hat{\mathbf{u}}_{\mathrm{bulk},i} = \sum\limits_{\substack{j=1\\j\neq i}}^{n_B}
  \frac{-R_{\mathrm{B}_j}^2(0)}
  {|\mathbf x_{\mathrm{B}_i}(0) - \mathbf x_{\mathrm{B}_j}(0)|^3}(\mathbf x_{\mathrm{B}_i}(0) - \mathbf x_{\mathrm{B}_j}(0))
  \label{eq:microjet_bulk}
\end{equation}
at the center $\ve x_{\mathrm{B}_i}$ of bubble $i$.
Eq.~\eqref{eq:microjet_bulk}
provides an estimation for the bulk flow direction and its strength which is
purely based on the initial geometrical arrangement.
The assumption of constant $\dot R_{\mathrm{B}_j}$ does not exactly hold 
for cloud collapses since the bubbles behind the collapse front compress
but remain at rest ahead of it.
Therefore, Eq.~\eqref{eq:microjet_bulk} characterizes only 
the flow velocity perpendicular to the radial direction
which is governed by the arrangement of bubbles along the collapse front.
To examine the influence of the bulk flow induced by the collapse
of the surrounding bubbles on the microjet direction, 
$\mathbf u_{\mathrm{tip},i}$ and $\hat{\mathbf u}_{\mathrm{bulk},i}$
are  projected onto a plane perpendicular to the radial direction.
The resulting velocity components are marked by the additional superscript $(\cdot)^{\perp}$
and are also schematically represented in Fig.~\ref{fig:microjet_sketch}. 
The angle between $\mathbf{u}^\perp_{\mathrm{tip},i}$
and
$\hat{\mathbf{u}}^\perp_{\mathrm{bulk},i}$ is denoted $\varphi_i$.
The PDF of $\varphi_i$ 
as well as scatter plots of  $\varphi_i$ versus $\theta_i$ and $\theta_i$ versus
the magnitude $\vert \hat{\mathbf{u}}^\perp_{\mathrm{bulk},i} \vert$ of the projected bulk velocity are shown in Figs.~\ref{fig:pdf_phi} and~\ref{fig:microjet_dtau}, respectively.
For $68\%$ of the bubbles, $\varphi_i$ is smaller than $45^\circ$,
which demonstrates that the microjets are inclined towards the  direction of the bulk liquid flow around the bubble.
This angle reduces with increasing inclination.
The mean value of $\varphi_i$ is $45^\circ$ for $\theta_i=10^\circ$ and $25^\circ$ for $\theta_i=40^\circ$.
Moreover, a positive correlation between the inclination angle $\theta_i$ and
the magnitude of the projected component of the bulk flow indicator $|\hat{\mathbf u}^\perp_{\mathrm{bulk},i}|$
is observed.
\begin{figure}[tb]
  \centering
\begin{tikzpicture}[baseline]
  \pgfplotsset{set layers}% using layers
  \begin{axis}[
    set layers,
    grid=major,
    width=0.49\textwidth,
    %style={font=\large},
    xmin=0.049150334753, xmax=61.2821487841,
    ymin=0.00041740014744, ymax=179.834406011,
    xlabel=$\theta\units{[deg]}$,
    ylabel=$\varphi\units{[deg]}$,
    clip mode=individual,
    ]
    \addplot[only marks, mark=*, mark size=1.0pt, color15, %mark options={solid,thick,fill=color4, opacity=0.6},
    line width= 0pt,
    opacity=0.3,
    ]
    table [x index=0,y index=1]
    {./pics/microjets/data/cloud12500_dtau_vs_theta_scatter.dat};
    \addplot+[nodes near coords, nodes near coords style={anchor=west,font=\small},
    only marks, point meta=explicit symbolic, mark=triangle*, mark options={color0}]
    table [meta=label] {
    x y label
    9.794704416973353 50.618368628603335 1
    49.42719293291682 22.931093393496162 2
    12.643623377410279 92.51382690282503 3
    };

    \addplot[color0, thin, solid, name path=A] table[x index=0,y index=2]
    {./pics/microjets/data/cloud12500_dtau_vs_theta_avg_tikz.dat};
    \addplot[color0, thin, solid, name path=B] table[x index=0,y index=3]
    {./pics/microjets/data/cloud12500_dtau_vs_theta_avg_tikz.dat};
    \addplot [fill=color15, fill opacity=0.1] fill between[of=A and B];
    \addplot[mark=none, color0, very thick, densely dashed] table [x index=0, y index=1]
    {./pics/microjets/data/cloud12500_dtau_vs_theta_avg_tikz.dat};
  \end{axis}
\end{tikzpicture}
  \hfill%
\begin{tikzpicture}[baseline]
  \pgfplotsset{set layers}% using layers
  \begin{axis}[
    set layers,
    grid=major,
    width=0.49\textwidth,
    %style={font=\large},
    xmin=0.00164190683823, xmax=0.463672461258,
    ymin=0.049150334753, ymax=61.2821487841,
    xlabel=$|\hat{\mathbf u}^\perp_\mathrm{bulk}|$,
    ylabel=$\theta\units{[deg]}$,
    clip mode=individual,
    ]
    \addplot[only marks, mark=*, mark size=1.0pt, color15, %mark options={solid,thick,fill=color4, opacity=0.6},
    line width= 0pt,
    opacity=0.3,
    ]
    table [x index=0,y index=1]
    {./pics/microjets/data/cloud12500_theta_vs_indicator_scatter.dat};
    \addplot+[nodes near coords, nodes near coords style={anchor=west,font=\small},
    only marks, point meta=explicit symbolic, mark=triangle*, mark options={color0}]
    table [meta=label] {
    x y label
    0.00500994489213711 9.794704416973353 1
    0.2926169766370557 49.42719293291682 2
    0.14763085938654397 12.643623377410279 3
    };

    \addplot[color0, thin, solid, name path=A] table[x index=0,y index=2]
    {./pics/microjets/data/cloud12500_theta_vs_indicator_avg_tikz.dat};
    \addplot[color0, thin, solid, name path=B] table[x index=0,y index=3]
    {./pics/microjets/data/cloud12500_theta_vs_indicator_avg_tikz.dat};
    \addplot [fill=color15, fill opacity=0.1] fill between[of=A and B];
    \addplot[mark=none, color0, very thick, densely dashed] table [x index=0, y index=1]
    {./pics/microjets/data/cloud12500_theta_vs_indicator_avg_tikz.dat};
  \end{axis}
\end{tikzpicture}
  \caption{Angle $\varphi_i$ between 
  $\mathbf{u}^\perp_{\mathrm{tip},i}$ and
  $\hat{\mathbf{u}}^\perp_{\mathrm{bulk},i}$ depending on inclination angle
  $\theta_i$ (left) and inclination angle depending on the magnitude 
  $|\hat{\mathbf u}^\perp_{\mathrm{bulk},i}|$ (right).}
  \label{fig:microjet_dtau}
\end{figure}

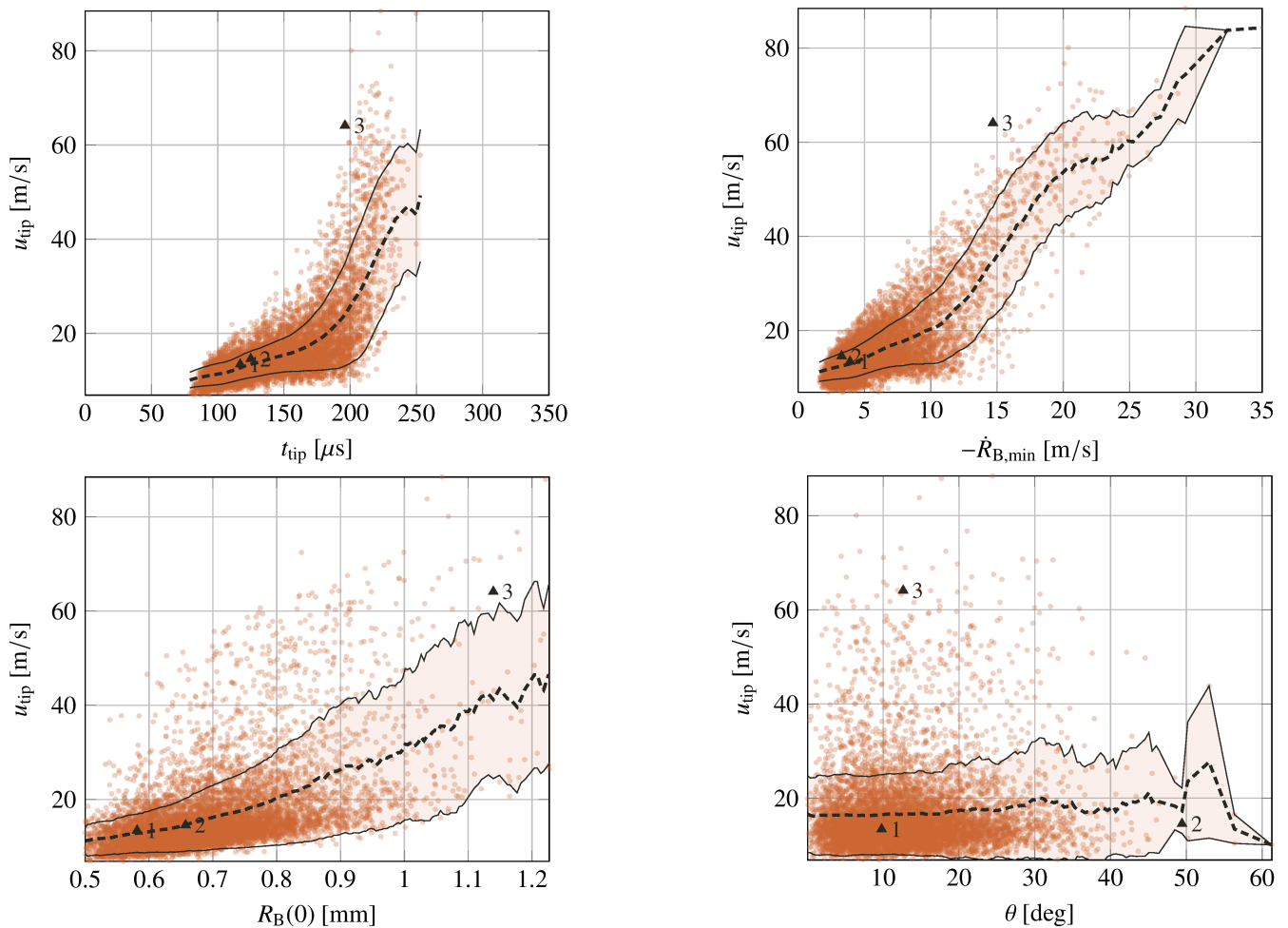
\begin{figure}[tb]
  \centering
\begin{tikzpicture}[baseline]
  \pgfplotsset{set layers}% using layers
  \begin{axis}[
    set layers,
    grid=major,
    width=0.49\textwidth,
    %style={font=\large},
    xmin=0.0, xmax=350.0,
    ymin=6.86789214352, ymax=88.4447909721,
    xlabel=$t_\mathrm{tip}\units{[\mu s]}$,
    ylabel=$u_\mathrm{tip}\units{[m/s]}$,
    clip mode=individual,
    ]
    \addplot[only marks, mark=*, mark size=1.0pt, color15, %mark options={solid,thick,fill=color4, opacity=0.6},
    line width= 0pt,
    opacity=0.3,
    ]
    table [x index=0,y index=1]
    {./pics/microjets/data/cloud12500_utip_vs_tiptime_scatter.dat};
    \addplot+[nodes near coords, nodes near coords style={anchor=west,font=\small},
    only marks, point meta=explicit symbolic, mark=triangle*, mark options={color0}]
    table [meta=label] {
    x y label
    117.00427342623004 13.41752672148342 1
    124.90996757665098 14.567002081676495 2
    196.06121493043952 64.10994451026495 3
    };

    \addplot[color0, thin, solid, name path=A] table[x index=0,y index=2]
    {./pics/microjets/data/cloud12500_utip_vs_tiptime_avg_tikz.dat};
    \addplot[color0, thin, solid, name path=B] table[x index=0,y index=3]
    {./pics/microjets/data/cloud12500_utip_vs_tiptime_avg_tikz.dat};
    \addplot [fill=color15, fill opacity=0.1] fill between[of=A and B];
    \addplot[mark=none, color0, very thick, densely dashed] table [x index=0, y index=1]
    {./pics/microjets/data/cloud12500_utip_vs_tiptime_avg_tikz.dat};
  \end{axis}
\end{tikzpicture}
  \hfill%
\begin{tikzpicture}[baseline]
  \pgfplotsset{set layers}% using layers
  \begin{axis}[
    set layers,
    grid=major,
    width=0.49\textwidth,
    %style={font=\large},
    xmin=0.0, xmax=35.0,
    ymin=6.86789214352, ymax=88.4447909721,
    xlabel=$-\dot R_{\mathrm{B},\min}\units{[m/s]}$,
    ylabel=$u_\mathrm{tip}\units{[m/s]}$,
    clip mode=individual,
    ]
    \addplot[only marks, mark=*, mark size=1.0pt, color15, %mark options={solid,thick,fill=color4, opacity=0.6},
    line width= 0pt,
    opacity=0.3,
    ]
    table [x index=0,y index=1]
    {./pics/microjets/data/cloud12500_utip_vs_maxrbdot_scatter.dat};
    \addplot+[nodes near coords, nodes near coords style={anchor=west,font=\small},
    only marks, point meta=explicit symbolic, mark=triangle*, mark options={color0}]
    table [meta=label] {
    x y label
    3.909142725152433 13.41752672148342 1
    3.274047101298358 14.567002081676495 2
    14.690519777949334 64.10994451026495 3
    };

    \addplot[color0, thin, solid, name path=A] table[x index=0,y index=2]
    {./pics/microjets/data/cloud12500_utip_vs_maxrbdot_avg_tikz.dat};
    \addplot[color0, thin, solid, name path=B] table[x index=0,y index=3]
    {./pics/microjets/data/cloud12500_utip_vs_maxrbdot_avg_tikz.dat};
    \addplot [fill=color15, fill opacity=0.1] fill between[of=A and B];
    \addplot[mark=none, color0, very thick, densely dashed] table [x index=0, y index=1]
    {./pics/microjets/data/cloud12500_utip_vs_maxrbdot_avg_tikz.dat};
  \end{axis}
\end{tikzpicture}
\\
\begin{tikzpicture}[baseline]
  \pgfplotsset{set layers}% using layers
  \begin{axis}[
    set layers,
    grid=major,
    width=0.49\textwidth,
    %style={font=\large},
    xmin=0.500004, xmax=1.22705,
    ymin=6.86789214352, ymax=88.4447909721,
    xlabel=$R_\mathrm{B}(0)\units{[mm]}$,
    ylabel=$u_\mathrm{tip}\units{[m/s]}$,
    clip mode=individual,
    ]
    \addplot[only marks, mark=*, mark size=1.0pt, color15, %mark options={solid,thick,fill=color4, opacity=0.6},
    line width= 0pt,
    opacity=0.3,
    ]
    table [x index=0,y index=1]
    {./pics/microjets/data/cloud12500_utip_vs_rb_scatter.dat};
    \addplot+[nodes near coords, nodes near coords style={anchor=west,font=\small},
    only marks, point meta=explicit symbolic, mark=triangle*, mark options={color0}]
    table [meta=label] {
    x y label
    0.58119 13.41752672148342 1
    0.657152 14.567002081676495 2
    1.13945 64.10994451026495 3
    };

    \addplot[color0, thin, solid, name path=A] table[x index=0,y index=2]
    {./pics/microjets/data/cloud12500_utip_vs_rb_avg_tikz.dat};
    \addplot[color0, thin, solid, name path=B] table[x index=0,y index=3]
    {./pics/microjets/data/cloud12500_utip_vs_rb_avg_tikz.dat};
    \addplot [fill=color15, fill opacity=0.1] fill between[of=A and B];
    \addplot[mark=none, color0, very thick, densely dashed] table [x index=0, y index=1]
    {./pics/microjets/data/cloud12500_utip_vs_rb_avg_tikz.dat};
  \end{axis}
\end{tikzpicture}
  \hfill%
\begin{tikzpicture}[baseline]
  \pgfplotsset{set layers}% using layers
  \begin{axis}[
    set layers,
    grid=major,
    width=0.49\textwidth,
    %style={font=\large},
    xmin=0.049150334753, xmax=61.2821487841,
    ymin=6.86789214352, ymax=88.4447909721,
    xlabel=$\theta\units{[deg]}$,
    ylabel=$u_\mathrm{tip}\units{[m/s]}$,
    clip mode=individual,
    ]
    \addplot[only marks, mark=*, mark size=1.0pt, color15, %mark options={solid,thick,fill=color4, opacity=0.6},
    line width= 0pt,
    opacity=0.3,
    ]
    table [x index=0,y index=1]
    {./pics/microjets/data/cloud12500_utip_vs_theta_scatter.dat};
    \addplot+[nodes near coords, nodes near coords style={anchor=west,font=\small},
    only marks, point meta=explicit symbolic, mark=triangle*, mark options={color0}]
    table [meta=label] {
    x y label
    9.794704416973353 13.41752672148342 1
    49.42719293291682 14.567002081676495 2
    12.643623377410279 64.10994451026495 3
    };

    \addplot[color0, thin, solid, name path=A] table[x index=0,y index=2]
    {./pics/microjets/data/cloud12500_utip_vs_theta_avg_tikz.dat};
    \addplot[color0, thin, solid, name path=B] table[x index=0,y index=3]
    {./pics/microjets/data/cloud12500_utip_vs_theta_avg_tikz.dat};
    \addplot [fill=color15, fill opacity=0.1] fill between[of=A and B];
    \addplot[mark=none, color0, very thick, densely dashed] table [x index=0, y index=1]
    {./pics/microjets/data/cloud12500_utip_vs_theta_avg_tikz.dat};
  \end{axis}
\end{tikzpicture}
  \caption{
  Microjet tip velocity depending on
  microjet initiation time $t_{\mathrm{tip},i}$,
  bubble compression rate $-\dot R_{\mathrm{B}_i,\mathrm{min}}$,
  bubble initial radius $R_{\mathrm{B}_i}(0)$ and
  inclination angle $\theta_i$ (from left to right and top to bottom).
  }
  \label{fig:microjet_utip}
\end{figure}

Fig.~\ref{fig:microjet_utip} displays scatter plots of the microjet velocity magnitude
depending on various quantities.
The velocity magnitude of the microjets increases with their time
of initiation.
For instance, the mean value amounts to
$10\units{m/s}$ for $t_{\mathrm{tip}}=80\units{\mu s}$
and to
$50\units{m/s}$ for $t_{\mathrm{tip}}=250\units{\mu s}$.
This behavior is consistent with the  acceleration of the cavitation collapse wave and the
growth of the pressure at the front.
One of the fastest microjets is observed for bubble ``3'' included in Fig.~\ref{fig:microjet_traj} and Table~\ref{tab:microjet_table}.
The scatter plot of the microjet velocity magnitude
versus the initial bubble radius $R_B(0)$ shows
that larger bubbles exhibit faster microjets.
The mean value rises from $20$ to $40\units{m/s}$
for bubbles with an initial radius between $0.5$ and $1.2\units{mm}$.
Another quantity relevant to the collapse strength of a bubble is the peak compression rate
$-\dot R_{\mathrm{B}_i,\min}$ 
which is
evaluated within the time interval $[t_{\mathrm{tip},i}, t_{\mathrm{imp},i}]$.
A positive correlation of the compression rate with the magnitude of the microjet velocity
is observed in Fig.~\ref{fig:microjet_utip}.
In contrast, the inclination angle $\theta_i$ does not affect
the magnitude of the microjet velocity.
The analyzed relations reveal that the microjet velocity is influenced by both
parameters of individual bubbles (e.g., the initial bubble radius)
and macroscopic parameters of the cloud collapse (e.g., the collapse front speed).
However, the overall large dispersion of these relations indicates the influence of
further factors such as the spatial configuration of the surrounding bubbles.
\begin{table}[tb]
  \caption{Microjet parameters of selected bubbles.}
  \begin{center}
    % bubble radial theta utip rb maxrbdot dtau indicator
    \begin{tabular}{l c c c c c c c}
      \hline
      \hline
      bubble\TTop\TBot & $r\units{[mm]}$ & $\theta\units{[deg]}$ & $u_\mathrm{tip}\units{[m/s]}$ & $R_\mathrm{B}(0)\units{[mm]}$ & $-\dot R_{\mathrm{B},\min} \units{[m/s]}$ & $\varphi\units{[deg]}$ & $|\hat{\mathbf u}^\perp_\mathrm{bulk}|$ \\
      \hline
      1\TTop & 41.9 & 9.8 & 13.4 & 0.58 & 3.9 & 50.6 & 0.005 \\
      2 & 41.4 & 49.4 & 14.6 & 0.66 & 3.3 & 22.9 & 0.293 \\
      3\TBot & 34.1 & 12.6 & 64.1 & 1.14 & 14.7 & 92.5 & 0.148 \\
      \hline
      \hline
    \end{tabular}
  \end{center}
  \label{tab:microjet_table}
\end{table}
 % \clearpage
%%%%%%%%%%%%%%%%%%%%%%%%%%%%%%%%%%%%%%%%%%%%%%%%%%%%%%%%%%%%%%%%%%%%%%%%%%%%%%%

%%%%%%%%%%%%%%%%%%%%%%%%%%%%%%%%%%%%%%%%%%%%%%%%%%%%%%%%%%%%%%%%%%%%%%%%%%%%%%%
%% 5. pressure pulse spectrum
% 5.1 identification of pressure pulses
% 5.2 pressure pulse rate
% 5.3 coverage rate
%
\section{Pressure pulse spectrum}
\label{sec:damage}
 
 While the propagation of the collapse wave through the cloud may also be well recovered by
 analytically simplified relations and homogeneous mixture approaches as shown in Sec.~\ref{sec:wave},
 predictions regarding the hydrodynamic aggressiveness
 of cloud cavitation collapse require capturing the pressure pulses generated by
 individual bubble collapses ~\cite{Kim:2014}.
 The pressure pulse spectrum describes the rate and strength of collapse-induced pressure
 pulses and may also be used to assess the erosive potential in case of wall-bounded cavitating flow.
 If the pressure pulses are continuously emitted close to a material surface,
 the resulting repeated loading acting on the surface causes
 failure of the material and mass loss.
 To experimentally capture collapse-induced pressure loads,
 high frequency pressure sensors placed at predefined locations are frequently used \cite{Franc:2011b,Singh:2013}.
 Alternatively, they are recovered from pitting tests with ductile
 materials ~\cite{Jayaprakash:2012b,Franc:2012a}.
 In contrast to experiments, numerical simulations provide spatially and temporally
 resolved information, enabling a more comprehensive picture.
 
 \subsection{Identification of pressure pules}
 
\begin{figure}[bt]
    \centering
\pgfkeys{/pgf/number format/.cd,1000 sep={\,}}
\begin{tikzpicture}[baseline]
\begin{axis}[
  grid=major,
  width=0.95\textwidth,
  %style={font=\large},
  width=16cm,
  height=5cm,
  ymin=0,ymax=350,
  xlabel=$t\units{[\mu s]}$,
  ylabel=$p\units{[MPa]}$,
  legend columns=2,
  legend style={at={(0.5,1.02)},anchor=south,draw=none,font=\small,
  /tikz/column 2/.style={column sep=6pt}},
  legend cell align=right
  ]

  %Pressure signal at two cell located at $r=6.1134$ and $0.0311\units{mm}$.
  \addplot[mark=*, mark size=1.0pt, line join=round, color0, thick, solid,restrict x to domain=290:345]
  table [x index=0, y index=1]
  {./pics/damage/data/cell_sensor_2032504_20.0.dat};
  \addlegendentry{$r=6.11\units{mm}$}
  
  \addplot[mark=*, mark size=1.0pt, line join=round, color4, thick, solid,restrict x to domain=290:345]
  table [x index=0, y index=1]
  {./pics/damage/data/cell_sensor_0.dat};
  \addlegendentry{$31.1\units{\mu m}$}

  \addplot[only marks, mark=o, mark size=3.0pt,  color8, thick, solid,restrict x to domain=290:345]
  table [x index=0, y index=1]
  {./pics/damage/data/cell_sensor_2032504_peaks.dat};  
  
  \addplot[only marks, mark=o, mark size=3.0pt,  color8, thick, solid,restrict x to domain=290:345]
  table [x index=0, y index=1]
  {./pics/damage/data/cell_sensor_0_peaks.dat};
  
%  # global peak 1.115648e+02
%  # time 3.099032e+02
  \addplot [mark=none, line join=round, color8, solid, thick]
  coordinates {(3.099032e+02,0) (3.099032e+02,1.115648e+02)};
  \node[anchor=west,inner sep=0, outer sep=0] (A) at (axis cs:313, 80){$p_\text{P}$};
  \node[inner sep=0, outer sep=0] (D) at (axis cs:3.099032e+02, 26){};
  \draw[-, solid, thick] (D) [out=0, in=190] to (A);
% # start 3.078300e+02
  \addplot [mark=none, line join=round, color8,thick, densely dashed]
  coordinates {(3.078300e+02,20) (3.078300e+02,205)};
% # end 3.109752e+02
 \addplot [mark=none, line join=round, color8,thick, densely dashed]
 coordinates {(3.109752e+02,20) (3.109752e+02,205)};
% # helper 3.094026e+02
 \node[anchor=south] (B) at (axis cs:3.094026e+02, 145){$\Delta t_\text{P}$};
 \node[inner sep=0, outer sep=0] (E) at (axis cs:3.078300e+02, 145){};
 \node[inner sep=0, outer sep=0] (F) at (axis cs:3.109752e+02, 145){};
 \draw[<->, solid, thick] (E)--(F);

 \node[anchor=west] (C) at (axis cs:290, 40){$p_\text{th}$};
 \addplot [mark=none, line join=round, color15, thick, dash dot dot] coordinates {(290,20) (345,20)};

\end{axis}
\end{tikzpicture}
     \caption{
    Pressure signal at two cells located at $r=6.11\units{mm}$ and $31.1\units{\mu m}$.
    Collected pressure peaks are
    marked by circles.  The threshold pressure is indicated by a horizontal line.
    Characteristic quantities of a pressure pulse are illustrated. 
    }
  \label{fig:cell_sensor}
 \end{figure}
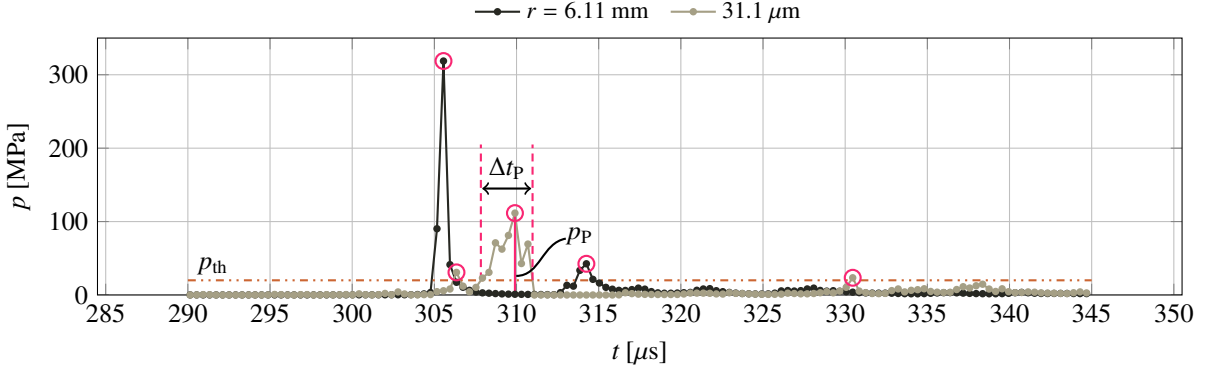

We evaluate the pressure pulse spectrum by identifying 
 individual pressure pulses  encountered in the center plane, i.e., the $xy$-plane at $z=0$,
 of the  cloud.
 This identification involves the following steps:
 \begin{itemize}
 \item Each grid cell of the center plane serves as a pressure sensor
          whose sampling frequency is $2.53\units{MHz}$,
          as illustrated in Fig.~\ref{fig:cell_sensor}.
 % dumpdt = 0.125 (Cubism units) = 0.39528470752104744 micro-second
 % 1/dumpdt(in sec) = 2529822.1281347037
 % which corresponds to 2.53MHz
          Data is collected up to the collapse time of the cloud, i.e., the time of minimum gas volume.
 \item For each cell, all pressure peaks are collected that are above a threshold pressure
          $p_\mathrm{th}=20\units{MPa}$; see also Fig.~\ref{fig:cell_sensor}.
          If several peaks are detected for a time interval with $p>p_\mathrm{th}$,
          only the highest peak value is kept,
          as also done for experimentally measured pressure signals~\cite{Singh:2013}. 
 \item At each sampling time, all cells exhibiting a peak pressure above $p_\mathrm{th}$
          are considered as potential locations of a pressure pulse.
 \item For a given sampling time, neighboring cells with a peak pressure
          above $p_\mathrm{th}$ are grouped together to form the footprint of a single pressure pulse
          acting at this time.
          The number of cells contributing to the footprint is denoted by~$n_{cf}$.
 \end{itemize}
 The center $\ve x_\mathrm{P}$ of the footprint of a pulse detected at time $t_\mathrm{P}$
 is given by the location of 
 the cell $n_i$ with the highest peak pressure value $p_\mathrm{P}=\max{p_{n_i}}$, where $i \in {1,...,n_{cf}}$.
 The extent of the footprint is defined as the equivalent diameter of the area covered
 by the $n_{cf}$ combined cells, i.e., $d_\mathrm{P}=2h\sqrt{n_{cf}/\pi}$. 
 The duration  $\Delta t_\mathrm{P}$ of
 the action of the pulse
 is given by the time period with  $p>p_\mathrm{th}$ at the cell with
 pressure $p_\mathrm{P}$; see also Fig.~\ref{fig:cell_sensor}.

Fig.~\ref{fig:pit_space} visualizes the pressure pulses detected in the
center plane during the collapse of the cloud.
Each pressure pulse is represented by
a circle with diameter $d_\mathrm{P}$ centered at $\ve x_\mathrm{P}$ and colored by $p_\mathrm{P}$.
\begin{figure}[!tbp]
  \centering
  \begin{minipage}[c]{0.49\textwidth}
  \centering
\begin{tikzpicture}[scale=1, trim axis left, trim axis right]
  \begin{axis}[
    width=\textwidth,
    scale=1.0,
    axis on top,
    grid=major,
    enlargelimits=false,
    xmin=-30, xmax=30, ymin=-30, ymax=30,
    xtick={-30,-20,...,30},
    ytick={-30,-20,...,30},
    xlabel=$x\units{[mm]}$,
    ylabel=$y\units{[mm]}$,
    ylabel near ticks]
    \addplotgraphicsnatural[xmin=-30, xmax=30, ymin=-30, ymax=30]
    {{./pics/damage/data/spatial_distribution_pits_pressure_coloring_closeup_red}.pdf};

  % marker
%  \node at (axis cs:6.0,1.0) {1};
%  %{\large 1};
%  \node at (axis cs:15.0,1.0) {2};
%  \node at (axis cs:24.0,1.0) {3};
  \node[draw,circle,inner sep=1.5pt] (A) at (axis cs:3.5,0.0) {1};
  \node[draw,circle,inner sep=1.5pt] (B) at (axis cs:10.5,0.0) {2};
  \node[draw,circle,inner sep=1.5pt] (C) at (axis cs:17.5,0.0) {3};
  \end{axis}
\end{tikzpicture}
     \caption{
    Pressure pulses visualized by colored circles with diameter $d_\mathrm{P}$.  
    Blue color corresponds to pressure values close
    to the threshold pressure and red color to highest pressures.
    }
    \label{fig:pit_space}
  \end{minipage}
  \hfill
  \begin{minipage}[c]{0.49\textwidth}
  \centering
\pgfkeys{/pgf/number format/.cd,1000 sep={\,}}
\begin{tikzpicture}[baseline]
\tikzset{mark size=1.5}
\begin{axis}[
    grid=major,
    width=0.9\textwidth,
    %style={font=\large},
    xlabel=$r\units{[mm]}$,
    ylabel=$n_\text{P}$,
    ymin=0,
    ymax=280,
    xmin=-2.5,xmax=47.5,
    try min ticks=6,
]

  \addplot [color15,fill=color15,fill opacity=0.6, thin,
    hist={
        %bins=30,
        bins=45,
        %bins=24,
        data min=0.0,
        data max=45.0
    }   
  ] table [y index=7] {./pics/damage/data/pits.dat};

\end{axis}
\begin{axis}[
    width=0.9\textwidth,
    %style={font=\large},
    hide x axis,
    axis y line*=right,
    ylabel=$\frac{\langle m_\text{P} \rangle}{\langle m_\text{P} \rangle_\text{max}}\text{ with } m \in \{p\text{,}d\text{,}\Delta t\}$,
    legend columns=3,
    legend style={at={(0.5,1.02)},anchor=south,draw=none,font=\small,
    /tikz/column 2/.style={column sep=6pt},
    /tikz/column 4/.style={column sep=6pt}},
    legend cell align=right,
    ymin=0,
    ymax=280,
    ytick={0,50,...,250},
    yticklabels={0,0.25,0.5,0.75,1.0,1.25},
    xmin=-2.5,xmax=47.5,
]

  \addplot[mark=triangle*, color0, thick] table [x index=0, y expr=\thisrowno{5}*200]
  {./pics/damage/data/radial_histogram_quantities.dat};
  \addlegendentry{$\frac{\langle p_\text{P} \rangle}{\langle p_\text{P} \rangle_\text{max}}$}
  
  \addplot[mark=*, color0, thick] table [x index=0, y expr=\thisrowno{4}*200]
  {./pics/damage/data/radial_histogram_quantities.dat};
  \addlegendentry{$\frac{\langle d_\text{P} \rangle}{\langle d_\text{P} \rangle_\text{max}}$}
  
   \addplot[mark=square*, color0, thick] table [x index=0, y expr=\thisrowno{6}*200]
  {./pics/damage/data/radial_histogram_quantities.dat};
  \addlegendentry{$\frac{\langle \Delta t_\text{P} \rangle}{\langle \Delta t_\text{P} \rangle_\text{max}}$}
  
%   \addplot[mark=none, color0, thick,dashed] table [x index=0, y expr=\thisrowno{7}*80]
%  {./pics/damage/data/radial_histogram_quantities.dat};

  \addplot [mark=none, color0, thick, dashed] coordinates {(0,0) (0,280)};
  \addplot [mark=none, color0, thick, dashed] coordinates {(7,0) (7,280)};
  \addplot [mark=none, color0, thick, dashed] coordinates {(14,0) (14,280)};
  \addplot [mark=none, color0, thick, dashed] coordinates {(21,0) (21,280)};
  \node[draw,circle,inner sep=2pt] (A) at (axis cs:3.5,250) {1};
  \node[draw,circle,inner sep=2pt] (B) at (axis cs:10.5,250) {2};
  \node[draw,circle,inner sep=2pt] (C) at (axis cs:17.5,250) {3};

\end{axis}
\end{tikzpicture}
     \caption{Histogram of pressure pulse distribution versus the distance
    from the center of the cloud and
    normalized average values for the peak pressure $\langle p_\mathrm{P} \rangle$, 
    diameter $\langle d_\mathrm{P} \rangle$ and duration $\langle \Delta t_\mathrm{P} \rangle$.}
    \label{fig:histogram}
  \end{minipage}
\end{figure}
 The highest pressure in pulses is  observed close to the
 center of the cloud.
 The histogram in Fig.~\ref{fig:histogram} shows the number of pressure pulses
 depending on the radial distance along with 
 the corresponding average values $\langle d_\mathrm{P} \rangle$,
 $\langle p_\mathrm{P} \rangle$
 and $\langle \Delta t_\mathrm{P} \rangle$.
 All average quantities are normalized by their maximum values,
 which are $\langle d_\mathrm{P} \rangle_\mathrm{max}=0.69\units{mm}$,
 $\langle p_\mathrm{P} \rangle_\mathrm{max}=67.06\units{MPa}$ and
 $\langle \Delta t_\mathrm{P} \rangle_\mathrm{max} = 1.63\units{\mu s}$, respectively.
 %
 % d_max  0.6910049085 p_max 67.0607066292 t_max 1.62754971
 %
 The average diameter remains approximately constant up to a radial distance of
 $r=17\units{mm}$ and then rapidly decreases.  As also indicated by
 Fig.~\ref{fig:pit_space}, the pressure achieves its highest value near the
 center, decreases as long as $r<7\units{mm}$ and continues without changing
 significantly. A similar behavior is observed for the average pulse duration.
 The number of pulses quickly increases from the center reaching its maximum
 close to $r=7\units{mm}$ and declines afterwards.
 Therefore, the center region is exposed to few but strong pressure pulses whereas
 the majority of weaker pulses occurs away from the center. 

\subsection{Pressure pulse rate}

 A collapsing cloud of bubbles causes a rich distribution of pressure pulses. 
 For further analysis, the center region, where almost all pulses are encountered,
 is divided into three circular rings of equal thickness
 $\Delta r=7\units{mm}$ as illustrated in Figs.~\ref{fig:pit_space} and~\ref{fig:histogram}.
 As seen in Fig.~\ref{fig:histogram},
 the center-most ring~$1$ contains the strongest and widest pressure pulses,
 ring~$2$ in the middle summarizes the
 first part of the plateau region into which most of the pulses fall, and the outer-most ring~$3$
 is determined
 by the second part of the plateau region where the pulses are only very loosely packed.

 The detected pressure pulses in each ring~$1$ to~$3$ can be classified according to their
 peak pressure, the diameter of their footprints and the duration of their action. 
 The results are presented in the form of cumulative histograms of the pressure pulse rate
 as a function of the aforementioned quantities. 
 The cumulative pressure pulse rate $\dot N_\mathrm{P}$ is defined
 as the number of pulses per unit time and unit area whose peak pressure,
 footprint or duration exceeds a given threshold value.
 The cumulative pressure pulse rate of the simulation
 is obtained by dividing the number of pressure pulses, counted based on
 the criterion given above,  by the exposure time
 and the area of the region of interest. The exposure time is equal to the collapse time of the cloud, 
 i.e., the complete sampling period.
 The region of interest is given by the area of the circular ring 
 for which $\dot N_\mathrm{P}$ is evaluated.

\begin{figure}[tb]
  \centering
\pgfkeys{/pgf/number format/.cd,1000 sep={\,}}
\begin{tikzpicture}[baseline]
\tikzset{mark size=2.0}
\begin{axis}[
%\begin{semilogyaxis}[
  ymode=log,
  grid=major,
  width=0.49\textwidth,
  %style={font=\large},
  %xmin=20, %xmax=20,
  ymin=1,
xlabel=$p_\text{P}\units{[MPa]}$,
ylabel=$\dot N_{\mathrm{P},p}\units{[\frac{1}{mm^2s}]}$,
%  yticklabel style={
%            /pgf/number format/fixed,
%            /pgf/number format/precision=2,
%            /pgf/number format/fixed zerofill
%        },
  legend columns=2,
  legend style={at={(0.5,1.02)},anchor=south,draw=none,font=\small,
  /tikz/column 2/.style={column sep=6pt}},
  legend cell align=left,
  try min ticks=6,
  ]
  
\addplot[only marks, mark=triangle, color0,mark options={solid,thick,fill=color0},]
table [x index=0,y index=1]
{./pics/damage/data/cumulative_pit_pressure_shell1.dat};

\addplot[color0, thick,dash dot dot]
table [x index=0,y index=4]
{./pics/damage/data/cumulative_pit_pressure_fitted_shell1.dat};

\addplot[only marks, mark=square, color15,mark options={solid,thick,fill=color15},]
table [x index=0,y index=1]
{./pics/damage/data/cumulative_pit_pressure_shell2.dat};

\addplot[color15, thick,densely dashed]
table [x index=0,y index=4]
{./pics/damage/data/cumulative_pit_pressure_fitted_shell2.dat};

\addplot[only marks, mark=o, color8,mark options={solid,thick,fill=color8},]
table [x index=0,y index=1]
{./pics/damage/data/cumulative_pit_pressure_shell3.dat};

\addplot[color8, thick,solid]
table [x index=0,y index=4]
{./pics/damage/data/cumulative_pit_pressure_fitted_shell3.dat};

%\addlegendentry[color0]{simulation}
%\addlegendentry[color0]{fit}

  \node[draw,circle,inner sep=2pt] (A) at (axis cs:250,60) {1};
  \node[draw,circle,inner sep=2pt] (B) at (axis cs:215,4.5) {2};
  \node[draw,circle,inner sep=2pt] (C) at (axis cs:130,1.9) {3};

\end{axis}
\end{tikzpicture}
   \hfill
\pgfkeys{/pgf/number format/.cd,1000 sep={\,}}
\begin{tikzpicture}[baseline]
\tikzset{mark size=2.0}
\begin{axis}[
%\begin{semilogyaxis}[
  ymode=log,
  grid=major,
  width=0.49\textwidth,
  xmin=0, %xmax=20,
  ymin=0,
xlabel=$d_\text{P}\units{[mm]}$,
ylabel=$\dot N_{\mathrm{P},d}\units{[\frac{1}{mm^2s}]}$,
%  yticklabel style={
%            /pgf/number format/fixed,
%            /pgf/number format/precision=2,
%            /pgf/number format/fixed zerofill
%        },
  legend columns=2,
  legend style={at={(0.5,1.02)},anchor=south,draw=none,font=\small,
  /tikz/column 2/.style={column sep=6pt}},
  legend cell align=left
  ]
  
\addplot[only marks, mark=triangle, color0,mark options={solid,thick,fill=color0}]
table [x index=0,y index=1]
{./pics/damage/data/cumulative_pit_diameter_shell1.dat};

\addplot[color0, thick,dash dot dot]
table [x index=0,y index=4]
{./pics/damage/data/cumulative_pit_diameter_fitted_shell1.dat};

\addplot[only marks, mark=square, color15,mark options={solid,thick,fill=color15}]
table [x index=0,y index=1]
{./pics/damage/data/cumulative_pit_diameter_shell2.dat};

\addplot[color15, thick,densely dashed]
table [x index=0,y index=4]
{./pics/damage/data/cumulative_pit_diameter_fitted_shell2.dat};

\addplot[only marks, mark=o, color8,mark options={solid,thick,fill=color8}]
table [x index=0,y index=1]
{./pics/damage/data/cumulative_pit_diameter_shell3.dat};

\addplot[color8, thick,solid]
table [x index=0,y index=4]
{./pics/damage/data/cumulative_pit_diameter_fitted_shell3.dat};

%\addlegendentry[color0]{simulation}
%\addlegendentry[color0]{fit}

  \node[draw,circle,inner sep=2pt] (A) at (axis cs:4,100) {1};
  \node[draw,circle,inner sep=2pt] (B) at (axis cs:4,6) {2};
  \node[draw,circle,inner sep=2pt] (C) at (axis cs:1.85,3.3) {3};
  
\end{axis}
\end{tikzpicture}
\\
\pgfkeys{/pgf/number format/.cd,1000 sep={\,}}
\begin{tikzpicture}[baseline]
\tikzset{mark size=2.0}
\begin{axis}[
%\begin{semilogyaxis}[
  ymode=log,
  grid=major,
  %style={font=\large},
  width=0.49\textwidth,
  %xmin=20, %xmax=20,
  ymin=1,
xlabel=$\Delta t_\text{P}\units{[\mu s]}$,
ylabel=$\dot N_{\mathrm{P},\Delta t}\units{[\frac{1}{mm^2s}]}$,
%  yticklabel style={
%            /pgf/number format/fixed,
%            /pgf/number format/precision=2,
%            /pgf/number format/fixed zerofill
%        },
  legend columns=2,
  legend style={at={(0.5,1.02)},anchor=south,draw=none,font=\small,
  /tikz/column 2/.style={column sep=6pt}},
  legend cell align=left
  ]
  
\addplot[only marks, mark=triangle, color0,mark options={solid,thick,fill=color0}]
table [x index=0,y index=1]
{./pics/damage/data/cumulative_pit_duration_shell1.dat};

\addplot[color0, thick,dash dot dot]
table [x index=0,y index=4]
{./pics/damage/data/cumulative_pit_duration_fitted_shell1.dat};

\addplot[only marks, mark=square, color15,mark options={solid,thick,fill=color15}]
table [x index=0,y index=1]
{./pics/damage/data/cumulative_pit_duration_shell2.dat};

\addplot[color15, thick,densely dashed]
table [x index=0,y index=4]
{./pics/damage/data/cumulative_pit_duration_fitted_shell2.dat};

\addplot[only marks, mark=o, color8,mark options={solid,thick,fill=color8}]
table [x index=0,y index=1]
{./pics/damage/data/cumulative_pit_duration_shell3.dat};

\addplot[color8, thick,solid]
table [x index=0,y index=4]
{./pics/damage/data/cumulative_pit_duration_fitted_shell3.dat};

%\addlegendentry[color0]{simulation}
%\addlegendentry[color0]{fit}

  \node[draw,circle,inner sep=2pt] (A) at (axis cs:4.2,260) {1};
  \node[draw,circle,inner sep=2pt] (B) at (axis cs:2.0,80) {2};
  \node[draw,circle,inner sep=2pt] (C) at (axis cs:1.55,5) {3};

\end{axis}
\end{tikzpicture}
   \hfill
\pgfkeys{/pgf/number format/.cd,1000 sep={\,}}
\begin{tikzpicture}[baseline]
\begin{axis}[
  grid=major,
  width=0.49\textwidth,
  %style={font=\large},
  xmin=-2, xmax=2,
  ymin=-2, ymax=2,
  zmax=38,
  xlabel=$t\units{[\mu s]}$,
  ylabel=$r\units{[mm]}$,
  zlabel=$P\units{[MPa]}$,
  ztick={0,15,30},
%        yticklabel style={
%            /pgf/number format/fixed,
%            /pgf/number format/precision=2,
%            /pgf/number format/fixed zerofill
%        },
%        scaled y ticks=false,
  legend columns=2,
  legend style={at={(0.5,1.02)},anchor=south,draw=none,font=\small,
  /tikz/column 2/.style={column sep=6pt}},
  legend cell align=left,
colorbar,
%colorbar style={
%ylabel= {$I_\text{C}$},
%},
%colormap={mine}{rgb(0pt)=(0,1,1); rgb(63pt)=(1,0,1)},
%colormap={mine}{rgb=(0,1,1); rgb=(1,0,1)},
colormap={mine}{rgb255=(166,226,46);rgb255=(249,38,114)},
colorbar horizontal,
%colorbar style={font=\small},
%colorbar style={
%at={(0,1.2)},anchor=north west},
domain = 1:2,
view = {45}{45}
  ]

  \addplot3[surf,z buffer=sort,shader=flat,draw=black]
  %\addplot3[surf,z buffer=sort,shader=flat]
  %\addplot3[surf,z buffer=sort,shader=interp]
  table[x index=0,y index=1,z index=2] {./pics/damage/data/char_load.dat};
  % \addlegendentry{}

\end{axis}
\end{tikzpicture}
   \caption{
  Cumulative histograms of pressure pulse rate versus peak pressure $p_\mathrm{P}$, 
  diameter $d_\mathrm{P}$ and duration $\Delta t_\mathrm{P}$ for all rings
  as well as representative pressure pulse $P$ for the center-most ring~$1$
  (from left to right and top to bottom).
  Symbols correspond to simulation data and lines to fitted curves.
  }
  \label{fig:cum_hist_pit_size}
\end{figure}
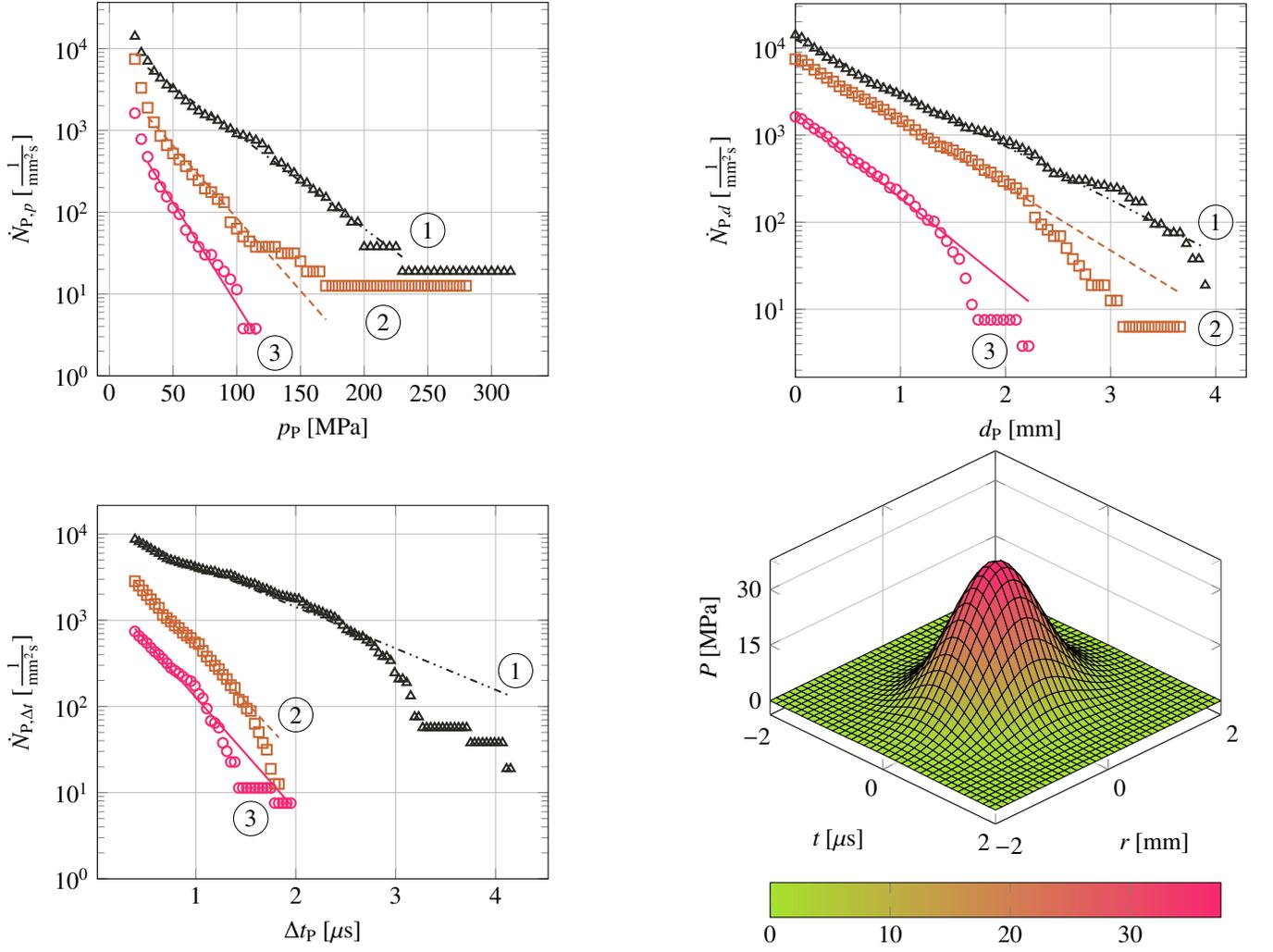

 Fig.~\ref{fig:cum_hist_pit_size} displays the cumulative pressure pulse rate versus
 the peak pressure $p_\mathrm{P}$, the diameter $d_\mathrm{P}$ and the duration
 $\Delta t_\mathrm{P}$. Data from all three rings are included.
It has been suggested  in~\cite{Franc:2011b,Singh:2013,Jayaprakash:2012b} that the cumulative
 pressure pulse rate can be approximated by an exponential function of the form
 \begin{equation}
 \label{eq:peak_rate}
 \dot N_{\mathrm{P},m}= \dot N_{\mathrm{P},m}^* \mathrm{e}^{- \frac{m_\mathrm{P}}{m_\mathrm{P}^*}},
 \end{equation}
 where $m \in \{p,d,\Delta t\}$.
 Exponent $-m_\mathrm{P}/m_\mathrm{P}^*$ may be additionally enhanced by a shape factor $\ell$
 leading to $-(m_\mathrm{P}/m_\mathrm{P}^*)^\ell$ %; see, e.g., 
 \cite{Jayaprakash:2012b,Singh:2013}. As done in \cite{Franc:2011b}, a simple exponential
 law (i.e., $\ell=1$) is applied since it already provides appropriate approximations to our data.
 The two parameters $\dot N_{\mathrm{P},m}^*$ and $m_\mathrm{P}^*$
 are a reference cumulative pressure pulse rate and a reference pressure, diameter
 or duration, respectively, that characterize the hydrodynamic aggressiveness of the collapsing cloud.
 %
 % see http://mathworld.wolfram.com/LeastSquaresFittingExponential.html
% To fit Eq.~\eqref{eq:peak_rate} to the data, we use a least-squares method of the from
% $\rho^2 = \sum y_i (\ln y_i - a - b x_i)^2$ where an exponential function $y=a\mathrm{e}^{bx}$
% and residual $\rho$ is assumed.
% which weights all data points equally.
 To fit Eq.~\eqref{eq:peak_rate}, which is of exponential form $y=a\mathrm{e}^{bx}$, to the data,
 a least squares method minimizing the residual $\sum y_i (\ln y_i - a - b x_i)^2$ is used.
 The fitted curves are also included in Fig.~\ref{fig:cum_hist_pit_size}, and
 the corresponding values for $\dot N_{\mathrm{P},m}^*$ and $m_\mathrm{P}^*$
 are provided in Tab.~\ref{tab:pit_height_fitting}.
\begin{table}[btp]
\caption{Regression parameters of the exponential function for the cumulative pressure pulse rate.}
\begin{center}
  \begin{tabular}{l c c c c c c}
    \hline
    \hline
     ring\TTop\TBot & $\dot N_{\mathrm{P},p}^*\units{[1/(mm^2 s)]}$ & $p_\mathrm{P}^*\units{[MPa]}$ & $\dot N_{\mathrm{P},d}^*\units{[1/(mm^2 s)]}$ & $d_\mathrm{P}^*\units{[mm]}$ & $\dot N_{\mathrm{P},\Delta t}^*\units{[1/(mm^2 s)]}$ & $\Delta t_\mathrm{P}^*\units{[\mu s]}$\\
    \hline
    $1$\TTop  & $12968$ & $37.56$ & $12551$ & $0.71$ & $12958$ & $0.90$ \\ 
    $2$  & $4996$ & $24.51$ & $7750$ & $0.59$ & $8729$ & $0.35$ \\ 
    $3$\TBot  & $2280$ & $17.48$ & $1794$ & $0.45$ & $2596$ & $0.33$ \\ 
    \hline
    \hline
  \end{tabular}
\end{center}
\label{tab:pit_height_fitting}
\end{table}
 All data sets are well captured by the exponential law
 for the smaller values of $m_\mathrm{P}$. 
 Both $\dot N_{\mathrm{P},m}^*$ and $m_\mathrm{P}^*$ decrease from
 the center-most ring~$1$ to the outer-most ring~$3$. 
 The present results for
 $d_\mathrm{P}$ and $p_\mathrm{P}$ compare well
  to the respective diagrams from other cavitation problems provided in
 \cite{Franc:2012a} (see Fig.~$5$ therein for the diameter) and \cite{Mihatsch:2015} (see Fig.~$15$
 therein for the pressure). Those diagrams also recover the exponential law
 for the smaller values of $m_\mathrm{P}$ and reveal some deviations for larger values.
 In particular, larger values of $m_\mathrm{P}$ correspond to rare events which are not fully 
 captured by only one collapse process. Furthermore, a maximum diameter
 has to be expected according to \cite{Kim:2014}  for $d_\mathrm{P}$ which explains the
 fast decrease of $\dot N_{\mathrm{P},d}$  for larger values.
 A similar fast decay is observed for $\dot N_{\mathrm{P},\Delta t}$ here.

The values for $p_\mathrm{P}^*$, $d_\mathrm{P}^*$ and $\Delta t_\mathrm{P}^*$
allow for recovering a representative pressure pulse $P$ for the present cloud collapse process.
Following~\cite{Singh:2013,Choi2014}, the encountered impact loads caused
by the collapse of cavitation bubbles
can be approximated by a Gaussian function in space and time as
\begin{equation}
\label{eq:char_pres}
  P= p_\mathrm{P}^*
                          \mathrm{e}^{- \left( \frac{t}{\Delta t_\mathrm{P}^*} \right)^2} 
                          \mathrm{e}^{- \left( \frac{r}{d_\mathrm{P}^*} \right)^2}.
\end{equation}
In \cite{Singh:2013}, the characteristic pressure and duration 
of Eq.~\eqref{eq:char_pres} were estimated by using a linear curve
fitted to the peak pressure versus the duration for a Gaussian function in time only.
Exemplarily, Fig.~\ref{fig:cum_hist_pit_size} displays the representative impact load
of ring~$1$.

\subsection{Coverage rate}

\begin{figure}[tb]
  \centering
\pgfkeys{/pgf/number format/.cd,1000 sep={\,}}
\begin{tikzpicture}[baseline]
\tikzset{mark size=2.0}
\begin{axis}[
  ymode=log,
  grid=major,
  width=0.49\textwidth,
  %style={font=\large},
  xmin=0, %xmax=20,
  ymin=0,
xlabel=$d_\text{P}\units{[mm]}$,
ylabel=$\beta\units{[\frac{1}{s}]}$,
%  yticklabel style={
%            /pgf/number format/fixed,
%            /pgf/number format/precision=2,
%            /pgf/number format/fixed zerofill
%        },
  legend columns=3,
  legend style={at={(0.6,1.02)},anchor=south,draw=none,font=\small,
  /tikz/column 2/.style={column sep=6pt}},
  legend cell align=right
  ]

  \addplot[only marks, mark=triangle, color0,mark options={solid,thick,fill=color0}]
  table [x index=0,y index=1]
  {./pics/damage/data/cumulative_pit_area_shell1.dat};

  \addplot[mark=none, color0, thick, dash dot dot] table [x index=0, y index=7]
  {./pics/damage/data/cumulative_pit_diameter_fitted_shell1.dat};
  %\addlegendentry{shell 1}

  \addplot[only marks, mark=square, color15,mark options={solid,thick,fill=color15}]
  table [x index=0,y index=1]
  {./pics/damage/data/cumulative_pit_area_shell2.dat};
  
  \addplot[mark=none, color15, thick,densely dashed] table [x index=0, y index=7]
  {./pics/damage/data/cumulative_pit_diameter_fitted_shell2.dat};
  %\addlegendentry{shell 2}
  
  \addplot[only marks, mark=o, color8,mark options={solid,thick,fill=color8}]
  table [x index=0,y index=1]
  {./pics/damage/data/cumulative_pit_area_shell3.dat};
  
  \addplot[mark=none, color8, thick] table [x index=0, y index=7]
  {./pics/damage/data/cumulative_pit_diameter_fitted_shell3.dat};
  %\addlegendentry{shell 3}
  
  \node[draw,circle,inner sep=2pt] (A) at (axis cs:4,1500) {1};
  \node[draw,circle,inner sep=2pt] (B) at (axis cs:4,70) {2};
  \node[draw,circle,inner sep=2pt] (C) at (axis cs:2.35,40) {3};

\end{axis}
\end{tikzpicture}
   \hfill
\pgfkeys{/pgf/number format/.cd,1000 sep={\,}}
\begin{tikzpicture}[baseline]
\begin{axis}[
  grid=major,
  width=0.49\textwidth,
  %style={font=\large},
  xmin=0, %xmax=20,
  ymin=0,
xlabel=$d_\text{P}\units{[mm]}$,
ylabel=$\text{PDF of }\beta\units{[\frac{1}{mm\,s}]}$,
%  yticklabel style={
%            /pgf/number format/fixed,
%            /pgf/number format/precision=2,
%            /pgf/number format/fixed zerofill
%        },
  legend columns=3,
  legend style={at={(0.6,1.02)},anchor=south,draw=none,font=\small,
  /tikz/column 2/.style={column sep=6pt}},
  legend cell align=right
  ]

  \addplot[mark=none, color0, thick, dash dot dot] table [x index=0, y index=10]
  {./pics/damage/data/cumulative_pit_diameter_fitted_shell1.dat};
  %\addlegendentry{shell 1}
  
  \addplot[mark=none, color15, thick,densely dashed] table [x index=0, y index=10]
  {./pics/damage/data/cumulative_pit_diameter_fitted_shell2.dat};
  %\addlegendentry{shell 2}
  
  \addplot[mark=none, color8, thick] table [x index=0, y index=10]
  {./pics/damage/data/cumulative_pit_diameter_fitted_shell3.dat};
  %\addlegendentry{shell 3}
  
  \node[draw,circle,inner sep=2pt] (A) at (axis cs:2.75,2600) {1};
  \node[draw,circle,inner sep=2pt] (B) at (axis cs:2,1750) {2};
  \node[draw,circle,inner sep=2pt] (C) at (axis cs:1.25,550) {3};

\end{axis}
\end{tikzpicture}
   \caption{
  Cumulative histogram of coverage rate (left) and its estimated PDF (right).
  Symbols correspond to simulation data and lines to fitted curves.
  }
  \label{fig:coverage}
\end{figure}
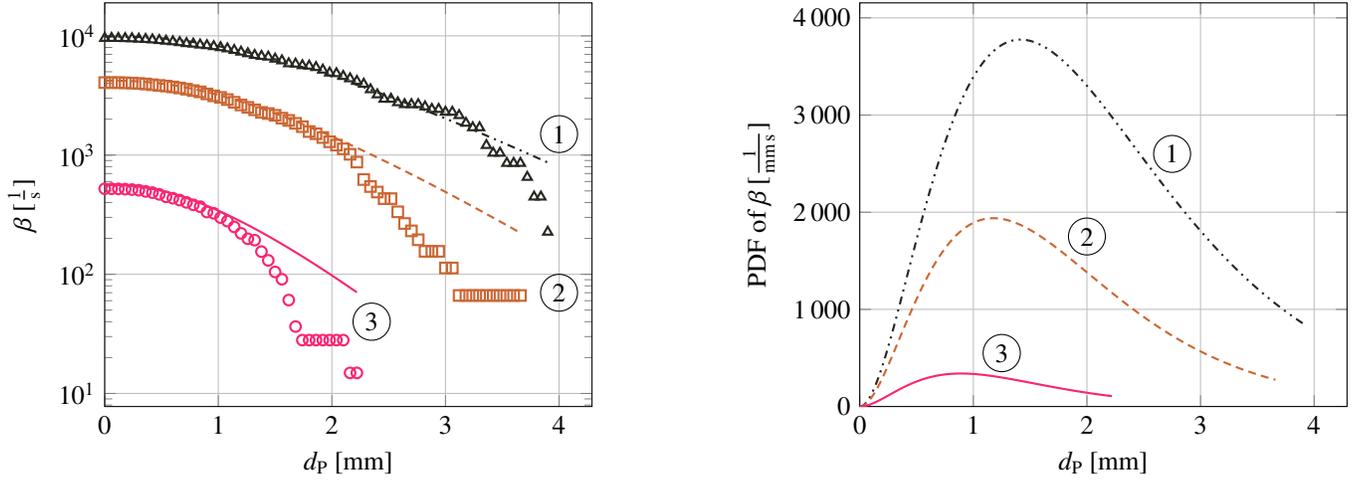

Using $\dot N_{\mathrm{P},d}^*=8/(\pi \delta^2 \tau)$ and
$d_\mathrm{P}^*=\delta/2$, Eq.~\eqref{eq:peak_rate}
may be rewritten as
\begin{equation}
\label{eq:pit_rate}
\dot N_\mathrm{P}= \frac{8}{\pi \delta^2 \tau} \mathrm{e}^{- \frac{2 d_\mathrm{P}}{\delta}},
\end{equation}
see~\cite{Franc:2012a}.
The values of the two parameters $\delta$ and $\tau$ are calculated from
$\dot N_{\mathrm{P},d}^*$ and $d_\mathrm{P}^*$ and are provided
in Tab.~\ref{tab:pit_diameter_fitting}.
\begin{table}[b]
 \caption{
 Characteristic parameters of the coverage rate.
 }
\begin{center}
  \begin{tabular}{l c c}
    \hline
    \hline
     ring\TTop\TBot & $\delta\units{[mm]}$ & $\tau\units{[ms]}$ \\
    \hline
    $1$\TTop  & $1.42$ & $0.10$ \\ 
    $2$  & $1.18$ & $0.24$ \\ 
    $3$\TBot  & $0.89$ & $1.79$ \\ 
    \hline
    \hline
  \end{tabular}
\end{center}
\label{tab:pit_diameter_fitting}
\end{table}
Parameter $\delta$ is a characteristic diameter of the footprint
and parameter $\tau$ a characteristic time. 
Both parameters are related to the cumulative coverage rate $\beta$ which is given by
\begin{equation}
\label{eq:beta}
\beta = \frac{1}{\tau} 
            \left( 1 + \frac{2 d_\mathrm{P}}{\delta} + \frac{2 d_\mathrm{P}^2}{\delta^2}  \right)
            \mathrm{e}^{- \frac{2 d_\mathrm{P}}{\delta}}
\end{equation}
and derived in~\cite{Franc:2012a}. 
The cumulative coverage rate $\beta$ is interpreted as the fraction of the center plane that
encounters pressure pulses with a footprint larger than a given $d_\mathrm{P}$.
When $d_\mathrm{P}$ approaches zero, $\beta$ goes to $1/\tau$. Hence,
$\tau$ denotes the coverage time and represents the time required to fully cover
the center plane by pressure pulses.
Fig.~\ref{fig:coverage} displays the cumulative coverage rate obtained from the simulation
together with the curves obtained from Eq.~\eqref{eq:beta} by inserting the values for
$\delta$ and $\tau$ given in Tab.~\ref{tab:pit_diameter_fitting}. The cumulative coverage rate
of the simulation is determined by, first, summing up the area of all impacts that
exhibit a footprint with a diameter larger than a specified threshold value and, second,
dividing by the exposure time and the area of the region of interest, as done for $\dot N_\mathrm{P}$.

The PDF of $\beta$ reads as
\begin{equation}
\label{eq:pdf_beta}
-\frac{\mathrm{d}\beta}{\mathrm{d} d_\mathrm{P}} = \frac{4}{\delta^3 \tau} d_\mathrm{P}^2
\mathrm{e}^{- \frac{2 d_\mathrm{P}}{\delta}}
\end{equation}
and enables an interpretation of $\delta$. 
It provides the contribution of the pressure pulses to the covered area
as a function of their diameter and is depicted in Fig.~\ref{fig:cum_hist_pit_size}.
The curves shown in Fig.~\ref{fig:cum_hist_pit_size}
are again obtained by inserting the parameters $\delta$ and $\tau$ 
given in Tab.~\ref{tab:pit_diameter_fitting} into Eq.~\eqref{eq:pdf_beta}.
The maximum value of the PDF of $\beta$
occurs for a pressure pulse with diameter $d_\mathrm{P}=\delta$.
Tab.~\ref{tab:pit_diameter_fitting} illustrates that $\delta$ decreases from the
center-most ring to the outer-most one.
Together with the increasing area of the outer rings,
more time is required to cover the entire ring by pressure pulses.
While $\tau$ for ring~$1$ and~$2$ is smaller than the collapse time of the cloud, a significantly
larger value is obtained for ring~$3$. Hence, ring~$1$ and~$2$ should be fully covered
by pressure pulses at least once.
As seen from Fig.~\ref{fig:pit_space}, this is the case for ring~$1$. For ring~$2$, some non-covered regions are
left towards its outer boundary, whereas many overlapping impacts occur towards its inner boundary, which
in turn contribute to cover the entire ring area. In contrast, mainly non-covered regions are observed for ring~$3$ consistent
with the large value for $\tau$.
 %\clearpage
%%%%%%%%%%%%%%%%%%%%%%%%%%%%%%%%%%%%%%%%%%%%%%%%%%%%%%%%%%%%%%%%%%%%%%%%%%%%%%%

%%%%%%%%%%%%%%%%%%%%%%%%%%%%%%%%%%%%%%%%%%%%%%%%%%%%%%%%%%%%%%%%%%%%%%%%%%%%%%%
%% 6. conclusions
%
\section{Conclusions}
\label{sec:conclu}

We have presented the results  from state-of-the-art simulations of the collapse of 
a spherical cloud of $12'500$ gas-filled bubbles, corresponding to a gas volume fraction of $4.9\%$.
This cloud composed by many small bubbles allows for proper averaging over the
global system and enables a large sample count for reliable statistics on the
scale of the bubbles.  To capture the dynamics of the bubbles, i.e., their
interactions and deformations, a diffuse interface finite volume method that
represents the bubbles on the computational grid has been applied.

Starting from a macroscopic point of view, we have examined the collapse
process which starts at the surface of the cloud and then propagates inward
focusing in the core of the cloud.  We have calculated spherical averages of
the gas-volume-fraction, pressure and velocity-magnitude fields and have
identified the collapse wave front. The collapse wave front advances in
accordance with simplified models such as M\o rch's ordinary differential
equation or homogeneous mixture approaches. In contrast to those simplified
models, the detailed simulation discloses the thickness of the collapse wave
front which is of the order of a few bubble diameters.
Furthermore, we have examined the bubbles individually. We have analyzed their
oscillation frequency and have used their deformation to recover the microjets.
We have shown that the oscillation frequency of the bubbles is governed by the
pressure of the collapse wave front at their location.  Our investigations have
revealed that the microjets do in general not exactly point towards the cloud
center. For the present cloud configuration, they are inclined to an angle up
to $50^{\circ}$ with respect to the radial direction.  Closer examinations have
demonstrated the correlation between this inclination and the bubble
distribution in the vicinity of the microjets. For the velocity at the tip of
the microjet, we have observed correlations with the radial location and the
size of the bubble from which the microjet has been extracted.
Eventually, we have evaluated the pressure pulse spectrum of the considered
cloud collapse process.  We have found that the region around the center
particularly suffers from many strong pressure pulses and have reconstructed
the corresponding characteristic pressure pulse. The pressure pulse rate is in
good agreement with an experimental law.
 %%%%%%%%%%%%%%%%%%%%%%%%%%%%%%%%%%%%%%%%%%%%%%%%%%%%%%%%%%%%%%%%%%%%%%%%%%%%%%%

%\vspace{-0.1cm}
\section*{Acknowledgment}

  %% INCITE, PRACE, CSCS
  We gratefully acknowledge a number of awards for computer time that made these
  large scale simulations possible. Computer time was provided by the Innovative and Novel
  Computational Impact on Theory and Experiment (INCITE) program under the
  project CloudPredict. This research used resources of the Argonne Leadership
  Computing Facility, which is a DOE Office of Science User Facility supported
  under Contract DE-AC02-06CH11357.  We acknowledge PRACE for awarding us
  access to  Fermi (CINECA, Italy)  with project Pra09\_2376 and
  Juqueen (J\"ulich Supercomputing Centre, Germany) with project PRA091.
  This work was also supported by a grant from the Swiss National
  Supercomputing Centre (CSCS) under project s500.  All provided
  computational resources are gratefully acknowledged.

  %\section*{References}
  \bibliographystyle{apsrev4-1}
{}

  %\listoffigures
  \end{document}